\documentclass[sn-standardnature]{sn-jnl}% Standard Nature Portfolio Reference Style
\jyear{2022}%

\theoremstyle{thmstyleone}%
\theoremstyle{thmstyletwo}%

\theoremstyle{thmstylethree}%

\begin{document}

\title{\vspace{-2.0cm} Combining acoustic bioprinting with AI-assisted Raman spectroscopy for high-throughput identification of bacteria in blood.}

\author*[1]{\fnm{Fareeha} \sur{Safir}}\email{fsafir@stanford.edu}
\author[2]{\fnm{Nhat} \sur{Vu}}\email{nhat@pumpkinseed.bio}
\author[3]{\fnm{Loza F.} \sur{Tadesse}}\email{lozat@mit.edu}
\author[4]{\fnm{Kamyar} \sur{Firouzi}}\email{kfirouzi@stanford.edu}
\author[5,6,7]{\fnm{Niaz} \sur{Banaei}}\email{nbanaei@stanford.edu}
\author[8]{\fnm{Stefanie S.} \sur{Jeffrey}}\email{ssj@stanford.edu}
\author*[9,10]{\fnm{Amr. A. E.} \sur{Saleh}}\email{aessawi@eng.cu.edu.eg}
\author[4,11]{\fnm{Butrus (Pierre)} T. \sur{Khuri-Yakub}}\email{pierreky@stanford.edu}
\author*[10,12]{\fnm{Jennifer A.} \sur{Dionne}}\email{jdionne@stanford.edu}

\affil*[1]{\orgdiv{Department of Mechanical Engineering}, \orgname{Stanford University}, \orgaddress{\city{Stanford}, \postcode{94305}, \state{CA}, \country{United States}}}

\affil[2]{\orgdiv{Pumpkinseed Technologies, Inc.},  \orgaddress{\city{Palo Alto}, \postcode{94306}, \state{CA}, \country{United States}}}

\affil[3]{\orgdiv{Department of Bioengineering}, \orgname{Stanford University School of Medicine and School of Engineering}, \orgaddress{\city{Stanford}, \postcode{94305}, \state{CA}, \country{United States}}}
\affil[]{\orgdiv{Present Address: Department of Electrical Engineering and Computer Science}, \orgname{University of California, Berkeley}, \orgaddress{\city{Berkeley}, \postcode{94720}, \state{CA}, \country{United States}}}

\affil[4]{\orgdiv{E. L. Ginzton Laboratory}, \orgname{Stanford University}, \orgaddress{\city{Stanford}, \postcode{94305}, \state{CA}, \country{United States}}}

\affil[5]{\orgdiv{Department of Pathology}, \orgname{Stanford University School of Medicine}, \orgaddress{\city{Stanford}, \postcode{94305}, \state{CA}, \country{United States}}}

\affil[6]{\orgdiv{Clinical Microbiology Laboratory}, \orgname{Stanford Health Care}, \orgaddress{\city{Palo Alto}, \postcode{94304}, \state{CA}, \country{United States}}}

\affil[7]{\orgdiv{Department of Infectious Diseases and Geographic Medicine}, \orgname{Stanford University School of Medicine}, \orgaddress{\city{Stanford}, \postcode{94305}, \state{CA}, \country{United States}}}

\affil[8]{\orgdiv{Department of Surgery}, \orgname{Stanford University School of Medicine}, \orgaddress{\city{Stanford}, \postcode{94305}, \state{CA}, \country{United States}}}

\affil[9]{\orgdiv{Department of Engineering Mathematics and Physics}, \orgname{Cairo University}, \orgaddress{\city{Cairo}, \postcode{12613}, \country{Egypt}}}

\affil[10]{\orgdiv{Department of Materials Science and Engineering}, \orgname{Stanford University}, \orgaddress{\city{Stanford}, \postcode{94305}, \state{CA}, \country{United States}}}

\affil[11]{\orgdiv{Department of Electrical Engineering}, \orgname{Stanford University}, \orgaddress{\city{Stanford}, \postcode{94305}, \state{CA}, \country{United States}}}

\affil[12]{\orgdiv{Department of Radiology, Molecular Imaging Program at Stanford (MIPS)}, \orgname{Stanford University School of Medicine}, \orgaddress{\city{Stanford}, \postcode{94035}, \state{CA}, \country{United States}}}

%\linenumbers
\maketitle

\section*{Abstract}

Identifying pathogens in complex samples such as blood, urine, and wastewater is critical to detect infection and inform optimal treatment. Surface-enhanced Raman spectroscopy (SERS) and machine learning (ML) can distinguish among multiple pathogen species, but processing complex fluid samples to sensitively and specifically detect pathogens remains an outstanding challenge. Here, we develop an acoustic bioprinter to digitize samples into millions of droplets, each containing just a few cells, which are identified with SERS and ML. We demonstrate rapid printing of 2 pL droplets from solutions containing \textit{S. epidermidis}, \textit{E. coli}, and blood;  when mixed with gold nanorods (GNRs), SERS enhancements of up to 1500x are achieved.We then train a ML model and achieve $\ge$99\% classification accuracy from cellularly-pure samples, and $\ge$87\% accuracy from cellularly-mixed samples. We also obtain $\ge$90\% accuracy from droplets with pathogen:blood cell ratios $<$1. Our combined bioprinting and SERS platform could accelerate rapid, sensitive pathogen detection in clinical, environmental, and industrial settings. 
\newline
\newline
\textbf{Keywords:} acoustic bioprinting, surface-enhanced Raman-Spectroscopy, machine learning, infectious disease, gold nanorods, bacteria

\section*{Main}

Reliable detection and identification of microorganisms is crucial for medical diagnostics, environmental monitoring, food production, biodefense, biomanufacturing, and pharmaceutical development. Samples can contain anywhere from $10^{6}$ colony-forming units (CFU)/mL to as few as 1-100 CFU/mL\cite{Kreger1980-eh,Werner1967-id,Gordon2010-mq}.Though \textit{in vitro} liquid culturing is typically used for pathogen detection, it is estimated that less than 2\% of all bacteria can be readily cultured using current laboratory protocols. Further, amongst that 2\%, culturing can take hours to days depending on the bacterial species \cite{Wade2002-ap,Bodor2020-zi,Hahn2019-my,Pedros-Alio2016-td}. In the case of diagnostics, broad spectrum antibiotics are often administered while waiting for culture results, leading to an alarming rise in antibiotic resistant bacteria\cite{Fleischmann2016-od}. We postulated that culture-free methods to detect pathogens in complex, multi-cellular samples might be possible by first digitizing samples into single-to-few-cellular droplets with bioprinting, then rapidly interrogating each droplet with Raman spectroscopy, and finally classifying the results using machine learning. 

Raman spectroscopy is a label-free, vibrational spectroscopic technique that has recently emerged as a promising platform for bacterial species identification\cite{Bantz2011-lr,Ho2019-en,Tian2022-mo,Tadesse2020-cd}. Since every cell species and strain has a unique molecular structure, they have a unique spectral fingerprint that can be used for identification\cite{Ho2019-en}. Compared to nucleic acid based tests such as polymerase chain reaction (PCR)\cite{Hoshino2004-vq,Rodriguez-Lazaro2003-bk,Greisen1994-hr} and protein based tests such as matrix-assisted laser desorption/ionization time-of-flight mass spectrometry (MALDI-TOF)\cite{Tsuchida2020-ht,Luethy2019-wf} and enzyme-linked immunoassay (ELISA)\cite{Dylla1995-bx,Sandstrom1986-fr}, Raman requires minimal-to-no use of reagents or labels, with relatively low-cost equipment and the potential for amplification-free detection\cite{Peters2004-zq,Klouche2008-dd,Murray2012-ik,Decuypere2016-yb}. Furthermore, Raman is a non-destructive technique, with excitation laser powers low enough for living cells\cite{Kneipp2006-up,Ou2020-hg} and negligible interference from water allowing for minimal sample preparation\cite{Atkins2017-oz}. Combined with plasmonic or Mie-resonant nanoparticles, Raman signals can be enhanced on average by 10\textsuperscript{5}-10\textsuperscript{6}, and up to 10\textsuperscript{10}\cite{Indrasekara2014-oo,Jackson2004-zf,Alonso-Gonzalez2012-ms}, allowing for rapid interrogation of cells. With these advantages, Raman has been successfully applied to genetic profiling\cite{De_Silva_Indrasekara2015-ym}, protein detection\cite{Guarrotxena2010-ca,Fabris2010-uc,Talamona2021-nl,Er2021-io}, and even single molecule detection \cite{Krug1999-pv,Xie2009-jz,Nie1997-vt}(Supplementary Note 1). More recent work has also shown exciting advances in Raman for cellular identification, including bacterial identification\cite{Ho2019-en,Tadesse2020-lm}, immune profiling\cite{Ramoji2021-ke,Pistiki2021-vu}, and in-vivo biopsies\cite{Balasundaram2021-it}. 

To advance Raman spectroscopy to clinical and industrial relevance, it must be combined with facile sample preparation methods. Nominally, the millions to billions of cells in milliLiter-scale volumes found in key target samples would need to be processed within seconds to minutes. Acoustic droplet ejection (ADE) is among the most promising droplet generation platforms for biological samples. In ADE, ultrasonic waves are focused at the fluid-air interface, giving rise to radiation pressure that ejects a droplet from the surface.  The diameter of the ejected droplet is inversely proportional to the frequency of the transducer, with 5 MHz and 300 MHz ultrasonic waves generating droplet diameters of 300 $\mu$m and 5 $\mu$m, respectively (Supplementary Fig. 1)\cite{Hadimioglu1992-sn,Elrod1989-wj}. Unlike other commercial piezo or thermal inkjet printers, the size, speed, and directionality of the ADE ejected droplets are completely controlled by the sound waves without the need for a physical nozzle\cite{Hadimioglu1992-sn}. As a nozzle-less technology, acoustic droplet ejection has an unparalleled advantage in handling biological samples; in particular, it eliminates clogging, sample contamination, and compromised cell viability or biomarker structure due to shear forces from the nozzle. Furthermore, ADE allows for high throughput droplet generation, processing fluids at rates of up to 25,000 droplets/s or approximately 50 nL/s for a single ejector head. Micro-electromechanical system (MEMS)-based arrays of 1024 ejector heads have been previously reported, showing potential for processing volumes over 180 mL in under an hour\cite{Hadimioglu2001-qi} as compared with the days required by existent microfluidic cell separation methods\cite{Shields2015-gw}. Additionally, as this platform relies on acoustic waves, these waves can propagate through a matched coupling media with minimal loss of acoustic energy while avoiding any direct contact between the sample and the transducer. This eliminates any cross-sample contamination and maintains sterility (Supplementary Note 2).

Here, we demonstrate a novel approach for rapid pathogen identification in complex, multi-cellular samples by combining Raman spectroscopy with acoustic droplet ejection. We develop a bioprinter to allow sub-5-picoLiter droplets, each consisting of a variety of cells printed with and without GNRs; thousands of droplets are printed within seconds (1kHz rates). To our knowledge, this is the first demonstration of stable and precise high-frequency (147 Mhz) acoustic printing of multi-component samples printed from both microscale biological entities (bacterial cells and RBCs) along with nanoscale particles (GNRs). We leverage this novel liquid-sample digitization method to facilitate high-throughput SERS identification of cells within individual droplets using advanced ML classification approaches. This approach allows us to sensitively and specifically detect individual cells within a complex liquid sample and gain insights about those cells.

We print samples of mouse red blood cells, suspended in an solution of aqueous ethylenediaminetetraacetic acid (EDTA), with spike-ins of gram-positive \textit{Staphylococcus epidermidis} (\textit{S. epi}) bacteria, and gram-negative \textit{Escherichia coli} (\textit{E. coli}), as well as gold nanorods (GNRs). Then, we collect Raman spectra from each printed droplet, using the optical signature to identify the cell constituents. We train machine-learning algorithms on samples printed from uniform cell types as well as mixed-cell samples to identify the droplet constituents. By optimizing our printing parameters, cell to nanorod concentrations, buffer solutions, and substrates, we achieve high Raman signal across cells while correctly identifying cell types in each droplet. We achieve cellular classification accuracies of $\ge$99\% from single cell-line prints and $\ge$87\% from mixed-pathogen samples, validated using scanning electron microscopy images of our droplets ton confirm the presence of particular cells. Furthermore, we identify key spectral bands for classification by determining wavenumber importance and confirm that these features correspond to biologically relevant components within our known cell lines. Our work lays a foundation for future SERS based bioprinting diagnostic platforms, paving the way for rapid, specific, sensitive, label-free, and amplification-free detection of live cells.

We built a Zinc Oxide 147 MHz transducer bonded to a quartz focusing lens with a focal distance of 3.5 mm. The transducer is encased in a stainless steel housing and mounted 3.5 mm above a machined stainless steel plate with a 1 mm diameter hole through which droplets are ejected downwards (Supplementary Fig. 2a). 200 $\mu$L of sample solution is pipetted between the transducer and this plate to fill the 3.5 mm focal distance of the transducer. The aperture is large enough to negate any nozzle-like effects, and the fluid is held in place against the transducer and the plate through surface tension (Fig. 1a). We position a motorized, programmable xy stage 1 mm beneath this plate, allowing for patterned ejection. The setup is monitored through a stroboscopic camera mounted opposite to an LED to evaluate droplet stability and ejection (Fig. 1b, Supplementary Fig. 2b, 3a, b, 4). After first experimenting with a range of frequencies and droplet diameters (Fig. 1c, Supplementary Fig. 1), we selected our 147 MHz transducer frequency with droplet diameters of $\sim$15 $\mu$m or $\sim$2.15 pL in volume, to match the order of magnitude of our cellular diameters. We found this volume allows us to print droplets with a number of cells in each droplet, while also maximizing Raman enhancement from GNR coating\cite{Micciche2018-ft,Talbot2012-jo}.

We synthesized GNRs with a longitudinal plasmon resonance of 960 nm, chosen to be used with an excitation wavelength of 785 nm to minimize background fluorescence. We synthesized GNRs to be close enough to our laser line to be excited by our laser, but red-shifted enough to minimize competitive extinction of the incident and Raman-scattered light\cite{Tadesse2020-lm, Sivapalan2013-fd, Van_Dijk2013-ni}(Fig. 1d, Supplementary Fig. 5, 6a, b). UV-vis absorption spectra and transmission and scanning electron micrographs (TEM and SEM) of the gold nanorod samples confirm the strong near-infrared plasmon resonance peak and reasonable sample monodispersity (Supplementary Fig. 5). All rods were coated in sodium oleate and hexadecyl(trimethyl)ammonium bromide (CTAB), which gives them a slight positive charge\cite{Tadesse2020-lm}, further increasing binding with our negatively charged bacteria\cite{Tadesse2020-lm,Berry2005-yg} and, to a lesser degree, the negatively charged RBCs\cite{Fernandes2011-ti,Hayashi2018-kx}.  

\begin{figure}[H]%
\centering
\includegraphics[width=0.9\textwidth]{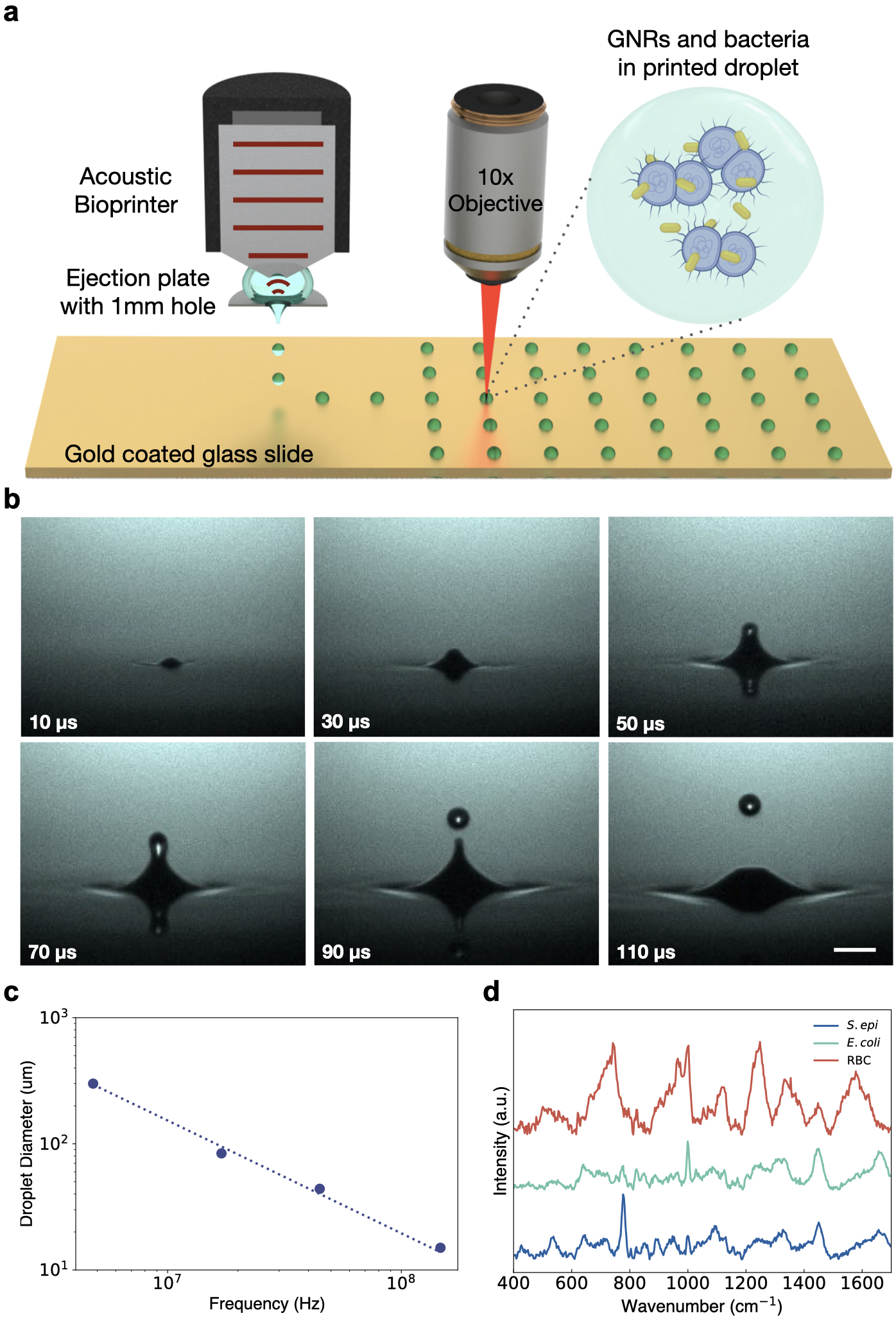}
\caption{(a) Schematic of acoustic printing platform and confocal Raman setup. Droplets  containing bacteria (purple) and nanorods (gold) suspended in EDTA solution are acoustically printed onto a glass slide coated in 200 nm of gold. (see also Supplementary Fig.2, 3, 4). (b) Stroboscopic images of the time evolution of upward droplet ejection at $\sim$3.5m/s from an open pool at an acoustic frequency of 44.75 MHz and a droplet ejection repetition rate of 1kHz. Images were captured with an exposure time of 40 ms, and as such, each frame is composed of 40 droplet ejections, highlighting ejection stability. Scale bar is 100 $\mu$m. (see also Supplementary Fig. 2). (c) Graph of droplet diameter versus ultrasound transducer resonant frequency. Droplets were printed with 4.8 MHz, 17 MHz, 44.75 MHz, and 147 MHz and had droplet diameters of 300 $\mu$m, 84 $\mu$m, 44 $\mu$m, and 15 $\mu$m respectively, highlighting the tunability of acoustic droplet ejection. (see also Supplementary Fig.1). (d) Raman spectra of dried cellular samples, including \textit{S. epi}, \textit{E. coli}, and red blood cells (RBCs) on  a gold coated slide.}\label{fig1}
\end{figure}

For this study, cells were suspended in a 1:9 volumetric mixture of EDTA and deionized water, diluted to a final concentration of 1e9 cells/mL. This solution was chosen to prevent hemolysis of our red blood cells (RBCs), while avoiding crystallization upon drying present in droplets printed from salt-based buffers (Supplementary Fig. 7). Furthermore, we have observed that the inclusion of EDTA provides a denser coating of GNRs on cell surfaces with few rods located elsewhere in the droplet (Supplementary Fig. 8)\cite{Sreeprasad2011-dz}; we hypothesize such coating is due to the interaction between the surface charge of our cells and the CTAB on the GNRS. Samples were printed on silane-treated, gold-coated glass substrates to minimize background spectra in the region of interest while further inducing coating of GNRs on our cells through their hydrophobicity (Supplementary Fig. 9, 10).

We tune the acoustic pulse width and input power into our transducer and ensure our printer is in focus for each sample to ensure that we can reliably and precisely print patterned grids of droplets containing bacteria and RBCs with GNRs and without GNRs, printed at ejection rates of 1 kHz, as shown in Fig. 2a. Grid prints of additional cell line mixtures can be found Supplementary Fig. 11.  Furthermore, we maintain cell viability during printing as demonstrated by the positive growth of cells printed directly onto agar-coated slides. Fig. 2b, for example, shows droplets of \textit{E. coli} bacteria grown 0, 12, 24, and 36 hours post printing, demonstrating the maintained viability of the cells after acoustic droplet ejection. 

\begin{figure}[H]%
\centering
\includegraphics[width=1\textwidth]{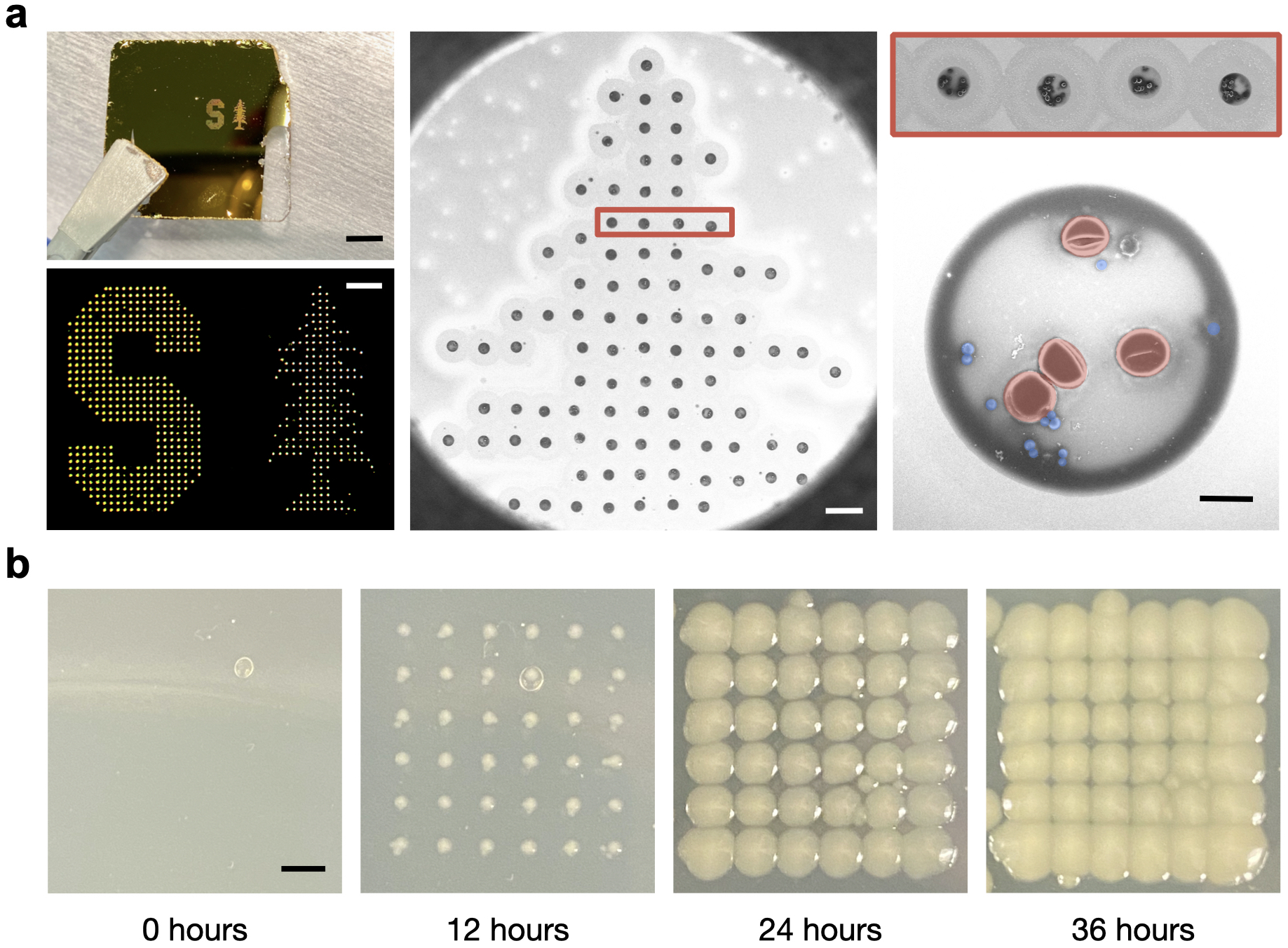}
\caption{\textbf{Patterned droplet ejection from cellular stock solution.} All droplets were ejected at 147 MHz. (a) Pattern printout of the Stanford University logo printed from droplets containing a 1:1 mixture of \textit{S. epi} bacteria and mouse RBCs onto a gold coated slide. The image on the left shows a photograph of print (top) with a scale bar of 4 mm. The brightfield image (bottom) was collected using a 5× objective lens and has a scale bar of 500 $\mu$m. Middle, SEM of the top portion of the tree region of the print with a scale bar of 100 $\mu$m. Right shows a single row of 4 droplets from the large area print, and then a magnified image of a single droplet with false coloring showing RBCs in red and \textit{S. epi} bacteria in blue. Scale bar is 5 $\mu$m. (b) Droplets containing E. coli bacteria were printed onto an agar coated slide and incubated at 37°C for upto 36 hours to demonstrate cellular viability of printed samples. 100 droplets were placed at each location to ensure each droplet contained cells. Scale bar is 2 mm. }\label{fig2}
\end{figure}

SERS spectra from our acoustically-printed droplets are collected using a 785 nm laser (Supplementary Fig. 12). We first print grids of droplets from 6 cellularly-pure samples: \textit{S. epi}, \textit{S. epi} with GNRs, \textit{E. coli}, \textit{E. coli} with GNRs, mouse RBCs, and mouse RBCs with GNRs (Fig. 3a, Supplementary Fig. 11). Fig. 3b shows a magnified SEM of the droplet printed with \textit{S. Epi} and GNRs and demonstrates that our cells are abundantly coated with GNRs. The normalized, average signal from 100 droplets of each cellular sample with GNRs and average signal from 15 droplets of each sample without GNRs are shown in Fig. 3c, with spectral acquisition times of 15 s for each droplet (Supplementary Fig. 13, 14, 15, 16). Note that little to no signal is observed with this collection for droplets without the nanorods. Relative signal intensities for non-normalized samples with data standard deviations can be found in Supplementary Fig. 17. While our work on cellular identification was performed using isolated RBCs, we demonstrate our platform’s ability to work on more complex samples by precisely printing droplets from mouse whole blood diluted with an anticoagulant (EDTA) and mixed with GNRs, without the need for any further sample pre-processing. We then collect Raman spectra from these droplets and show that we maintain the spectral peaks found in our pure RBC sample with the presence of additional peaks as would be expected of this more complex sample. Spectra and SEMs can be found in Supplementary Fig. 18.

Our data shows significant Raman signal enhancement from the sample sets with nanorods compared to the controls, estimated at between 300 - 1500x. For more precise classification of our droplet mixtures, we start by reducing the dimensionality of our spectra from 508 wavenumbers to 24 components using PCA in order to prevent classifier oversampling due to our dataset having more features than samples. We show that the first 24 principal components account for \textgreater90\% of our sample variance (Supplementary Fig. 19), and we still see clear sample differentiation between each dataset and cell type on a 2-component t-distributed stochastic neighbor embedding projection (t-SNE) after PCA (Fig. 3d, Supplementary Fig. 20). We then use a random forest classifier for our multiclass analysis from our complex samples. We tune our classifier hyperparameters using a cross-validated grid search to generate optimized parameters. Inputting these parameters into our classifier, we take 100 spectra from each of our 3 classes of cellular samples with GNRs and perform a stratified K-fold cross validation of our classifier’s performance across 10 splits and demonstrate $\ge$99\% classification accuracy across all samples (Fig. 3e).  

\begin{figure}[H]%
\centering
\includegraphics[width=1\textwidth]{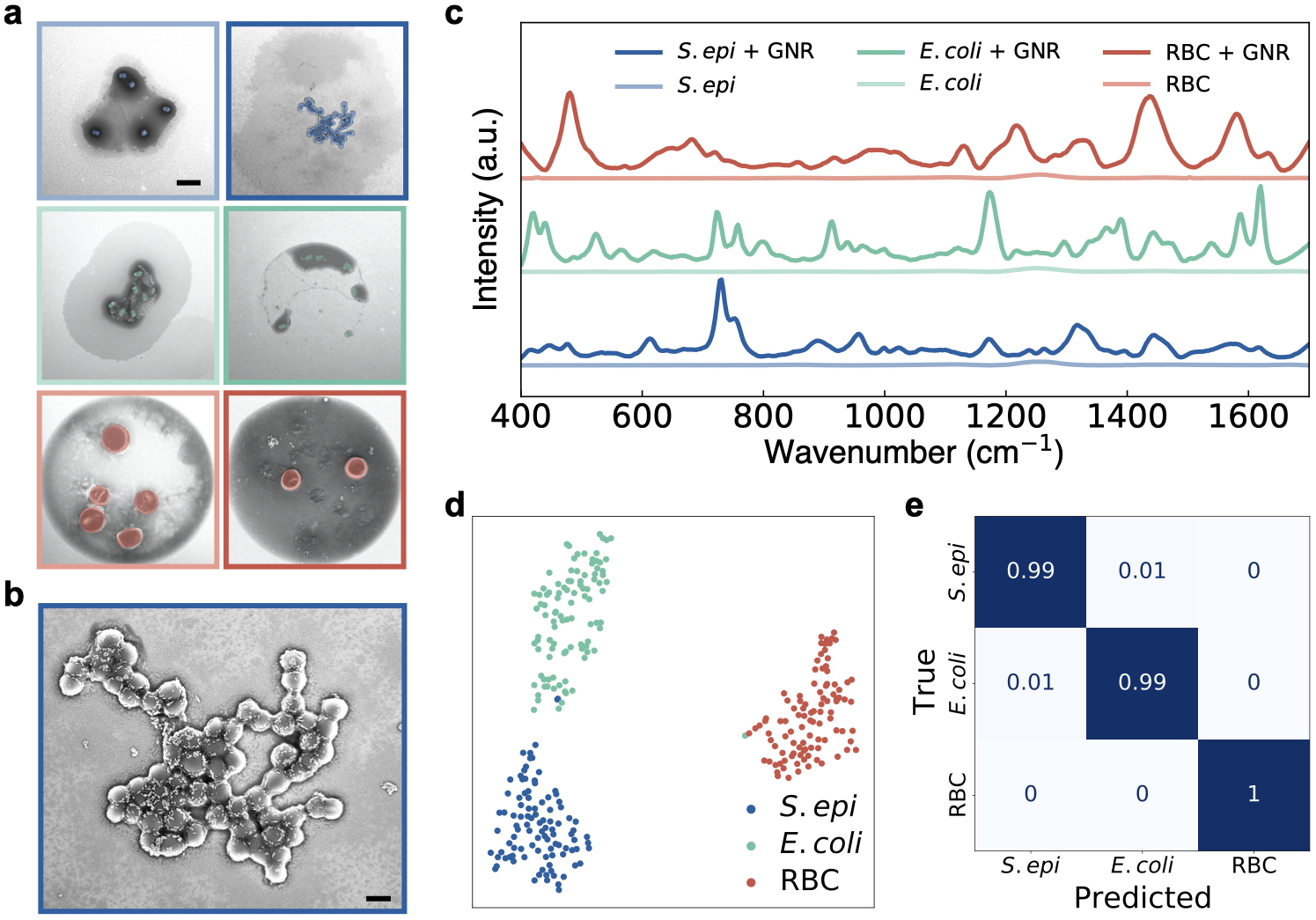}
\caption{\textbf{Spectral identification of cells printed with GNRs.} (a) SEMs showing single droplets printed from varying cellular samples suspended in our EDTA solution at a concentration of 1e9 cells/mL. Left column shows samples without GNRs, and the  right column shows cells printed with GNRs. From top to bottom, droplets contain: \textit{S. epi}, \textit{E. coli}, and RBCs with false coloring added to highlight the cells. The scale bar is 5 $\mu$m. (b) Magnified SEM of a droplet containing \textit{S. epi} coated with GNRs from Fig. 3a. SEM highlights that the bacteria are coated with GNRs with very few rods dispersed in the rest of the droplet. Scale bar 2 $\mu$m. (c) Mean SERS spectra of 100 measurements each taken from single droplets printed from three cell lines (\textit{S. epi}, \textit{E. coli}, and RBCs) mixed with GNRs. (d) 2-component t-SNE projection across all 300 Raman spectra acquired from droplets printed with GNRs. Data is plotted after performing a 24-component PCA for dimensionality reduction. Plots show distinct clustering of our cell lines. (e) Normalized confusion matrix generated using a random forest classifier on the 300 spectra collected from single cell-line droplets of \textit{S. epi}, \textit{E. coli}, and mouse RBCs mixed with GNRs. Samples were evaluated by performing a stratified K-fold cross validation of our classifier’s performance across 10 splits, showing $\ge$99\% classification accuracy across all samples.}\label{fig3}
\end{figure}

We demonstrate that we can accurately classify droplets printed at 147 MHz from complex, clinically-relevant cellular mixtures. We print arrays of droplets from 200 $\mu$L of solution formed from equal mixtures of \textit{S. epi}  and RBCs, \textit{E. coli} and RBCs, and  \textit{S. epi}, \textit{E. coli}, and RBCs, all diluted to a final concentration of 1e9 cells/mL of each cell type in our aqueous EDTA solution and mixed with GNRs (Fig. 4a).  We collect single-droplet SERS spectra from our mixture printouts, identically to that of our single cell-line droplets, using a 785 nm laser with a 15 s acquisition time. We then evaluate 100 spectra each of all six classes of our samples, the three single-cellular samples presented in Fig. 3 and our 3 mixture classes. We reduce the dimensionality of our samples to 30 components using PCA, sufficient to account for \textgreater90\% of our sample variance (Supplementary Fig. 21), and plot a 2-component t-SNE projection to show clear clustering between each dataset (Fig. 4b, Supplementary Fig. 22). We then re-tune our classifier hyperparameters with our new data, evaluate our samples using a random forest classifier with a stratified K-fold cross-validation as previously described, and demonstrate  $\ge$87\% classification accuracy across all samples (Fig. 4c).

To verify that our classifier is using physiologically meaningful spectral bands for prediction, we compute the feature importance at each wavenumber and validate that high importance bands correspond to specific biological components and vibrations in our cells. To identify these meaningful bands, we start by repeatedly splitting our 600 spectra into random 80:20 train/test splits and train a model on each training set. For the test set, we iterate through the wavenumbers and at each iteration, perturb the spectrum by modulating the amplitude with a Voigt distribution. After each perturbation, we recalculate the classification accuracy, compare the updated results with our baseline accuracy, and determine the importance for each wavenumber -  the greater the decrease in accuracy due to our perturbation, the more important the wavenumber. We split our samples using a stratified shuffle split and repeat 10 times. Each wavenumber of each spectrum in the test set is perturbed 5 times and all results are averaged to determine our final feature importance. We plot a heatmap highlighting the relative wavenumber importance overlaid with a plot of the mean and standard deviation of the perturbed classification accuracy (Fig. 4d, Supplementary Fig. 23, 24). We further plot the normalized, average signal from 100 droplets of each cellular sample with GNRs. Relative signal intensities for non-normalized samples with data standard deviations can be found in Supplementary Fig. 25. We note that the key spectral bands highlighted by our algorithm match peaks in our spectra and that these distinct peak wavenumbers represent bands previously reported in literature of dried and liquid SERS of our cell-lines including \textit{S. epi}, \textit{E. coli}, and RBCs (Fig. 4e, Supplementary Fig. 26)\cite{Tadesse2020-lm,Su2015-wa,Moghtader2018-lh,Witkowska2019-ox,Wang2010-hr,Sivanesan2014-er,Choi2020-fu,Zhou2014-ns,Drescher2013-ac,Premasiri2012-vb,Reokrungruang2019-qj,Paccotti2018-eo}. We specifically note that peaks at 732.5 and 1330 cm\textsuperscript{-1} from our \textit{S. epi}-containing samples are attributed to purine ring-breathing modes\cite{Choi2020-fu} and the Adenine part of the flavin derivatives or glycosidic ring mode of polysaccharides\cite{Sivanesan2014-er}; peaks at 755 and 1450 cm\textsuperscript{-1} from our \textit{E. coli}-containing samples are attributed to Tryptophan ring breathing\cite{Paccotti2018-eo} and CH\textsubscript{2}/CH\textsubscript{3} deformation of proteins and lipids\cite{Witkowska2019-ox}; and peaks at 482 and 1224 cm\textsuperscript{-1} from our RBC-containing samples are attributed to $\gamma$12 out of plane deformation of porphyrin, a main component of hemoglobin\cite{Drescher2013-ac}, and $\nu$13 or $\nu$42 valence\cite{Premasiri2012-vb}. Further peak assignments can be found in Supplementary Table 1.

\begin{figure}[H]%
\centering
\includegraphics[width=0.56\textwidth]{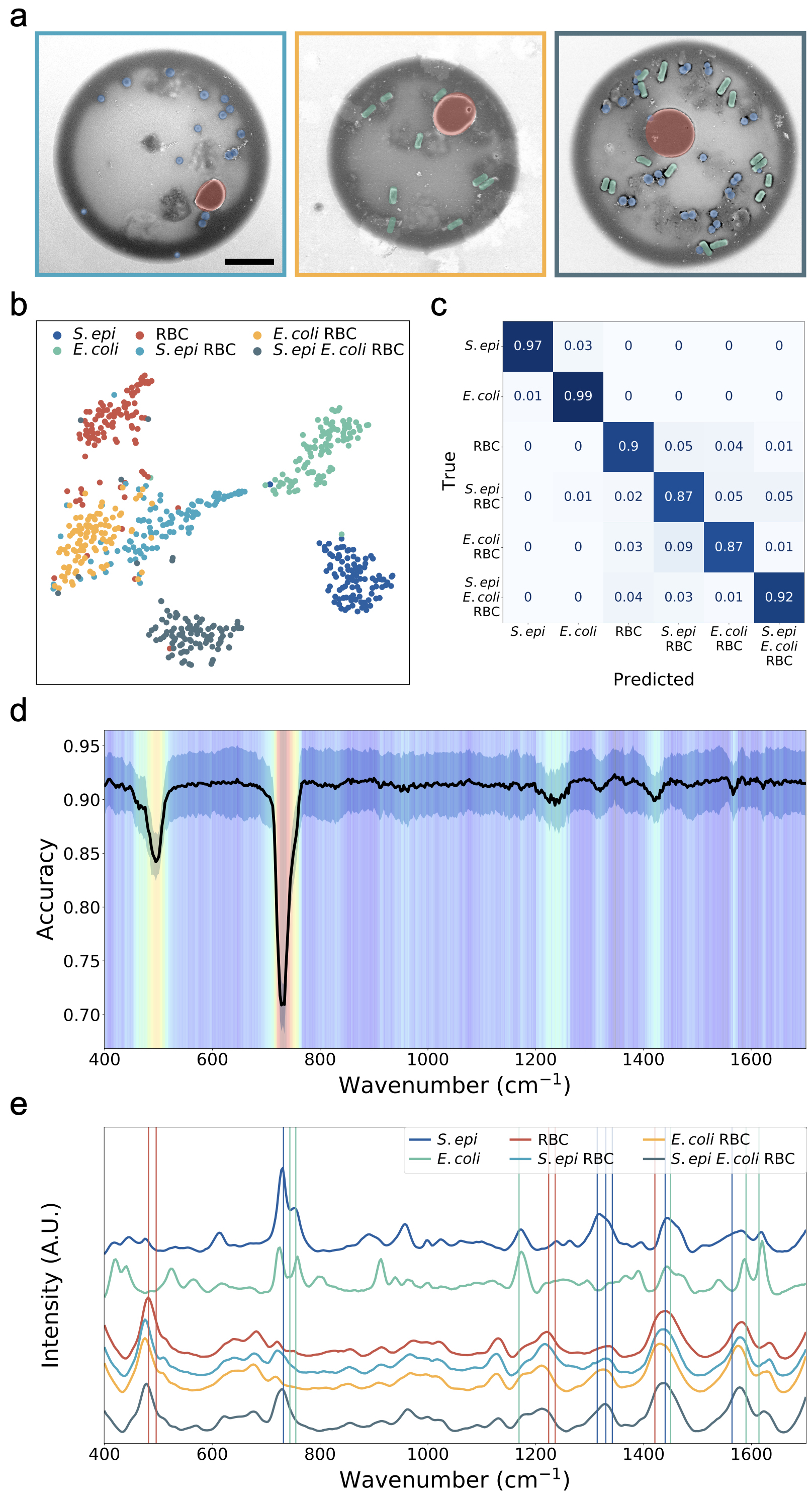}
\caption{(a) False-color SEMs of droplets printed from (left to right): an equal mixture of \textit{S. epi} bacteria and RBCs, \textit{E. coli} bacteria and RBCs, and \textit{S. epi}, \textit{E. coli}, and RBCs all diluted to 1e9 cells/mL in aqueous EDTA and mixed with GNRs. The scale bar is 5 $\mu$m. (b) 2-component t-SNE projection across all 600 Raman spectra acquired from 100 droplet measurements each, taken from single droplets printed from three cell lines (\textit{S. epi}, \textit{E. coli}, and RBCs) and three mixtures (\textit{S. epi} and RBCs, \textit{E. coli} and RBCs, and \textit{S. epi}, \textit{E. coli}, and RBCs) mixed with GNRs. Data is plotted after performing a 30-component PCA for dimensionality reduction. Plots show clustering of our cell lines with the most overlap between droplet mixture samples. (c) Normalized confusion matrix generated using a random forest classifier on the 600 spectra collected from single cell-line droplets of \textit{S. epi}, \textit{E. coli}, and mouse RBCs mixed with GNRs, and our 3 cell mixtures. Samples were evaluated by performing a stratified K-fold cross validation of our classifier’s performance across 10 splits, showing $\ge$87\% classification accuracy across all samples. (d) Heatmap highlighting feature extraction performed to determine relative weight of spectral wavenumbers in our Random Forest classification. Heatmap is overlaid with a plot of the mean and standard deviation of the classification accuracy (black) calculated across all trials. Wavenumbers with lower accuracies are shown to be critical features as random perturbations are highly correlated with decreases in classification accuracy. (e) Plot of the mean SERS spectra of 100 measurements each, taken from single droplets printed from three cell lines (\textit{S. epi}, \textit{E. coli}, and RBCs)  and three mixtures (\textit{S. epi} and RBCs, \textit{E. coli} and RBCs, and  \textit{S. epi}, \textit{E. coli}, and RBCs) mixed with GNRs. Wavenumbers reporting to biological peaks found in SERS spectra of \textit{S. epi}, \textit{E. coli}, and RBCs are plotted as blue, green, and red vertical lines, respectively. Peak assignments can be found in Supplementary Table 1.}\label{fig4}
\end{figure}

While our Raman analysis was on samples printed with high concentrations of blood and bacteria, much lower cell concentrations should be detectable. For example, we can analyze droplets that contain a smaller number of \textit{S. epi} bacteria than RBCs. As shown in the Supporting Information (Supplementary Fig. 27), droplets that have pathogen:RBC ratios less than 1 still exhibit high classification accuracies of 90\%. Therefore, we can detect bacterial signal from a printed droplet even with low cell counts. Furthermore, Raman interrogation can also be performed within each droplet. As shown in Supplementary Figs. 28 and 29, we collect spectral maps across a single droplet. Even in these mappings, our ML classifier can identify the cell type of each individual spectra and predict the cellular makeup of each droplet. Additionally, we demonstrate that our machine learning algorithm for wavenumber importance can determine relevant feature bands by individual sample classes. We show that different bands carry differing weights for the classification of each sample class, with the largest differences present between bacterial and blood cells. With these results, we propose a vision for rapid, hyperspectral Raman imaging that allows for bacterial identification without the need for full spectroscopic analysis of each droplet. We propose to separately image an array of printed droplets at the bands that have the greatest feature importance for blood and bacteria, respectively. Droplets containing only red blood cells would “light up” or have high intensity at one wavelength while those few droplets containing a mixture of bacteria and red blood cells would “light up” at a different wavelength, allowing for rapid identification of droplets containing bacteria using this hyperspectral Raman imaging technique (Supplementary Fig. 30).

We have demonstrated a rapid platform for acoustic printing-based droplet SERS of biological samples. Our system enables rapid digitization of cells from fluid samples in picoliter droplets with minimal sample contamination through nozzle-free acoustic printing at kilohertz ejection rates. As a result of our choice in printer frequency, cell stock solution, and slide surface treatment, our platform generates droplets containing cells uniformly coated in GNRs. Our results show that we can stably print samples of cells with and without GNRs and can demonstrate clear signal enhancements of up to 1500x from the addition of our GNRs. Furthermore, from these droplets, we demonstrate single-droplet Raman interrogation and cellular identification in 15 seconds.  We show that we generate these consistent Raman spectra from gram-positive and gram-negative bacteria as well as from RBCs and can differentiate spectra. Finally, we demonstrate that we can identify distinct cell types present in droplets printed from a mixture of cell lines using machine learning algorithms. 

Our work could advance Raman-based clinical research, clinical diagnostics, and disease management. Minimally invasive, fluid-based biomarker detection is gaining traction for the development of new point-of-care systems. A reliable and automated biological acoustic printer coupled with SERS nanoparticles and Raman hyperspectral imaging could be used to separate, count, and identify various cell lines allowing for rapid, specific, and label-free cellular analysis. Furthemore, ADE-based SERS could be designed with an array of ejector heads to rapidly split large patient sample volumes, or a single-ejector could provide detailed analysis of a small volume, minimizing the use of expensive reagents. As such, ADE-based SERS could enable culture-free cellular identification and monitoring from samples with low concentrations or from samples with species that are difficult to culture, including circulating tumor cells (CTCs) for cancer screening and monitoring\cite{Zhang2020-er,Arya2013-mp,Jeffrey2019-ww},  CD4 levels for HIV monitoring\cite{Zhang2020-er,Glynn2013-er}, and strain specific identification of slow-growing \textit{Mycobacterium tuberculosis} for treatment planning\cite{Stockel2017-or,Khan2018-dl,Kaewseekhao2020-iz}. Additionally, given that acoustic printing is nozzle-free and contactless, ADE-based SERS could facilitate easy multiplexing of various patient samples or other relevant media as the ejector can easily scan across a number of different sample wells without risking contamination. Lastly, given the versatility of our substrates, colloidal GNRs, and printing platform, our system is not limited to processing cells but could easily be modified for use in detecting other biomarkers including small molecules and proteins, coupled with surface chemistry for labeled detection of nucleic acids, and used for low-volume interrogation of pharmaceutical samples in drug-development. Our work in integrating SERS cellular interrogation with acoustic bioprinting and machine learning provides a foundation for further research into rapid, cellular-based diagnostics, and paves the way for reliable, low-cost point-of-care diagnostics.

\section*{Author Information}
\textbf{Corresponding Authors}
\newline
\newline
\textbf{Fareeha Safir} - Department of Mechanical Engineering, Stanford University School of Engineering, Stanford, California 94305, United States
\newline
\textbf{Amr A. E. Saleh} - Department of Engineering Mathematics and Physics, Faculty of Engineering, Cairo University, Giza 12613, Egypt
\newline
\textbf{Jennifer Dionne} - Department of Materials Science and Engineering, Stanford University School of Engineering, Stanford, California 94305, United States;  Department of Radiology, Molecular Imaging Program at Stanford (MIPS)Stanford University School of Medicine, Stanford, California 94305, United States
\newline
\newline
\textbf{Authors}
\newline
\newline
\textbf{Nhat Vu} - Pumpkinseed Technologies, Inc., Palo Alto, CA 94306, United States
\newline
\textbf{Loza F. Tadesse} - Department of Electrical Engineering and Computer Science, University of California, Berkeley, Berkeley, CA 94720, United States
\newline
\textbf{Kamyar Firouzi} - E. L. Ginzton Laboratory, Stanford, CA 94305, United States 
\newline
\textbf{Niaz Banaei} - Department of Pathology, Stanford University School of Medicine, Stanford, CA 94305, United States; Clinical Microbiology Laboratory, Stanford Health Care, Palo Alto, CA 94304, United States; Department of Infectious Diseases and Geographic Medicine, Stanford University School of Medicine, Stanford, CA 94305, United States 
\newline
\textbf{Stefanie S. Jeffrey} - Department of Surgery Stanford University School of Medicine, Stanford, CA 94305, United States 
\newline
\textbf{Butrus (Pierre) Khuri-Yabub} - E. L. Ginzton Laboratory, Stanford, CA 94305, United States; Department of Electrical Engineering, Stanford University School of Engineering, Stanford, CA 94305, United States 

\section*{Competing interests}

The authors declare no competing financial interests.

\section*{Supporting Information}

The Supporting Information is available free of charge at http://pubs.acs.org.

\begin{itemize}
  \item Materials and Methods
  \item Supplementary Note 1: Gold Nanorods for SERS applications
  \item Supplementary Note 2: Acoustic printing for handling biological samples
  \item Supplementary Figures S1-S31: supplementary schematics of experimental setup, gold nanorod characterization, Raman spectra of control samples, mean and standard deviation of acquired Raman spectra, Raman spectra of mouse whole blood samples, t-SNE plots of data before and after PCA dimensionality reduction, feature validation, classification of samples with a greater number of red blood cells than bacteria, intradroplet Raman mapping, proposed vision for Raman hyperspectral imaging, and plot of spectral preprocessing
  \item Table S1: Tentative band assignments of the SERS spectra of S. epi, E. coli, RBCs as reported in literature  
\end{itemize}

\section*{Acknowledgments}

The authors gratefully acknowledge funding from the Stanford Catalyst for Collaborative Solutions (funding ID 132114), the Chan Zuckerberg Biohub Investigator Program, the NIH-NCATS-CTSA (Grant number UL1TR003142), the Gates Foundation (OPP 1113682), the National Science Foundation (grant number 1905209), the NIH New Innovator Award (1DP2AI152072-01), and from seed funds from the Stanford Center for Innovation in Global Health. Part of this work was performed at the Stanford Nano Shared Facilities (SNSF) and the Soft \& Hybrid Materials Facility (SMF) which are supported by the National Science Foundation and National Nanotechnology Coordinated Infrastructure under awards ECCS-2026822 and ECCS-1542152. The authors also thank Dr. Jack Hu for help with gold-coating substrates, Babatunde Ogunlade for help with piranha cleaning substrates, Rich Chin and Dr. Juliet Jamtgaard for assistance with SEM sample coating, and Hongquan Li, Dr. Jack Hu, Dr. Halleh Balch, Dr. Jeong Kim, and Shoaib Meenai for insightful discussions.  

\section*{Author Contributions}

F.S., L.T., A.A.E.S., B.K-Y., and J.A.D. conceived and designed the experiments. F.S., K.F., A.A.E.S, and B.K-Y. designed, developed, and characterized the acoustic bioprinter. F.S. synthesized the gold nanorods, cultured the cells, printed samples, and collected Raman spectra of printed droplets. N.V. and F. S. wrote and implemented algorithms for spectral data pre-processing and classification. J.A.D., A.A.E.S, and B.K-Y supervised the project along with S.S.J and N.B. on relevant portions of the research. All authors contributed to the preparation of the manuscript.

% \newpage
% \begin{figure}
% \centering
% \includegraphics[width=8.25cm]{TOC.pdf}
% \caption{TOC Graphic}
% \end{figure}

\clearpage

%%===========================================================================================%
%% If you are submitting to one of the Nature Portfolio journals, using the eJP submission   %
%% system, please include the references within the manuscript file itself. You may do this  %
%% by copying the reference list from your .bbl file, paste it into the main manuscript .tex %
%% file, and delete the associated \verb+\bibliography+ commands.                            %
%%===========================================================================================%%
\bibliography{main_bibliography}% common bib file


\end{document}

\title{\vspace{-2.0cm} Supplementary Information: Combining acoustic bioprinting with AI-assisted Raman spectroscopy for high-throughput identification of bacteria in blood.}

\author*[1]{\fnm{Fareeha} \sur{Safir}}\email{fsafir@stanford.edu}
\author[2]{\fnm{Nhat} \sur{Vu}}\email{nhat@pumpkinseed.bio}
\author[3]{\fnm{Loza F.} \sur{Tadesse}}\email{lozat@mit.edu}
\author[4]{\fnm{Kamyar} \sur{Firouzi}}\email{kfirouzi@stanford.edu}
\author[5,6,7]{\fnm{Niaz} \sur{Banaei}}\email{nbanaei@stanford.edu}
\author[8]{\fnm{Stefanie S.} \sur{Jeffrey}}\email{ssj@stanford.edu}
\author*[9,10]{\fnm{Amr. A. E.} \sur{Saleh}}\email{aessawi@eng.cu.edu.eg}
\author[4,11]{\fnm{Butrus (Pierre)} T. \sur{Khuri-Yakub}}\email{pierreky@stanford.edu}
\author*[10,12]{\fnm{Jennifer A.} \sur{Dionne}}\email{jdionne@stanford.edu}

\affil*[1]{\orgdiv{Department of Mechanical Engineering}, \orgname{Stanford University}, \orgaddress{\city{Stanford}, \postcode{94305}, \state{CA}, \country{United States}}}

\affil[2]{\orgdiv{Pumpkinseed Technologies, Inc.},  \orgaddress{\city{Palo Alto}, \postcode{94306}, \state{CA}, \country{United States}}}

\affil[3]{\orgdiv{Department of Bioengineering}, \orgname{Stanford University School of Medicine and School of Engineering}, \orgaddress{\city{Stanford}, \postcode{94305}, \state{CA}, \country{United States}}}
\affil[]{\orgdiv{Present Address: Department of Electrical Engineering and Computer Science}, \orgname{University of California, Berkeley}, \orgaddress{\city{Berkeley}, \postcode{94720}, \state{CA}, \country{United States}}}

\affil[4]{\orgdiv{E. L. Ginzton Laboratory}, \orgname{Stanford University}, \orgaddress{\city{Stanford}, \postcode{94305}, \state{CA}, \country{United States}}}

\affil[5]{\orgdiv{Department of Pathology}, \orgname{Stanford University School of Medicine}, \orgaddress{\city{Stanford}, \postcode{94305}, \state{CA}, \country{United States}}}

\affil[6]{\orgdiv{Clinical Microbiology Laboratory}, \orgname{Stanford Health Care}, \orgaddress{\city{Palo Alto}, \postcode{94304}, \state{CA}, \country{United States}}}

\affil[7]{\orgdiv{Department of Infectious Diseases and Geographic Medicine}, \orgname{Stanford University School of Medicine}, \orgaddress{\city{Stanford}, \postcode{94305}, \state{CA}, \country{United States}}}

\affil[8]{\orgdiv{Department of Surgery}, \orgname{Stanford University School of Medicine}, \orgaddress{\city{Stanford}, \postcode{94305}, \state{CA}, \country{United States}}}

\affil[9]{\orgdiv{Department of Engineering Mathematics and Physics}, \orgname{Cairo University}, \orgaddress{\city{Cairo}, \postcode{12613}, \country{Egypt}}}

\affil[10]{\orgdiv{Department of Materials Science and Engineering}, \orgname{Stanford University}, \orgaddress{\city{Stanford}, \postcode{94305}, \state{CA}, \country{United States}}}

\affil[11]{\orgdiv{Department of Electrical Engineering}, \orgname{Stanford University}, \orgaddress{\city{Stanford}, \postcode{94305}, \state{CA}, \country{United States}}}

\affil[12]{\orgdiv{Department of Radiology, Molecular Imaging Program at Stanford (MIPS)}, \orgname{Stanford University School of Medicine}, \orgaddress{\city{Stanford}, \postcode{94035}, \state{CA}, \country{United States}}}

%\linenumbers
\maketitle

\renewcommand\figurename{Supplementary Fig.}% defined as per springer style 

\section*{Methods}

\subsubsection*{Gold nanorod synthesis and characterization}\label{subsec1}

Hexadecyl(trimethyl)ammonium bromide (CTAB) and sodium oleate (NAOL) coated gold nanorods were synthesized following previously described protocols\cite{Ye2013-us}. The nanorods were cleaned by centrifuging 1.5 mL aliquots twice at (9000 rpm, 20 min), allowing for one wash after synthesis as this has been shown to be adequate to maintain cell viability while preventing nanorod aggregation\cite{Tadesse2020-lm}. Samples were concentrated down to 10 µL to be mixed with cell samples and diluted to a final volume of 200 $\mu$L. Absorption spectra were recorded using a Cary 5000 UV-vis-NIR spectrometer. Scanning electron microscopy images were taken using FEI Magellan 400 XHR Scanning Electron Microscope (SEM). Transmission electron microscopy images were taken using FEI Tecnai G2 F20 X-TWIN Transmission Electron Microscope (TEM). 

\subsubsection*{Scanning electron microscopy (SEM) of printed samples}\label{subsec2}

For scanning electron microscope (SEM) imaging, printed samples were imaged after completion of all Raman Spectroscopy. Samples were prepared by evaporating a $\sim$10 nm layer of 60:40 gold to palladium to allow for better visualization of cells under electron beam illumination. SEM images were taken using FEI Magellan 400 XHR Scanning Electron Microscope.

\subsubsection*{Bacteria culturing and preparation}\label{subsec3}

\textit{E. coli}, ATCC 25922, and \textit{S. epidermidis}, ATCC 12228, were grown from frozen stocks on Trypticase Soy Agar 5\% Sheep Blood 221239 BD plates. A single colony was seeded in 10 mL Lysogeny broth (LB) culture medium and incubated at 37°C shaking at 300 rpm for 15 hrs using Thermo Scientific MaxQ 4450 incubator. 1.5 mL of culture was washed with water three times at 6000 rpm for 3 min using a mySPINTM 6 Mini Centrifuge. Samples were then concentrated down to 100 µL volumes. The cell count was collected using a Bright-Line Hemacytometer using a 1:5000 dilution of the cell culture stock solution. Stock solutions contained on average $\sim$1e10 cells/mL. 

\subsubsection*{Preparation of red blood cell solutions}\label{subsec4}

For data collected with purified mouse red blood cells:  CD-1 (1CR) purified Mouse Red Blood Cells (RBCs) K2EDTA Gender Unspecified Pooled samples MSE00RBK2-0104095, were purchased from BioIVT in 5mL volumes. RBCs were diluted in a 1:9 v/v mixture of Invitrogen UltraPure 0.5 M EDTA, Invitrogen 15575020, to a final dilution of 1:5000 and cell counts were collected using a Nexcelom Cellometer X2 cell counter. 

For data collected with mouse whole blood: CD-1 (1CR) Mouse Whole Blood K2EDTA Gender Unspecified Pooled samples MSE00WBK2-0000627, were purchased from BioIVT in 5mL volumes. The whole blood were diluted in a 1:9 v/v mixture of Invitrogen UltraPure 0.5 M EDTA, Invitrogen 15575020, to a final dilution of 1:5000 and cell counts were collected using a Bright-Line Hemacytometer. Whole blood contained $\sim$1e10 cells/mL RBCs. 

\subsubsection*{Preparation of mixtures for printing}\label{subsec5}

Printing was completed using 200 $\mu$L of solution. All samples were diluted to a final volume of 200 $\mu$L in a 1:9 v/v mixture of Invitrogen UltraPure 0.5 M EDTA, Invitrogen 15575020, and Millipore water, unless otherwise noted. For samples with cells and no nanorods, a single concentrated cell solution or a mixture of cell solutions was diluted in aqueous EDTA to a final concentration of 1e9 cells/mL of each cell type in a given mixture. This concentration was chosen to ensure a majority of printed droplets contained at least 1 cell. For samples of cells mixed with nanorods, cleaned, concentrated nanorod solution was first mixed with concentrated cell stock solution, for our final concentration of 1e9 cells/mL per cell type, and then subsequently diluted with our aqueous EDTA solution. All solutions are mixed by inverting our microcentrifuge tubes a minimum of 10 times. 

EDTA was chosen as our sample buffer to avoid crystallization upon drying seen with PBS(Supplementary Fig. 6). On top of that, EDTA provides two further advantages for our printed samples. When EDTA-containing droplets dry on a hydrophobic substrate, a central region of aggregated EDTA, cells, and GNRs dries in a much smaller area than that of a full droplet, forcing the cells and GNRs into a much smaller volume, ensuring better coverage of the cells with GNRs (Supplementary Fig. 9).  Furthermore, we hypothesize that the EDTA induces aggregation among the GNRs due to the electrostatic interaction between any residual CTAB on our GNRs and the EDTA\cite{Sreeprasad2011-dz}, as demonstrated in Supplementary Fig. 7. We hypothesize that this clustering, when coupled with the addition of cells, allowed for greater quantities of nanorods to coat the cells, and led to the creation of SERS “hot spots” amongst the aggregated GNRs, providing strong enhancements\cite{Micciche2018-ft,Radziuk2015-ct}. 

Finally, we show that the addition of the EDTA and nanorods adds minimal Raman background noise (Supplementary Fig. 8, 9, 14), making it appropriate for our work in Raman cellular identification. Finally, to further minimize coffee-ring effects from nanorods upon droplet drying, we used vapor deposition to coat our gold-coated slides with a hydrophobic silane layer (3-Aminopropyl)triethoxysilane (APTMS) which allows for a more close packing of our GNRs, providing greater and more uniform enhancement on our cells\cite{Kuang2014-dv,Yang2014-wd,Zhang2015-aq}. 

\subsubsection*{Fabrication of silanized, gold-coated glass slides}\label{subsec6}

The gold substrates used in this work were prepared by evaporating a 5 nm adhesion layer of titanium, followed by 200 nm of gold at a rate of 1 Angstrom/second using a KJ LEsker e-beam evaporator onto piranha cleaned borosilicate glass slides. The gold-coated glass slides were then cleaned with an oxygen plasma, using a Diener Pico Oxygen Plasma Cleaner, for 3 min at 100 W power and $\sim$2 mbar of pressure, and silanized with 3-(aminopropyl)trimethoxysilane, APTMS, using vapor deposition in order to make the surface more hydrophobic and allow for greater aggregation of the gold nanorods on the cells\cite{Zhang2015-aq,Gao2019-wy,Shin2017-kt}. Slides were placed in a 1 liter flask in the presence of 100 µL of APTMS, Sigma-Aldrich 281778-5ML. The flask was then placed in a water bath at 40°C and allowed to react for 1 hr, after which the slides were removed from the flask and placed on a hot plate heated to 40°C for 10 min to allow for the evaporation of loosely bound molecules. We demonstrate that this APTMS layer also provides minimal Raman background noise, making it a great candidate for quick and easy substrate modification for biological Raman analysis (Supplementary Fig. 8). 

\subsubsection*{Acoustic printing}\label{subsec7}

Acoustic printing was completed using our custom-built ultrasonic, immersion transducer with a center frequency of 147 MHz and a focal distance of 3.5 mm (unless otherwise noted) as determined using a network analyzer, Hewlett Packard 8751A, and through pulse echo measurements taken on an oscilloscope, Keysight InfiniiVision DSOX3054A. The transducer was bound to a quartz, spherical focusing lens. The transducer was mounted on x,y,z manual translation stages, facing downwards, held 3.5 mm above a 303 stainless steel ejection plate with a 1 mm hole. For printing experiments, fluid was pipetted into the gap between the tip of the focusing lens and the ejection plate, held in place through surface tension. During printing experiments, droplets were ejected downwards through this 1 mm  hole onto our chosen substrates (Supplementary Fig. 2).

To generate our droplets, our transducer was powered by a waveform generator, Keysight 33600A Series Trueform Waveform Generator. The waveform generator was connected to a synthesized RF signal generator, Fluke 6062A, which in turn is connected to a power amplifier, Minicircuits ZHL-03-5WF+. Our waveform generator produces a square-wave burst with a repetition frequency of 1 kHz, when operating continuously, at our desired pulse width of 5.5 $\mu$s and at a voltage of 1.5 volts, enough to trigger our RF synthesizer. The RF synthesizer generates a sinusoidal wave at 147 MHz and at our desired voltage, which then gets amplified before reaching the transducer. Droplets printed from deionized water were ejected with 0.096 $\mu$J of energy, droplets printed from samples of \textit{S. epi} and \textit{E. coli} with and without GNRs were printed with 0.139 $\mu$J of energy, and droplets printed with RBCs with and without GNRs and from mixtures of RBCs, \textit{S. epi}, \textit{E. coli}, and GNRs were all printed with 0.386 $\mu$J of energy. 

To ensure stable ejection, we monitored ejection using a camera, Allied Vision Guppy Pro F-125 1/3" CCD Monochrome Camera, coupled with a 20x objective pointed at the bottom of our ejection plate. This camera was mounted opposite a strobing LED, also triggered by our waveform generator. We also monitored the acoustic echo using an inline oscilloscope, Keysight InfiniiVision DSOX3054A. To set up our printer, we pipette in 200 µL of fluid, turn on power to our transducer, and ensure that the transducer is in focus by manually adjusting the focal distance of our transducer until we maximize the echo as observed on the oscilloscope. We then vary the output voltage of the RF synthesizer until we stably eject a single droplet without any additional satellite droplets, as observed through our camera feed. We were then ready to pattern print arrays of droplets (Supplementary Fig. 3, 4).

\subsubsection*{Pattern printing}\label{subsec8}

Pattern printing was completed using a custom 3D printed substrate holder mounted to two perpendicularly stacked Thorlabs DDS100M 100mm brushless DC linear translation stages controlled by two Thorlabs K-Cube brushless DC servo drivers. Our substrate is mounted $\sim$1 mm below our ejection plate to minimize droplet translation before it reaches the substrate. A MATLAB, Mathworks, 2018b, script was used to pattern print droplets onto our substrate by controlling both our motorized stages and our waveform generator to trigger droplet ejection at specific substrate locations. 

\subsubsection*{Raman spectroscopy}\label{subsec9}

Raman spectra was collected using the Horiba XploRa confocal Raman microscope. The excitation wavelength for all measurements was 785 nm. The Raman shift from 400 cm\textsuperscript{-1} to 2000 cm\textsuperscript{-1} was collected using 600 gr/mm grating. For baseline Raman spectra shown in Fig. 1,  laser light was directed to and Raman scattered light was collected from the sample using a 100x LWD, 0.6 NA objective with spot size of 0.83 $\mu$m, with laser power at the sample of $\sim$6.71 mW, and acquisition time of 180 s. For spectra collected from each entire droplet, laser light was directed to and Raman scattered light was collected from the sample using a 10x, 0.25 NA objective with spot size of 2 $\mu$m, with laser power at the sample of $\sim$10.6 mW, and acquisition time of 15 s (Supplementary Fig. 11). Bacterial NR mixtures were measured within $\sim$2 hr of sample preparation.

\subsubsection*{Spectral data processing}\label{subsec10}

 Python (Jupyter Notebook) was used to process spectral data. For spectra pre-processing, samples were first thresholded to a minimum intensity of 150 a.u. to remove any spectra with weak signal that likely were collected on the substrate without the presence of cells. We then transformed our data by taking log\textsubscript{10}(y)\cite{Panda2010-il,Rodrigues2021-cq} and smoothed the spectra using wavelet denoising\cite{Ehrentreich2001-xn,Ramos2005-tt}. To perform our smoothing, we used the \textit{denoise\_wavelet} function from the scikit-image Python library: denoise\_wavelet(y, method='BayesShrink', mode='soft', wavelet\_levels=1, wavelet='coif3', rescale\_sigma='True'). We then performed a baseline removal by using a polynomial fit with degree 10. The specific package used and code line is: peakutils.baseline(y, deg=10, max\_it =1000, tol=0.0001). Note, the need for a higher degree polynomial arises from a typical instrumental background that is difficult to fit with lower degree fits. Following this baseline correction, Spectra were then individually normalized across all wavenumbers by subtracting the spectral mean and dividing by the standard deviation using the NumPy Python library\cite{Harris2020-wd}, where y is the array of intensity values across all wavenumbers for each spectra: (y - numpy.mean(y))/numpy.std(y) (Supplementary Fig. 31).

For classification of samples, we further pre-process data by reducing dimensionality of our spectra from 508 to the number of components necessary to account for 90\% of our sample variance using the PCA algorithm from Scikit-learn\cite{Pedregosa2011-rk} (Supplementary Fig. 19, 21). Classification was performed using a Random Forest Classifier. We first tuned our classifier hyperparameters using a cross-validated grid search to generate optimized parameters. To do this we use Scikit-learn \textit{StratifiedShuffleSplit}\cite{Pedregosa2011-rk} function to randomly split our sample 20 times into an 80:20 train:test split and created a parameter grid for our number of estimators: \{50, 100, 150, 200, 250, 300\}, max features: \{auto, sqrt, log2\}, and max depth: \{2, 5, 10, 15, 20\}. We created our Random Forest Classifier using Scikit-learn, with a min\_samples\_split=2, and then performed our grid search using Scikit-learn \textit{GridSearchCV}, with refit = True, n\_jobs = 3, and verbose = 190. We then perform a stratified K-fold cross validation (Scikit-learn \textit{StratifiedKFold}\cite{Pedregosa2011-rk} with shuffle=True) of our classifier’s performance across 10 splits using these optimal parameters. Finally, we use the Scikit-learn \textit{confusion\_matrix}\cite{Pedregosa2011-rk} function to plot our results. Intermediate t-SNE projections were plotted using Scikit-learn \textit{manifold.TSNE} with a perplexity = 10\cite{Pedregosa2011-rk} (Supplementary Fig. 20, 22). 

Raman wavenumber importance was performed using Voigt profile perturbation across all spectral wavenumbers. To achieve this, all spectra were first preprocessed as described above. Our spectra of interest (600 spectra across all 6 cellular classes) were partitioned into an 80:20 train/test split using Scikit-learn \textit{StratifiedShuffleSplit}\cite{Pedregosa2011-rk} with 10 splits. We reduce the dimensionality of our training set to 10 components using the PCA algorithm from the Scikit-learn\cite{Pedregosa2011-rk}. We train our Random Forest Classifier on our training spectra using optimized hyperparameters determined using a cross-validated grid search as previously described. We iteratively perturb our test set at each wavenumber to determine the relative importance of each wavenumber to accurate spectra classification. To do this, we iterate over each wavenumber in each normalized spectrum in our test set (120 spectra per split). For each wavenumber, we perturb our test spectra with a Voigt profile curve\cite{Alsmeyer2004-ro,Sundius1973-xk}, varying the intensity of the Voigt function 5 times at each wavenumber for each spectrum to get a large sample set. To generate our Voigt profiles, we first take all spectra in our entire sample set (600 spectra) and shift the intensity at a given wavelength (w) to guarantee that the intensities are positive. We then randomly shuffle all intensities and randomly select one to be used to generate our Voigt profile (voigt\_intensity). We generate this profile with our half-width at half-maximum (HWHM) of the Lorentzian profile, $\gamma$ = 2, and the standard deviation of the Gaussian profile, $\sigma$ = $\alpha$ / np.sqrt(2*np.log(2)), where $\alpha$ = 5. From here, we create our voigt\_profile = np.real(wofz((x - w + 1j*$\gamma$)/$\sigma$/np.sqrt(2))) / $\sigma$/np.sqrt(2*np.pi), where x is the entire range of wavenumbers, and scale this distribution to range from [0,1]. We utilize the \textit{wofz} function from the SciPy Python library to implement the Faddeeva function as the Voigt profile is related to the real part of the Faddeeva function. We also utilize various mathematical functions from the NumPy Python library to generate our profile. Voigt profile width was chosen to match peak widths seen in our spectra. To perturb our spectra with this profile, we take each spectra and shift intensities by the minimum, so that all intensities are positive (pos\_spectrum). We then perturb each wavenumber by our Voigt profile to generate a modified spectrum = pos\_spectrum*(1-voigt\_profile) + voigt\_intensity*voigt\_profile. We transform this perturbed spectrum with our established PCA and classify it using our trained Random Forest Classifier. We then plot a confusion matrix for each wavenumber using Scikit-learn \textit{confusion\_matrix}\cite{Pedregosa2011-rk} and generate an accuracy and f1 score using Scikit-learn \textit{classification\_report}\cite{Pedregosa2011-rk}, across all 6000 trials per wavenumber. Finally, we use our confusion matrix to generate the per class performance. See Supplementary Fig. 23 for more.

\section*{Supplementary Note 1: Gold nanorods for SERS applications}

SERS is a phenomenon that provides Raman intensity signal enhancements of on average of 10\textsuperscript{5}-10\textsuperscript{6}, with localized hotspots providing enhancements of 10\textsuperscript{8}-10\textsuperscript{10} \cite{Indrasekara2014-oo,Jackson2004-zf,Alonso-Gonzalez2012-ms}. Commonly, SERS utilizes metallic substrates that, through their plasmonic and chemical effects (such as charge transfer ability), enhance both the electric field from the incident light and the Raman scattered light from the sample, resulting in fourth order enhancement in the local electric field $\vert$E$\vert$\textsuperscript{4} \cite{Langer2020-ql,Felidj2003-xs,Lombardi2012-np,Andreou2015-ja,Pilot2019-iu,Schatz2006-tm,Latorre2016-ww}. SERS typically relies on metallic substrates to provide these enhancements. For biosensing applications, it is important for these substrates to provide large enhancements while being tunable, reproducible, stable, and inexpensive \cite{Bantz2011-lr}. As such, colloidal nanoparticles have gained traction as one of the primary forms of metallic SERS substrates \cite{Pilot2019-iu,Orendorff2006-ko,Dasary2009-fr,Joseph2011-gk,Stoerzinger2011-cw,Ha2019-ie}. In the realm of biosensing, gold and silver nanorods have been the primary metals used for SERS substrate synthesis due to their chemical stability and low toxicity \cite{Pilot2019-iu,Bantz2011-lr,Sharma2012-dj,Kunzmann2011-yd,Shukla2005-mc}. Particularly ideal for biological sensing, nanoparticles with sharp tips, such as nanocubes, nanostars \cite{Wen2020-fj}, nanopyramids, and nanorods, provide large Raman spectral enhancement factors, with nanorods providing the best balance of stability, reproducibility, tunability, cost, and enhancement \cite{Langer2020-ql,Reguera2017-fb}. Furthermore, due to their pervasiveness, gold nanorod (GNR) synthesis and properties are well documented allowing for reproducibility and easy tunability of optical properties through choice of particle size and shape \cite{Orendorff2006-ko,AnNisa2020-mh,Jain2006-ed,Lin2016-ex,Sharma2009-hx,Ye2013-us}.  Finally, advances in SERS substrates such as nanoparticle-on-mirror (NPoM) constructs \cite{Langer2020-ql,Baumberg2019-yy}, nanogaps and nanoholes \cite{Bantz2011-lr,Yu2008-yv,Reilly2007-wc}, graphene based nanodot arrays \cite{Zhang2016-ge}, and core-shell alloys \cite{Kumar2007-ef,Yin2011-fm,Jackson2003-lk} are paving the way for future Raman-based biosensing applications. 

\section*{Supplementary Note 2: Acoustic printing for handling biological samples}

Acoustic printing works by using ultrasonic waves to eject a droplet from a free surface of fluid. A radio frequency (RF) burst signal is used to excite a transducer at its resonant frequency, generating ultrasonic waves that exert force on the fluid surface \cite{Hadimioglu1992-sn,Elrod1989-wj}. When the focus of the transducer is aligned with the liquid-air interface and the intensity of the acoustic field is high enough, the generated radiation pressure will overcome the surface tension and the sound wave gives rise to a mound of fluid from the surface \cite{Elrod1989-wj,Chu1982-ps}. If the energy of the incident wave exceeds the threshold energy, a droplet breaks free from the fluid surface at a velocity of a few meters per second due to the Rayleigh-Taylor instability \cite{Hadimioglu1992-sn,noauthor_undated-tu}. The droplet diameter has been shown to closely match the diffraction-limited focal width at the liquid-air interface, and as such, the droplet diameter is inversely proportional to the frequency of the transducer, with 5 MHz and 300 MHz ultrasonic waves generating droplet diameters of 300 $\mu$m and 5 $\mu$m, respectively \cite{Elrod1989-wj}. (see Fig. 1a). ADE droplet ejection has been well characterized and has great tunability for handling a variety of biological samples \cite{Fang2012-ov}. Furthermore, the focused ultrasonic waves completely control the size, speed, and directionality of the ejected droplet and allow for ADE from an open liquid surface. Given that the dimensions of this open liquid surface are much larger than the diameter of the focal spot size, ADE is considered a nozzle-less technology \cite{Hadimioglu2016-nz}. This holds true for our downwards setup utilizing an ejection plate, given that our focal spot size is $\sim$2 orders of magnitude smaller than the 1 mm diameter hole \cite{Hadimioglu1992-sn,Elrod1989-wj}. As a nozzle-less technology, ADE has unparalleled advantage in biological sample handling as compared with other commercial piezo or thermal inkjet printers that rely on physical flow focusing. In particular, ADE eliminates system clogging and compromised cell viability or biomarker structure due to shear forces generated by nozzles. Additionally, ADE relies on ultrasonic waves to generate droplets, as such, the transducer never has to contact the ejection medium, but rather can propagate through a matched coupling medium, eg. through the bottom of an acoustically “transparent” multiwell plate, with minimal loss of acoustic energy, mitigating risks of sample contamination and loss of sterility \cite{Hadimioglu2016-nz}. ADE has also gained traction for versatility in setup and ability for high-throughput droplet generation. For a single acoustic ejector, the limiting factor for droplet ejection rate is the dissipation of capillary waves propagating radially outwards on the fluid surface after ejection \cite{Hadimioglu1992-sn,Eisenmenger1959-gt}. Advances in ADE have led to improvements in fabrication methods of the focusing lenses and ejector arrays including: spherical lenses in silicon, spherical PZT shells, and fresnel acoustic lenses \cite{Hadimioglu1992-sn,Hadimioglu2001-qi}. These advancements have lead to the development of high-throughput ejector arrays greater than 1000 print heads and ejection rates of 25 kHz, allowing for ejection of a 10 mL of fluid in under an hour \cite{Hadimioglu2001-qi}. Furthermore, these advances have expanded the utility of ADE for biological samples handling to include cellular acoustic printing \cite{Fang2012-ov,Gu2015-ks,demirci2007-qj}, biological crystallography \cite{Roessler2016-mw,Fuller2017-wx,Soares2011-cw}, high-throughput screening (HTS) of biological agents \cite{Rasmussen2016-sx}, and for sample preparation in MALDI \cite{Aerni2006-am}, highlighting the vast potential for biological analysis with ADE. 

\clearpage

\begin{figure}[H]%
\centering
\includegraphics[width=1\textwidth]{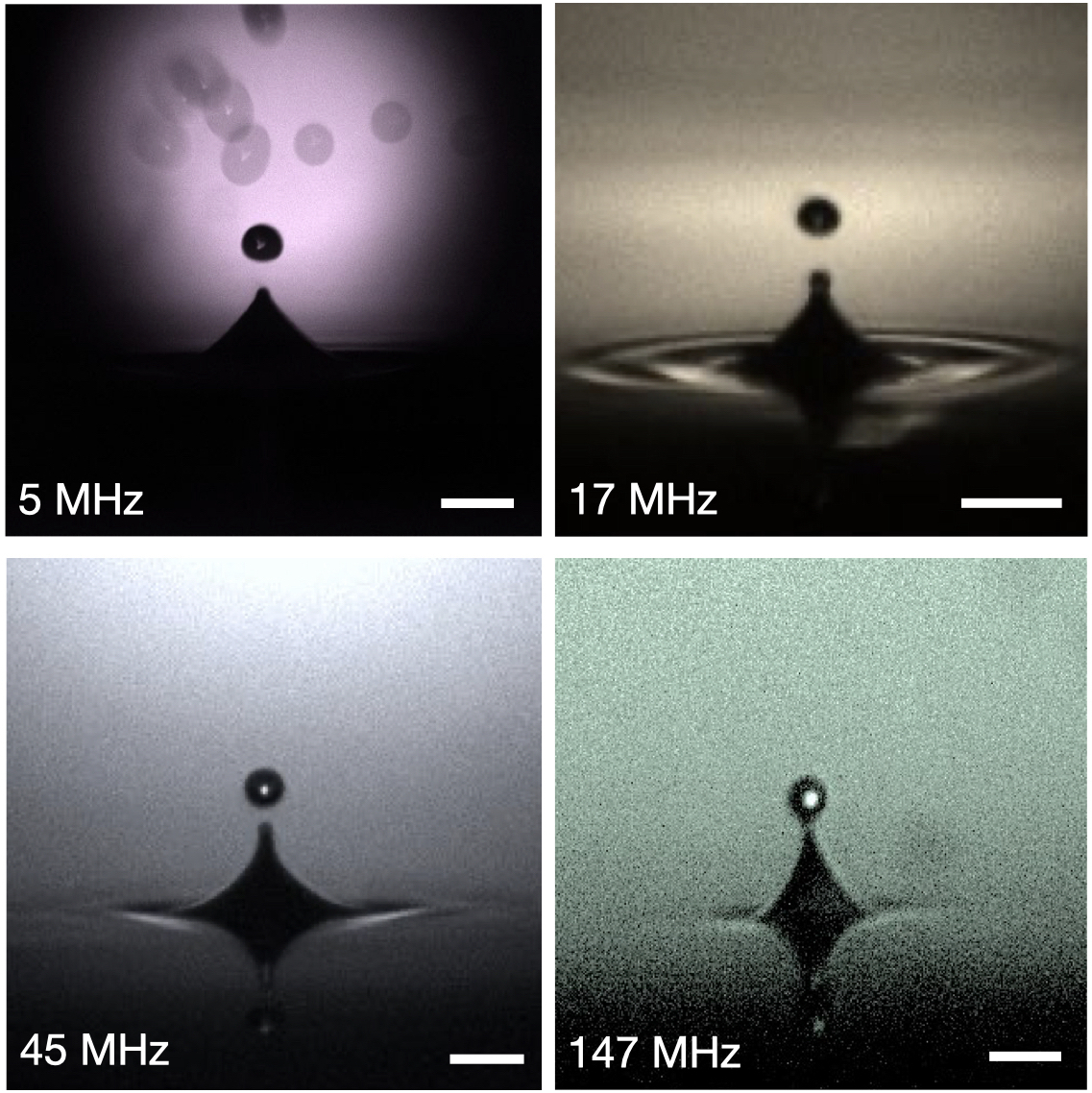}
\caption{Photographs of droplets printed with a range of acoustic frequencies. Droplets were printed with 4.8 MHz, 17 MHz, 44.75 MHz, and 147 MHz and had droplet diameters of 300 $\mu$m, 84 $\mu$m, 44 $\mu$m, and 15 $\mu$m respectively, highlighting the tunability of acoustic droplet ejection.  Scale bars are 500 $\mu$m, 200 $\mu$m, 100 $\mu$m, and 25 $\mu$m respectively.}\label{fig}
\end{figure}

\begin{figure}[H]%
\centering
\includegraphics[width=1\textwidth]{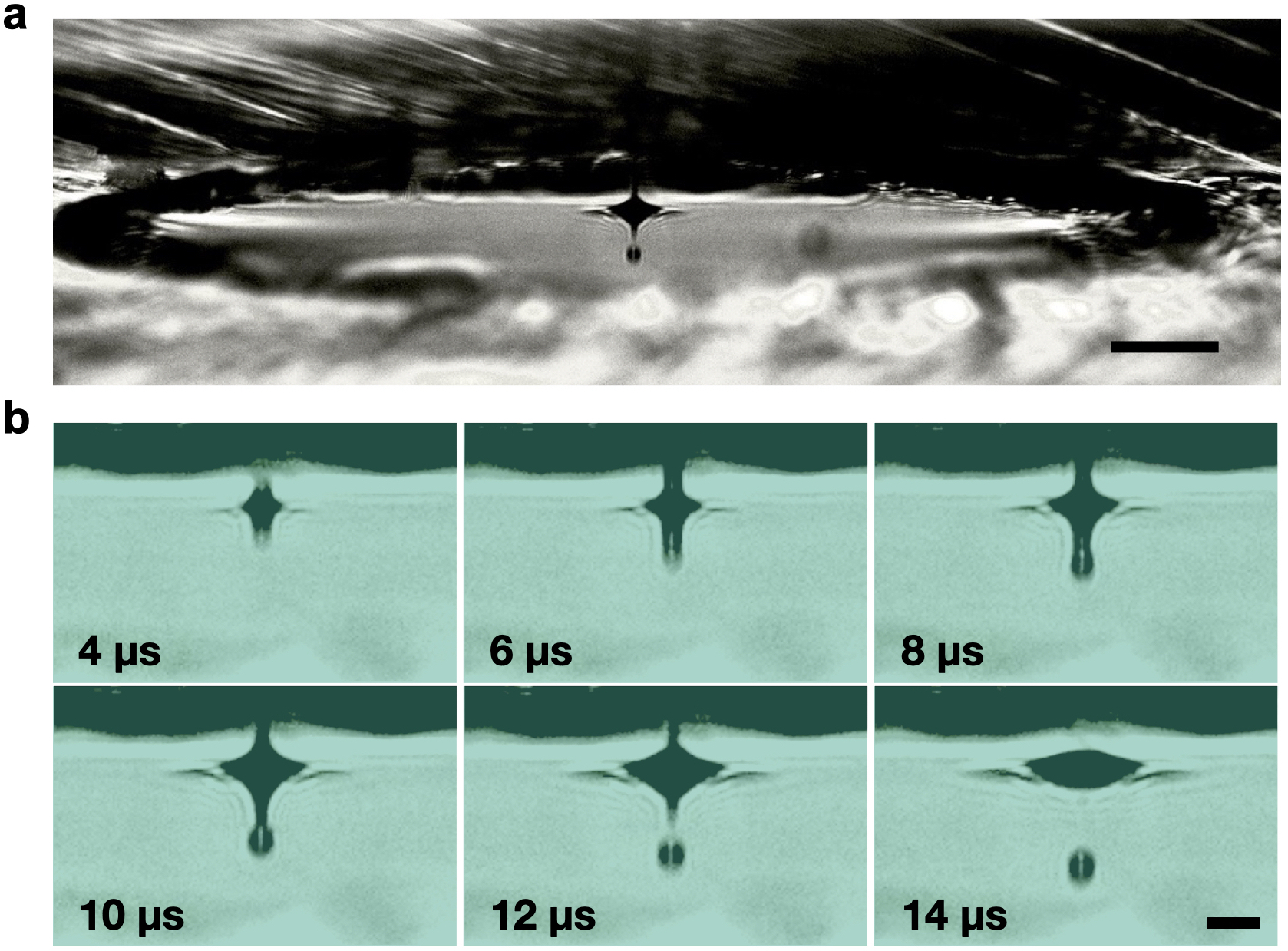}
\caption{\textbf{a,} Stroboscopic photograph showing a droplet being ejected downwards through the 1 mm hole on our ejection plate. Droplet was ejected from a pool of deionized water using a transducer operating at 147 MHz, Photo was taken with a 12 $\mu$s phase delay after the burst was triggered. Scale bar is 100 $\mu$m. \textbf{b,} Stroboscopic images of the time evolution of downward droplet ejection through the 1 mm hole at an acoustic frequency of 147 MHz. Droplet shown here is 15 $\mu$m in diameter and $\sim$2 pL in volume. Droplet was ejected with 0.096 $\mu$J of energy with a pulse width of 5.5 $\mu$s, and was ejected downwards at $\sim$3.5 m/s. Scale bar is 100 $\mu$m. All images were captured with an image exposure time of 40 ms and a droplet ejection rate of 1 kHz. As such, each frame is composed of 40 droplet ejections.}\label{fig}
\end{figure}

\begin{figure}[H]%
\centering
\includegraphics[width=1\textwidth]{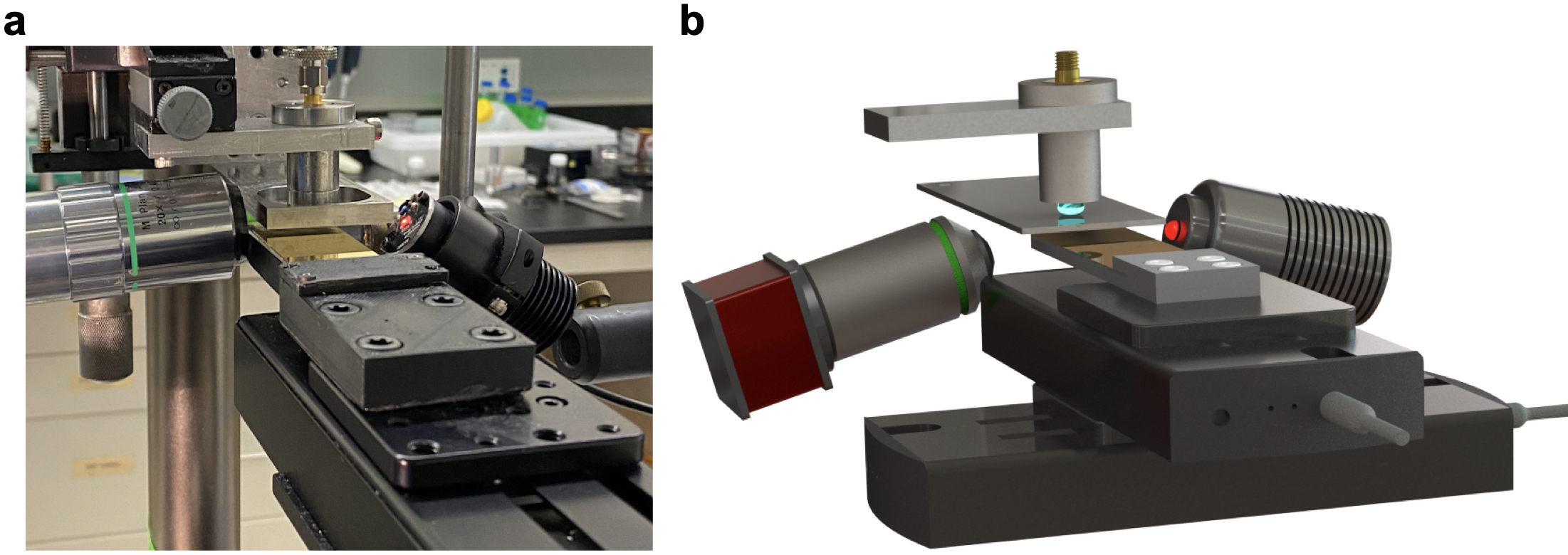}
\caption{Photo \textbf{a,} and rendering \textbf{b,} showing acoustic printing setup, including camera with 20x objective, baseplate with 1 mm diameter hole, strobing LED, gold-coated glass slide mounted onto a motorized XY stage, acoustic transducer, and printing fluid (teal).}\label{fig}
\end{figure}

\begin{figure}[H]%
\centering
\includegraphics[width=1\textwidth]{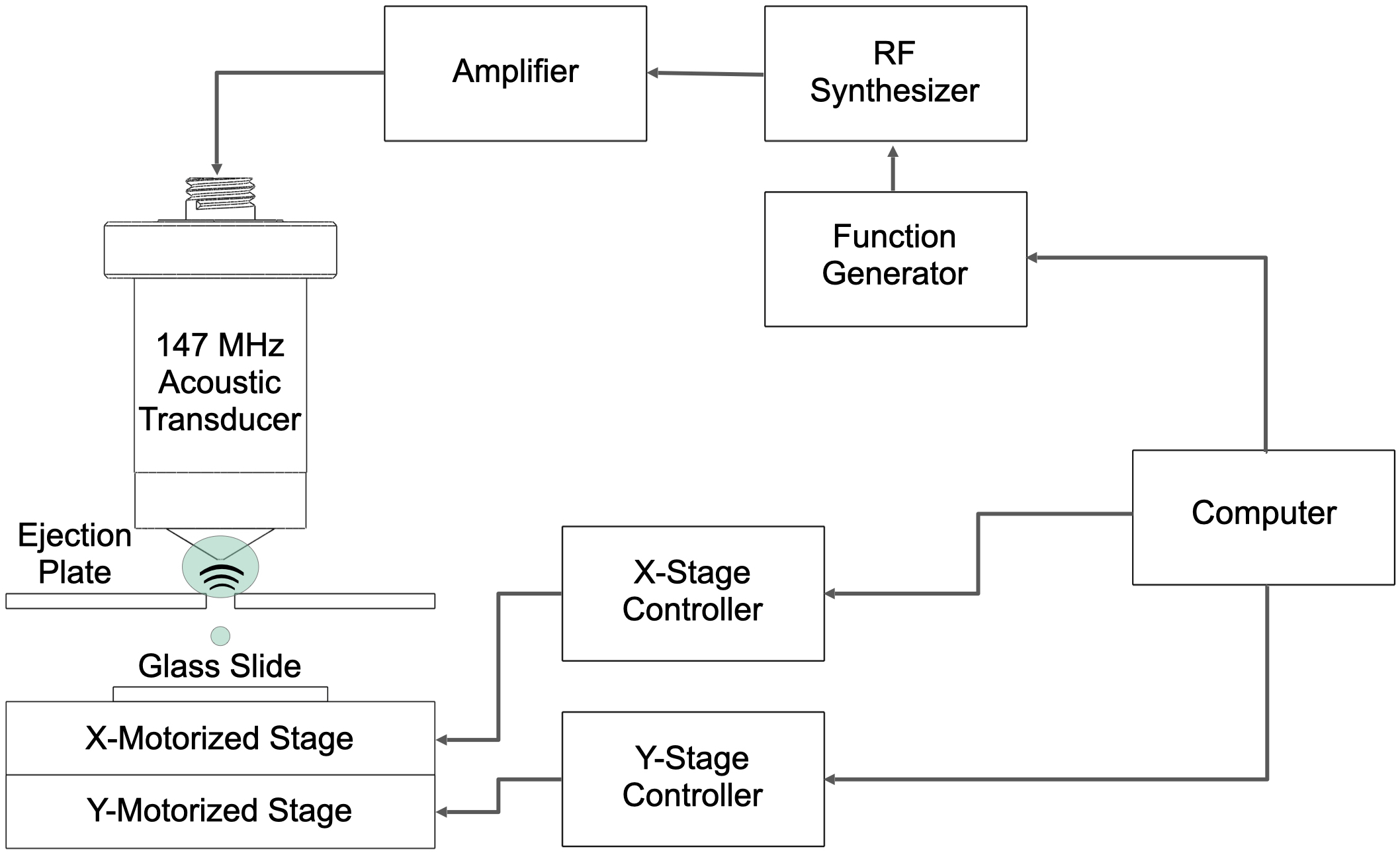}
\caption{Schematic showing the acoustic droplet ejection setup. The printing fluid (teal) rests between the focused acoustic transducer and the ejection plate with the 1 mm hole, held in place through surface tension. The droplets are ejected downwards onto a gold-coated glass slide mounted onto a motorized XY stage (stacked single axis stages). The burst signals to the transducer are generated from a function generator, routed through an RF synthesizer, and finally through a power amplifier before reaching the transducer. Ejection and movement of the mounted slide are controlled synchronously using MATLAB code.}\label{fig}
\end{figure}

\begin{figure}[H]%
\centering
\includegraphics[width=1\textwidth]{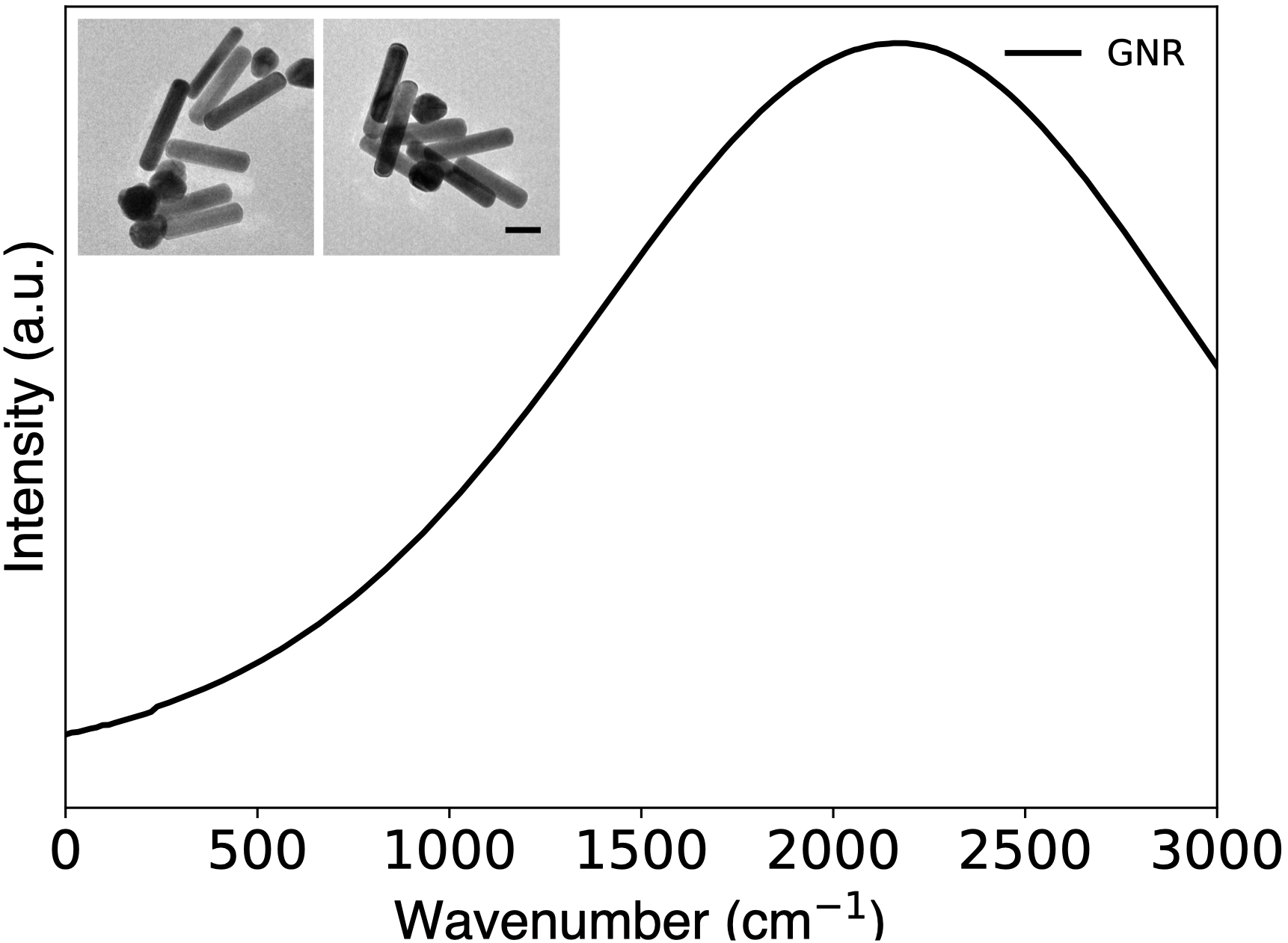}
\caption{Absorption spectra of GNRs used for Raman spectroscopy. Inset shows TEM of GNRs. Scale bar is 50 nm.}\label{fig}
\end{figure}

\begin{figure}[H]%
\centering
\includegraphics[width=1\textwidth]{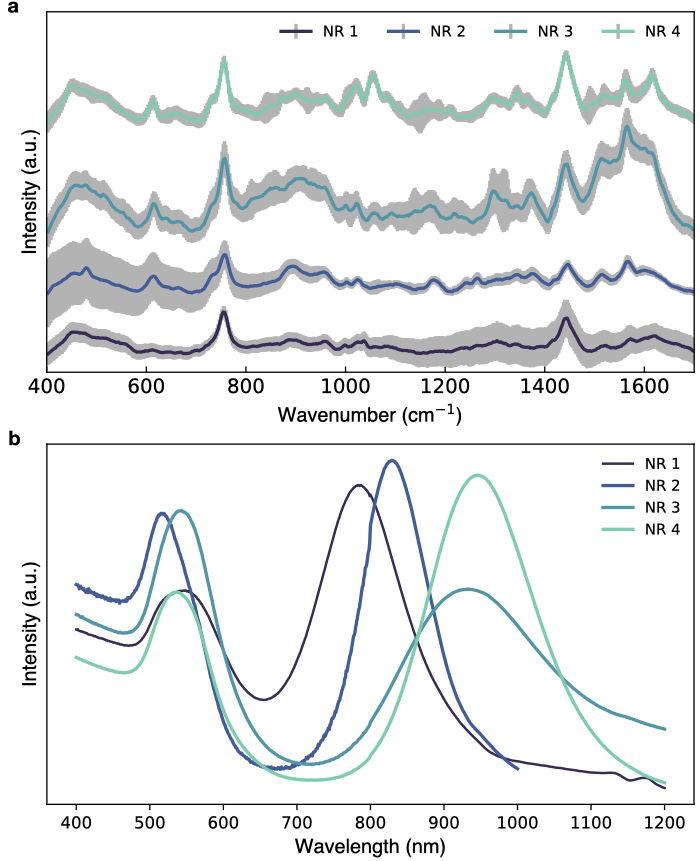}
\caption{Bacterial interrogation across multiple nanorod syntheses and resonance frequencies. We synthesized 4 different batches of GNRs and evaluated bacterial droplets with each batch. \textbf{a,} UV-Vis measurements of the 4 rods showing a range of resonance frequencies between 770 nm and 960 nm. Our chosen nanorods are those listed as NR4. \textbf{b,} Average spectral intensities and standard deviations collected from droplets printed with each of our four GNR batches mixed with \textit{S. epi} bacteria diluted to a final concentration of 1e9 cells/mL in a 1:9 mixture v/v of Invitrogen UltraPure 0.5M EDTA, Invitrogen 15575020, and Mili-Q purified water onto a silanized, gold-coated glass slide. Spectra were collected with a 10x objective lens with a 0.25 NA and $\sim$10.6 mW power. Each droplet was interrogated with a time study of 5 time points, with each exposure lasting 15 seconds. 10 droplets were analyzed from each GNR batch for a total of 50 data points across each group. The data shows the primary \textit{S. epi} peaks around 731 and 1317 cm\textsuperscript{-1}, with varying max intensities, highlighting both the robustness of our system for spectral collection as well as the potential for improvements through tuning of GNR aspect ratio.}\label{fig}
\end{figure}

\begin{figure}[H]%
\centering
\includegraphics[width=1\textwidth]{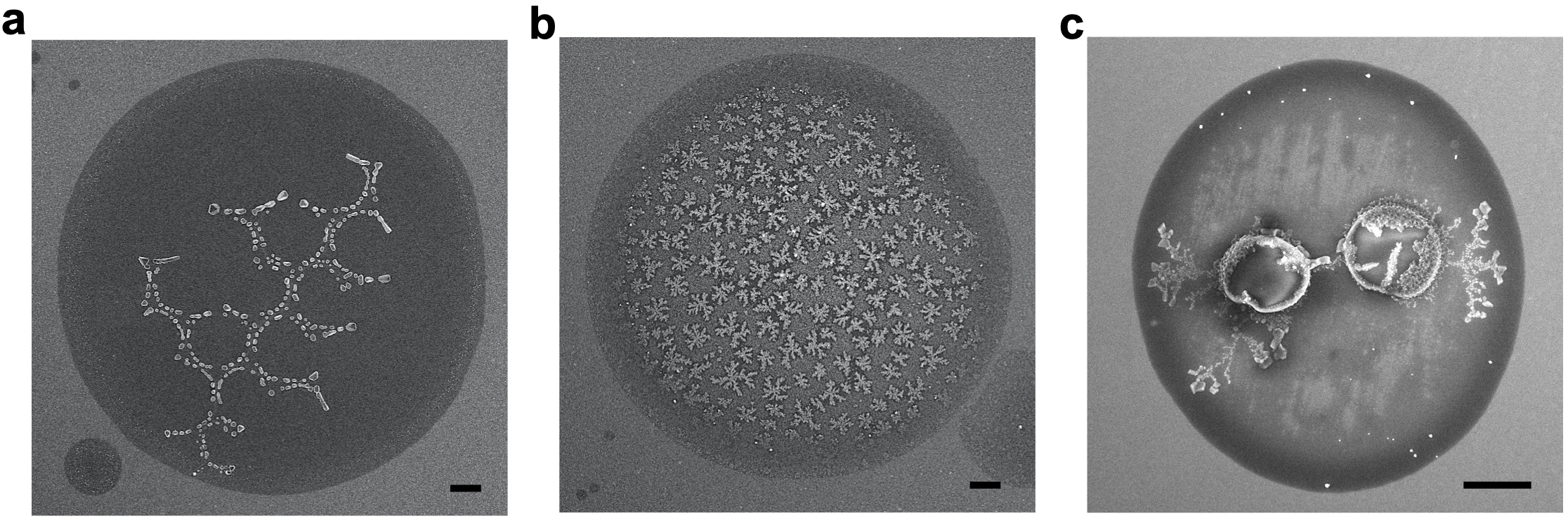}
\caption{Crystallization of saline upon drying in acoustically printed droplets printed at 147 MHz. Droplets \textbf{a,} and \textbf{b,} were printed from a 10\% v:v phosphate buffered saline (PBS) solution. Droplet \textbf{c,} was printed with a mixture of mouse RBCs suspended in 10\% v:v  PBS solution. Scale bars are 10, 10, 5 $\mu$m, respectively.}\label{fig}
\end{figure}

\begin{figure}[H]%
\centering
\includegraphics[width=1\textwidth]{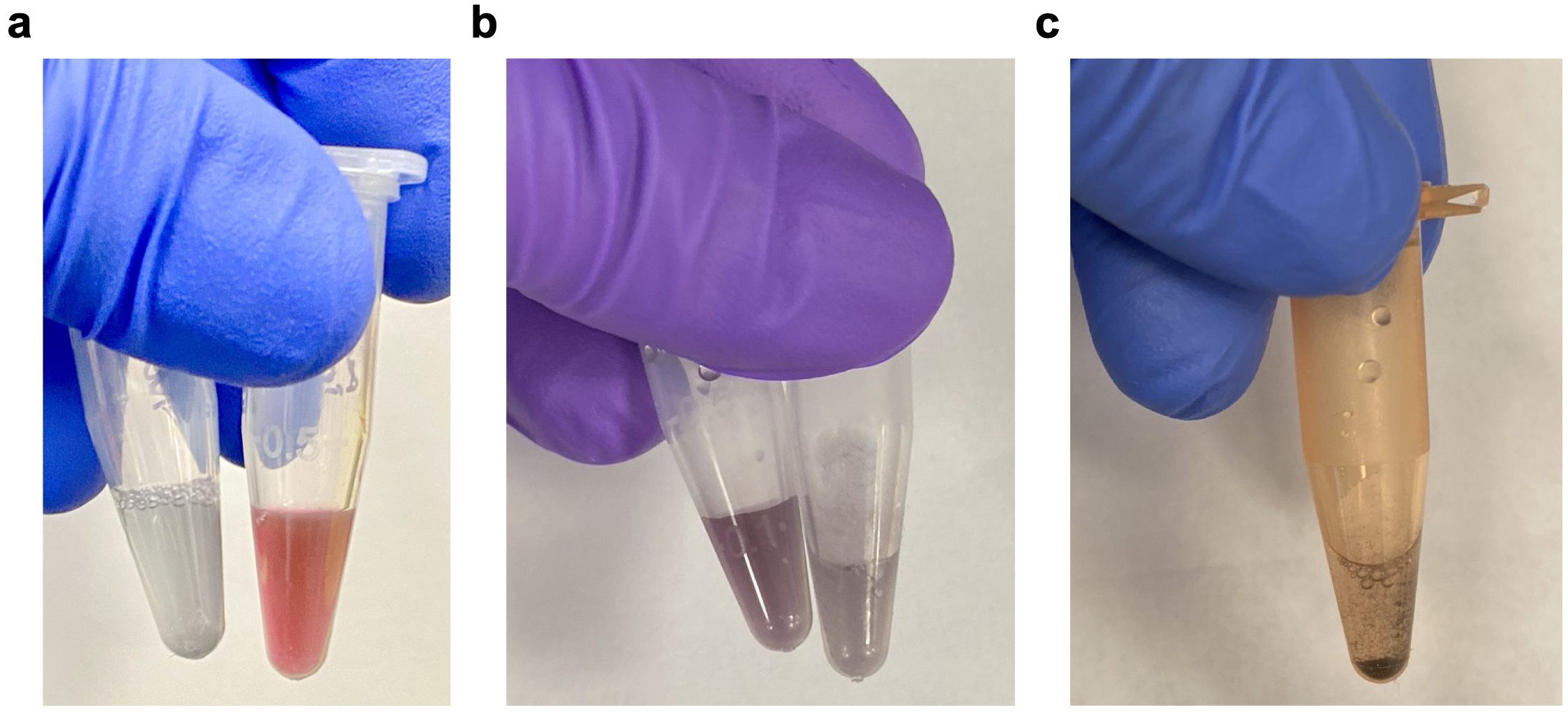}
\caption{Studying the effect of the EDTA on the nanorod dispersion in fluid. \textbf{a,} Photograph showing colorimetric comparison of gold nanorods (GNR) in (left) a 1:9 mixture of Invitrogen UltraPure 0.5 M EDTA, Invitrogen 155750, and Milli-Q purified water and (right) Milli-Q purified water only. \textbf{b,} Photograph showing colorimetric comparison of GNRs mixed with \textit{S. epi} bacteria at a concentration of 1e9 cells/mL. The image shows the gold nanorods and bacteria in (left) Milli-Q purified water only and (right) in a 1:9 ratio v/v of EDTA solution and Milli-Q purified water. \textbf{c,} Photograph showing the settling of solution of GNRs, \textit{S. epi}  at a concentration of 1e9 cells/mL, and a 1:9 ratio v/v of EDTA solution to Milli-Q purified water. Photograph was taken 5 min after the solution was mixed together. We note that these images highlight that, as expected, the EDTA appears to aggregate the GNRs into clusters. As the \textit{S. epi} bacteria seem to cause clustering regardless of the presence of the EDTA due to their surface charge, the difference between the sample with and without the EDTA is less noticeable than in the samples of only GNRs.}\label{fig}
\end{figure}

\begin{figure}[H]%
\centering
\includegraphics[width=1\textwidth]{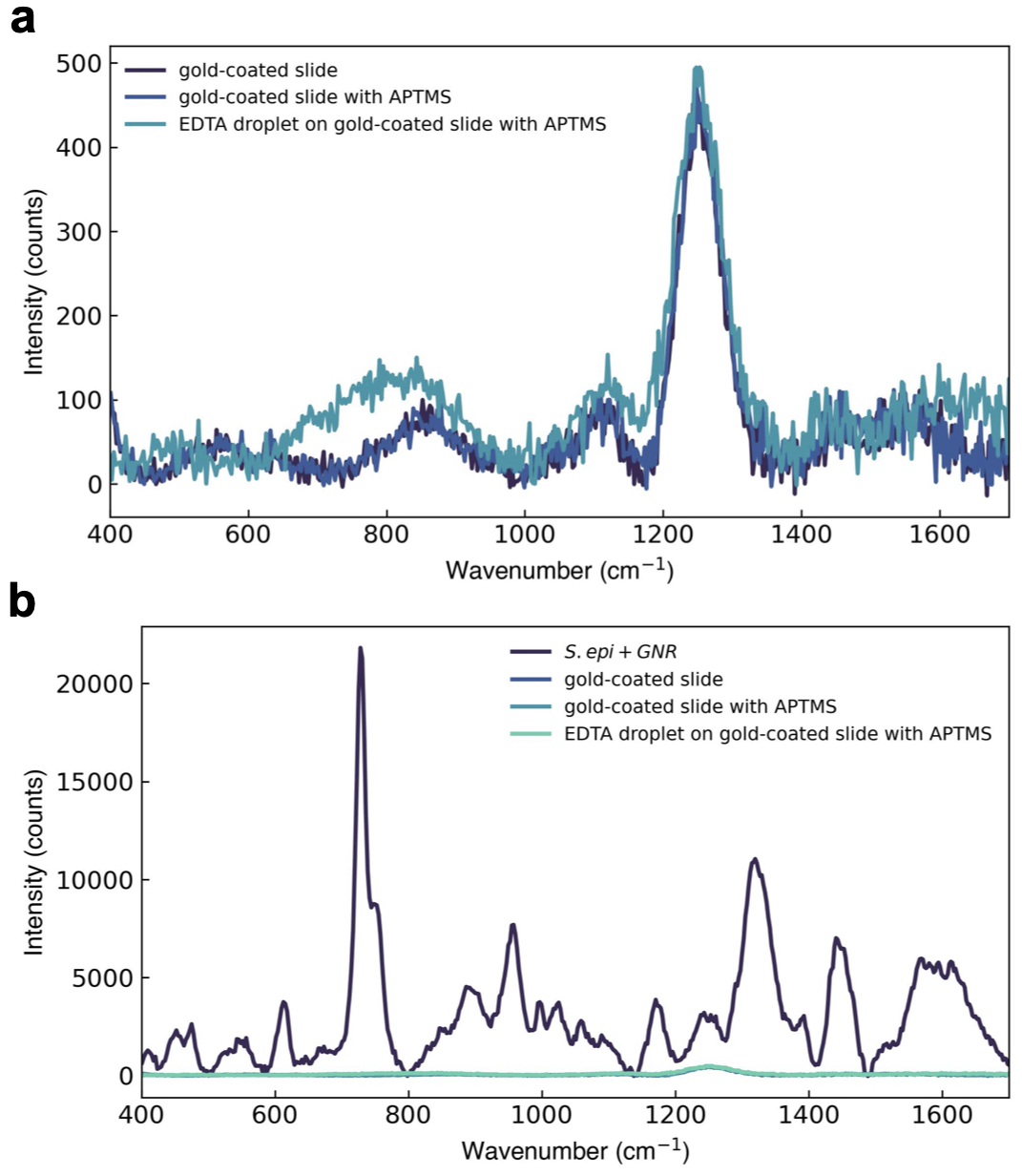}
\caption{Raman of background signals. \textbf{a,} Spectra were collected of a gold-coated glass slide, a gold-coated glass slide with an APTMS silane layer, and of a droplet printed onto a gold-coated glass slide with APTMS. Droplets were printed from Invitrogen UltraPure 0.5M EDTA, Invitrogen 15575020, mixed in a 1:9 ratio v/v with Mili-Q purified water. The spectra show that most of the background signal comes from the gold substrate with little additional background from our APTMS deposition and the additional EDTA used in our cell solutions. \textbf{b,} Identical spectra to that shown in \textbf{a} overlaid with a spectrum taken of \textit{S. epi} bacteria and GNRs suspended in EDTA solution at a concentration of 1e9 cells/mL. The plot highlights that the spectral signal intensity from our bacteria is much higher than that of the background. All spectra were collected with a 10x objective lens with a 0.25 NA and $\sim$10.6 mW power for 15 s.}\label{fig}
\end{figure}

\begin{figure}[H]%
\centering
\includegraphics[width=1\textwidth]{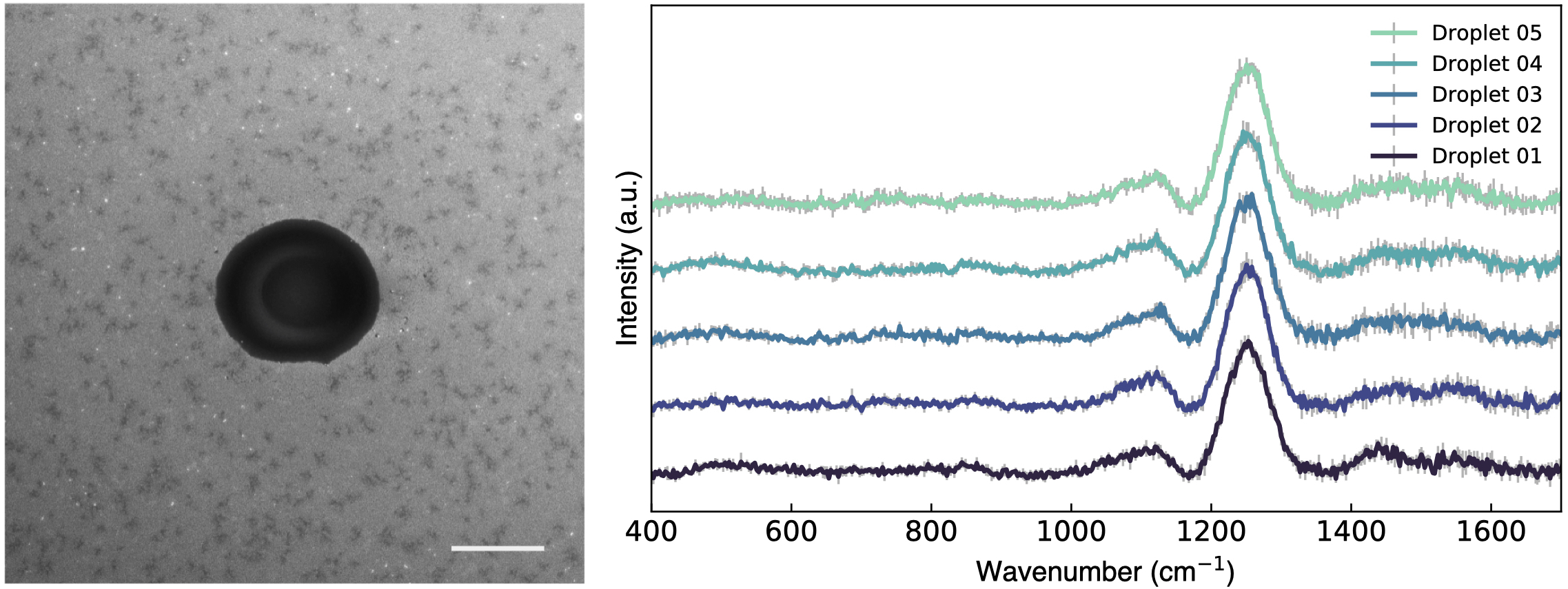}
\caption{Droplets were printed from cellular dilution mixture without any cells. Droplets were printed from Invitrogen UltraPure 0.5 M EDTA, Invitrogen 15575020, mixed in a 1:9 ratio v/v with Mili-Q purified water onto a silanized, gold-coated glass slide. Spectra were collected with a 10x objective lens with a 0.25 NA and $\sim$10.6 mW power for 15 s. The SEM clearly shows minimal spread of the EDTA solution onto the gold-coated slide. The spectra show minimal, and consistent background signals from the EDTA solution. Scale bar is 5 $\mu$m.}\label{fig1}
\end{figure}

\begin{figure}[H]%
\centering
\includegraphics[width=1\textwidth]{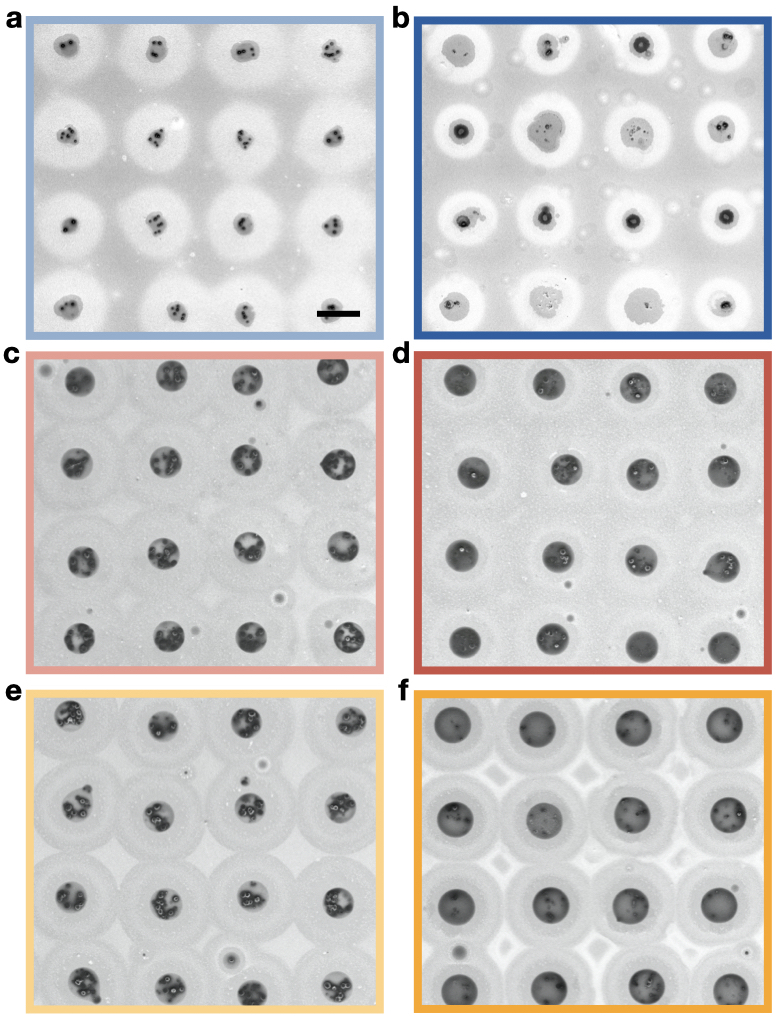}
\caption{SEMs of grids printed with cell and GNR mixtures. SEMs show 16 droplets imaged out of a grid of over 400 droplets. Mixtures were printed from cells with and without GNRs diluted in a 1:9 mixture of EDTA solution and Milli-Q water to a final concentration of 1e9 cells/mL. All grids were acoustically printed using a transducer operating at 147 MHz with a 5.5 $\mu$s pulse width burst signal. SEMs show \textbf{a,} \textit{S. epi} \textbf{b,} \textit{S. epi} mixed with GNRs both ejected with 0.096 $\mu$J of acoustic energy and \textbf{c,} mouse RBCs \textbf{d,} mouse RBCs mixed with GNRs \textbf{e,} 1:1 mixture of \textit{S. epi} and mouse RBCs \textbf{f,} 1:1 mixture of \textit{S. epi} and mouse RBCs with GNRs all printed with 0.139 $\mu$J of acoustic energy. The lighter and darker circles in each photo highlight the outer edge of the droplet as well as the smaller volume formed from the dried EDTA mixture containing our cells and GNRs in the center of the droplet. Scale bar is 50 $\mu$m.}\label{fig}
\end{figure}

\begin{figure}[H]%
\centering
\includegraphics[width=1\textwidth]{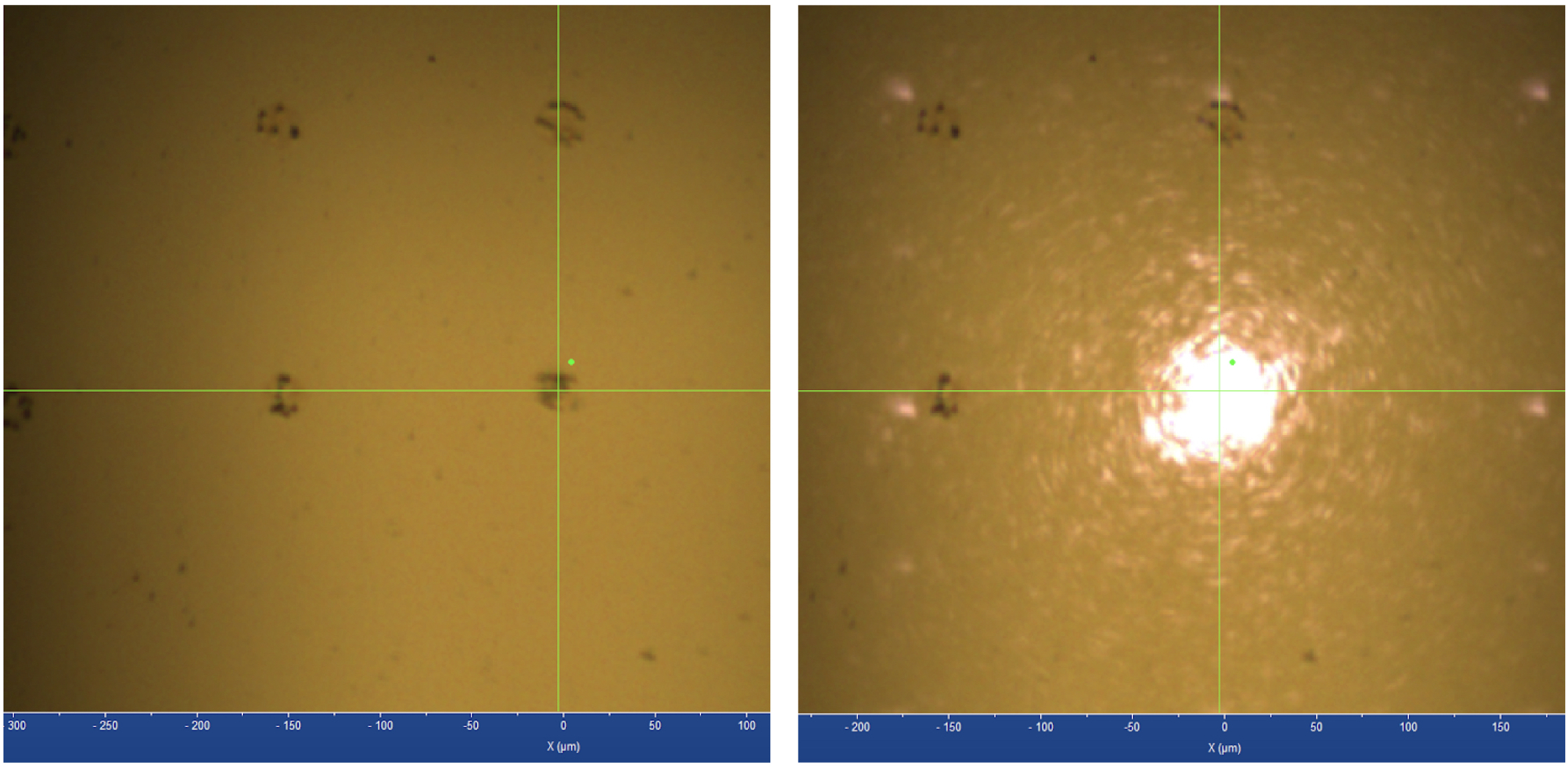}
\caption{Raman focal spot size. Images show screen shots from Horiba XploRA Raman confocal microscope UI. The image on the left shows an array of droplets printed onto an APTMS silanized, gold-coated slide. Droplets were printed from a solution of gold nanorods and \textit{S.epi}  bacteria at a concentration of 1e9 cells/mL, suspended in a solution of Invitrogen UltraPure 0.5 M EDTA, Invitrogen 15575020, mixed in a 1:9 ratio v/v with Milli-Q water. The image on the right shows the same array of droplets, with the 10x objective with the 785 nm laser turned on, operating at 25\% laser power or $\sim$10.6 mW of power. This laser spot size is $\sim$2 $\mu$m.}\label{fig}
\end{figure}

\begin{figure}[H]%
\centering
\includegraphics[width=1\textwidth]{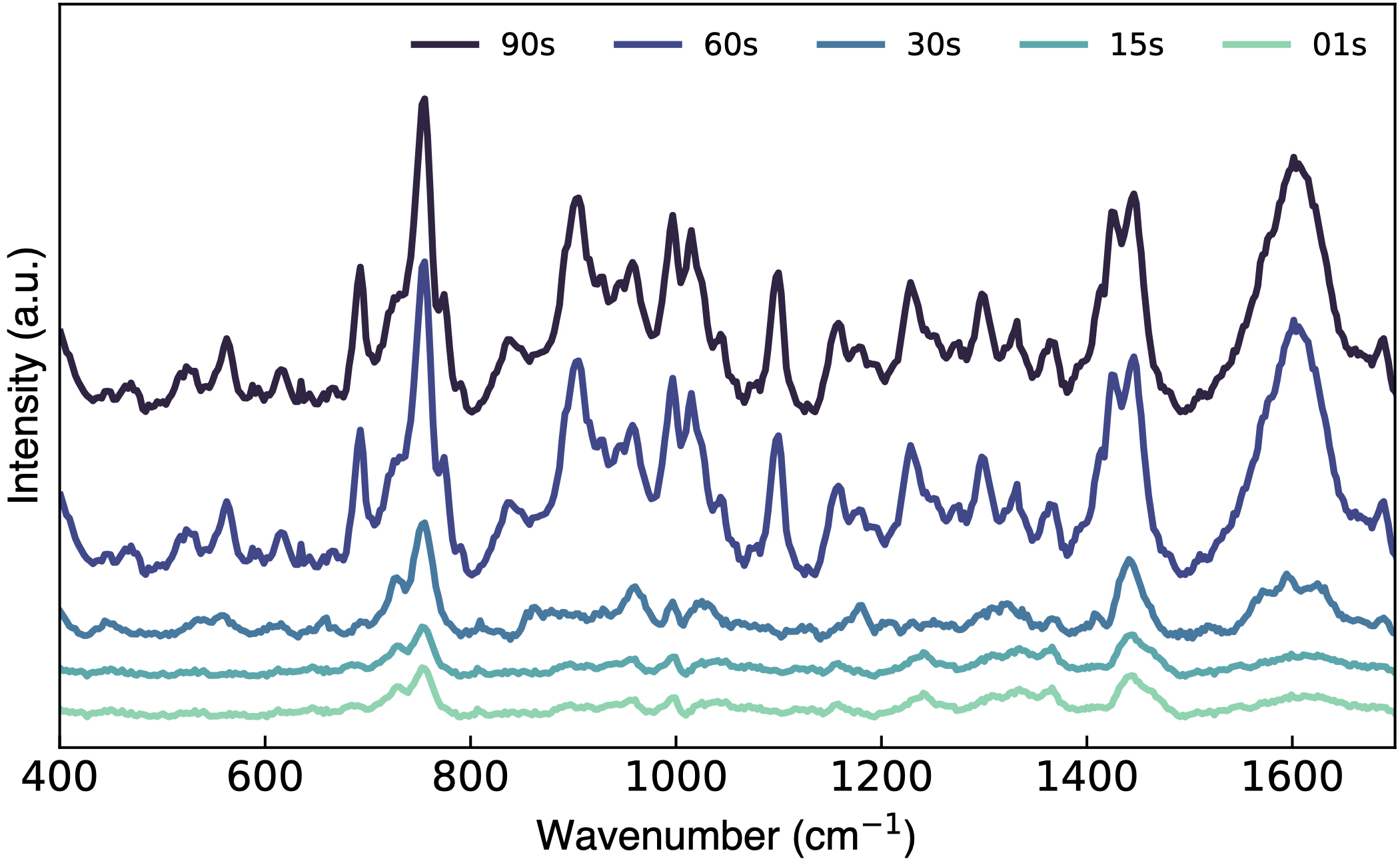}
\caption{Intensity study of single droplet with \textit{S. epi} bacteria and GNRs taken at 1, 15, 30, 60, and 90 seconds. Droplets were printed from a mixture of GNRs and \textit{S. epi} bacteria at a final concentration of 1e9 cells/mL diluted in a 1:9 v/v of Invitrogen UltraPure 0.5 M EDTA, Invitrogen 15575020, and Mili-Q purified water onto a silanized, gold-coated glass slide. Spectra were collected with a 10x objective lens with a 0.25 NA and $\sim$10.6 mW power. The spectra show increasing signal intensity and signal complexity with each longer exposure time.  This highlights that our time selection of 15 s is well below the time at which our sample gets damaged by the laser power. This analysis guided our choice of a 15 s acquisition time to balance gathering clear, distinct spectra with choosing a fast enough acquisition time to show potential for clinical translation.}\label{fig}
\end{figure}

\begin{figure}[H]%
\centering
\includegraphics[width=1\textwidth]{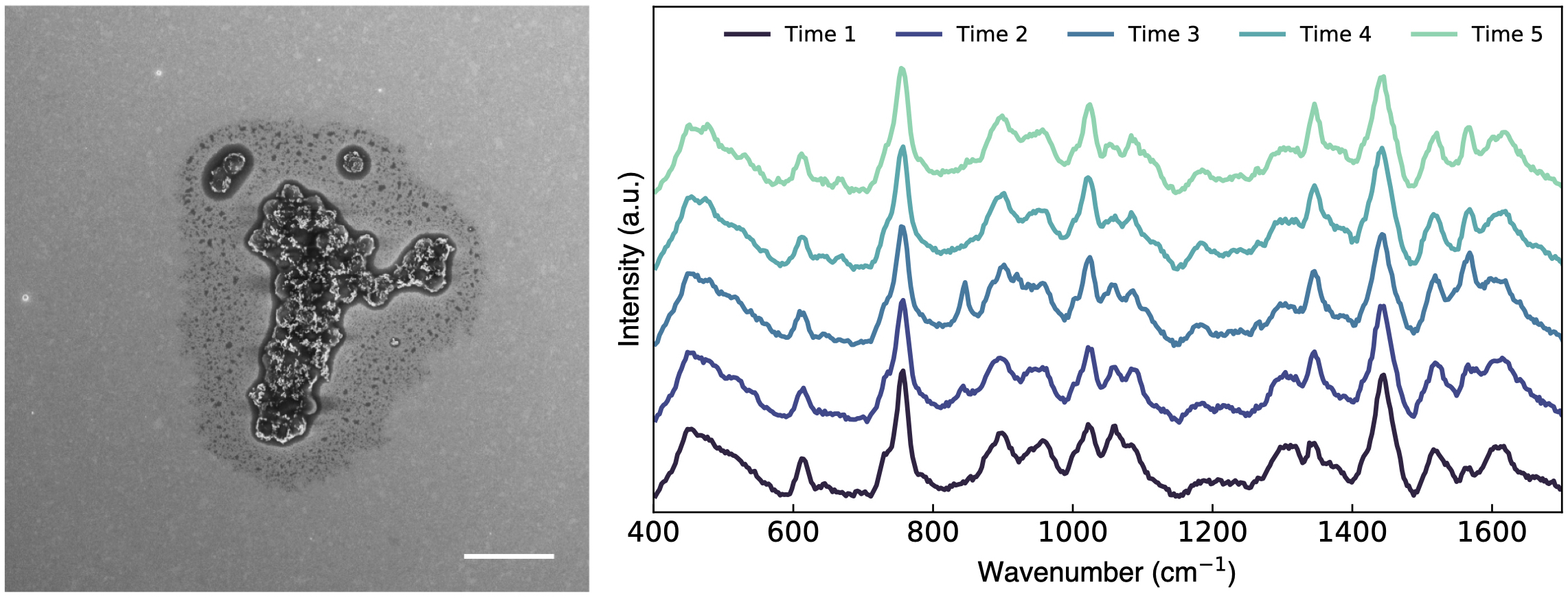}
\caption{Time study of single droplet with \textit{S. epi} bacteria and GNRs across multiple time points. Droplets were printed from a mixture of GNRs and \textit{S. epi} bacteria at a final concentration of 1e9 cells/mL diluted in a 1:9 v/v of Invitrogen UltraPure 0.5 M EDTA, Invitrogen 15575020, and Mili-Q purified water onto a silanized, gold-coated glass slide. Spectra were collected with a 10x objective lens with a 0.25 NA and $\sim$10.6 mW power. Each spectrum was collected for 15 s in a time series lasting a total of 75 seconds. The SEM clearly shows a cluster of bacteria coated in GNRs. The spectra show minimal variation over the 75 second duration, showing that our acquisition time of 15 seconds does not damage the cells. Scale bar is 5 $\mu$m.}\label{fig}
\end{figure}

\begin{figure}[H]%
\centering
\includegraphics[width=1\textwidth]{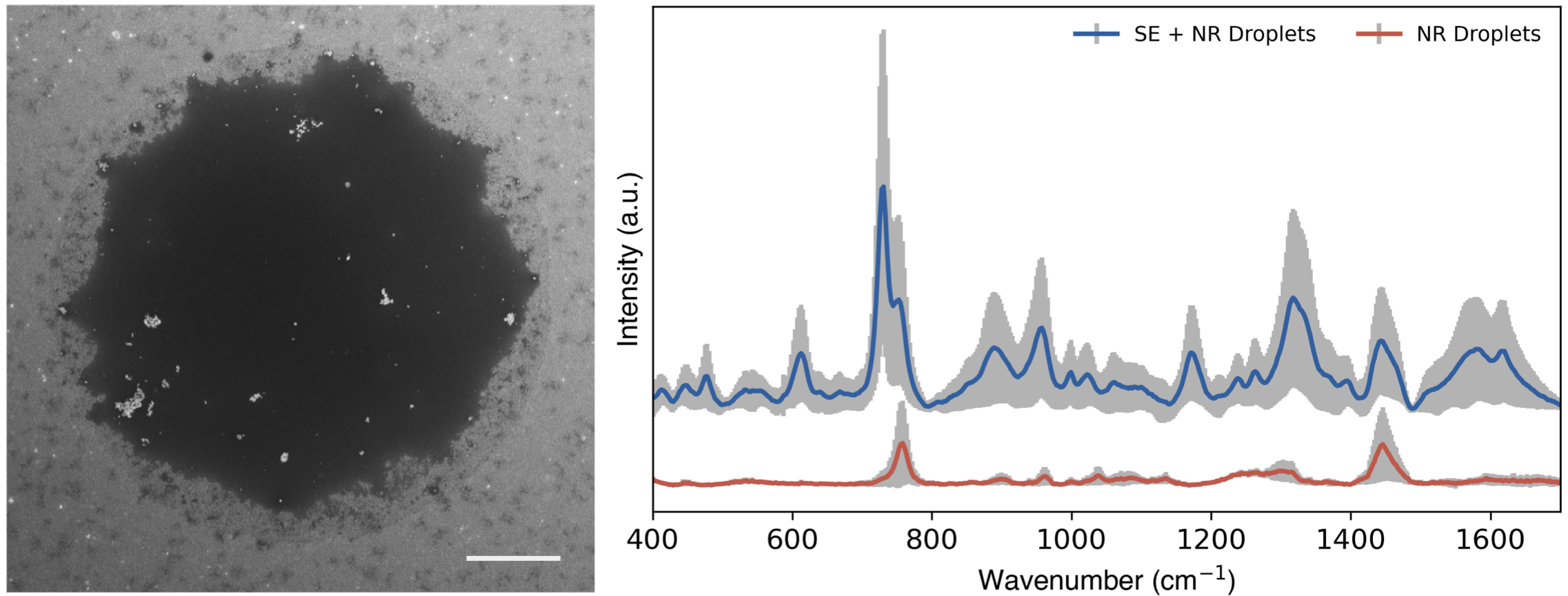}
\caption{Background signal from gold nanorods (GNR). Droplets were printed from a sample containing GNRs suspended without any cells in a 1:9 v/v mixture of Invitrogen UltraPure 0.5 M EDTA, Invitrogen 15575020, and Mili-Q purified water onto a silanized, gold-coated glass slide. Spectra were collected of each droplet with a 10x objective lens with a 0.25 NA and $\sim$10.6 mW power for 15 s. The SEM shows a droplet containing a few clusters of GNRs clearly distinguishable, highlighting both the presence of the GNRs and the absence of a coffee-ring of nanorods. The plot shows the mean and standard deviation of 100 droplets printed from a sample of \textit{S. epi} bacteria with GNRs in EDTA solution (in blue, identical to that from Fig. 3), and the mean and standard deviation of 20 droplets printed from the GNR solution without cells. The spectra show that while the GNRs have a background signal, hypothesized to be from any remaining CTAB present in the solution after rinsing the rods, the cells have a clearly distinguishable signal separate from that of the GNRs, similar to our baseline \textit{S. epi}  spectra. Scale bar is 5 $\mu$m.}\label{fig}
\end{figure}

\begin{figure}[H]%
\centering
\includegraphics[width=1\textwidth]{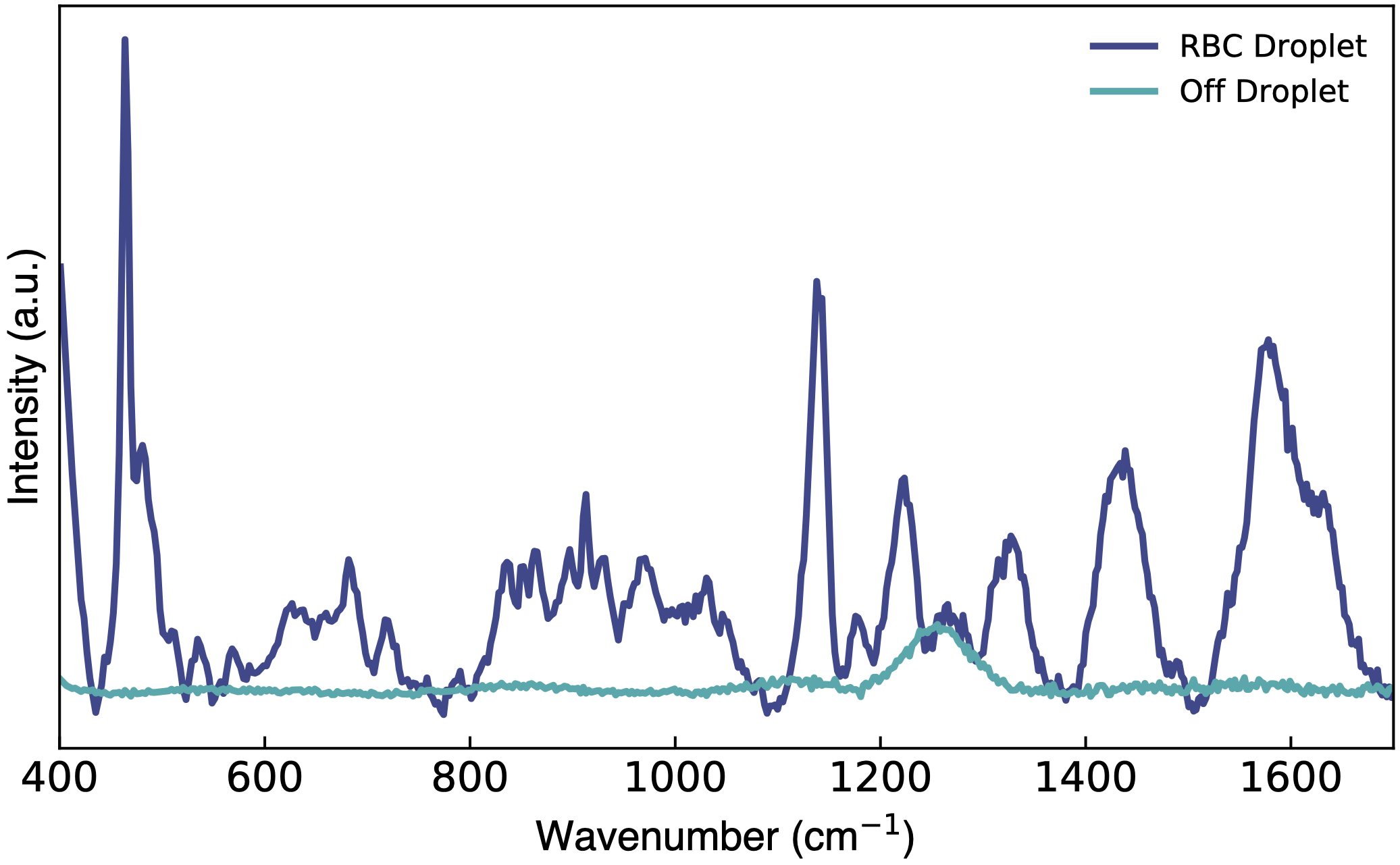}
\caption{Spectra were collected from a sample of droplets printed from GNRs mixed with mouse RBCs at a final concentration of 1e9 cells/mL diluted in a 1:9 ratio v/v of  Invitrogen UltraPure 0.5 M EDTA, Invitrogen 15575020, and Mili-Q purified water onto a silanized, gold-coated glass slide. Spectra were collected with a 10x objective lens with a 0.25 NA and $\sim$10.6 mW power for 15 s. The first spectra is taken while the focal spot is centered on the droplet while the other is taken when on the silanized gold substrate to the side of the droplet highlighting that our signal is coming directly from the droplet and not from any background material on the slide.}\label{fig}
\end{figure}

\begin{figure}[H]%
\centering
\includegraphics[width=1\textwidth]{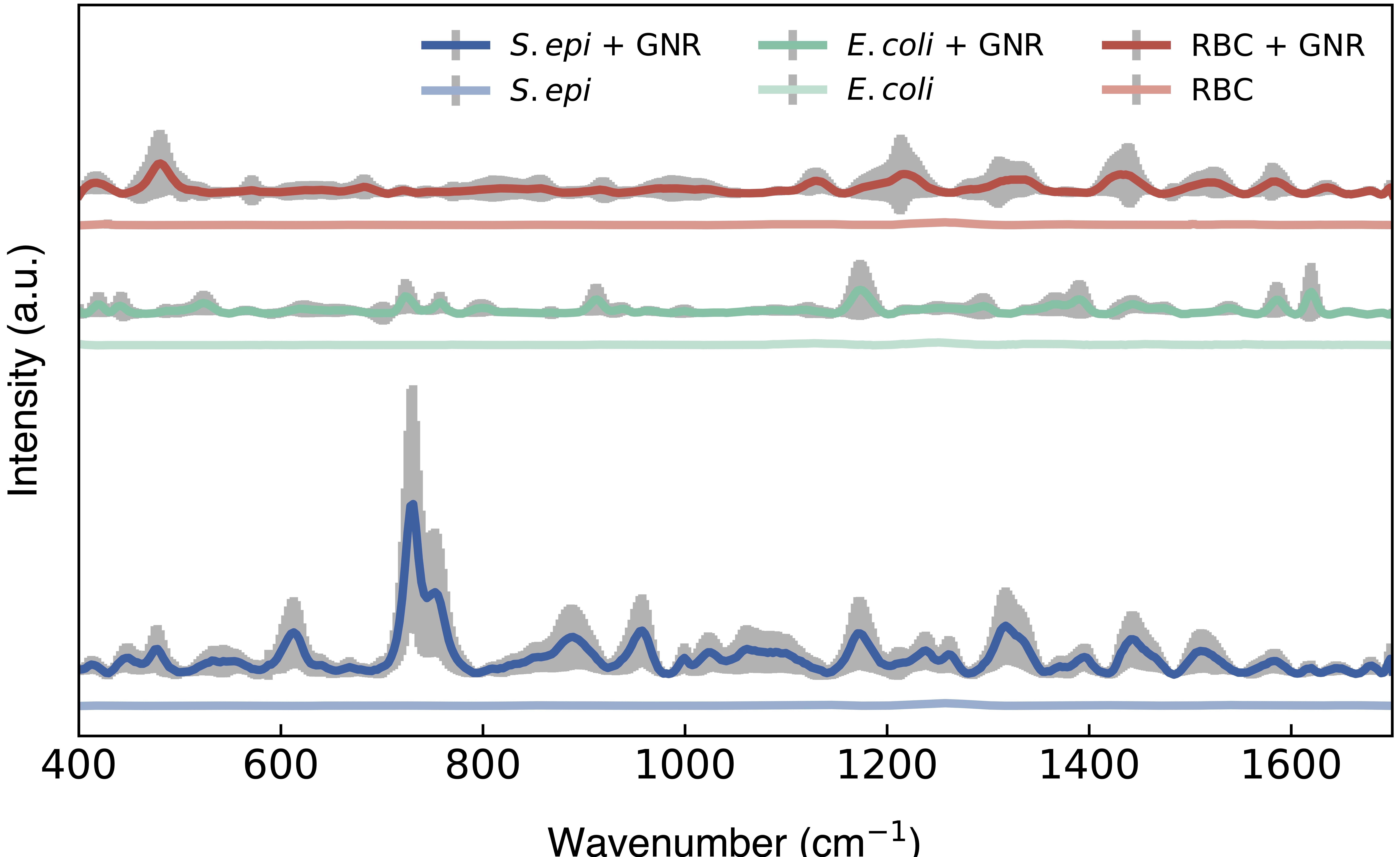}
\caption{Plot showing the mean and standard deviation of SERS spectra taken from droplets printed from three cell lines (\textit{S. epi}, \textit{E. coli}, and RBCs) with and without GNRs.  Spectra were collected from 100  and 15 droplets, respectively. The plots highlight both the enhancements generated with the presence of GNRs as well as the variations in peak spectra intensity due to the variations in surface charge density on each cell line and subsequently the cells’ varying attraction to the positive surface charge of the GNRs, resulting from the CTAB surfactant on their surface \cite{Tadesse2020-lm}.}\label{fig}
\end{figure}

\begin{figure}[H]%
\centering
\includegraphics[width=1\textwidth]{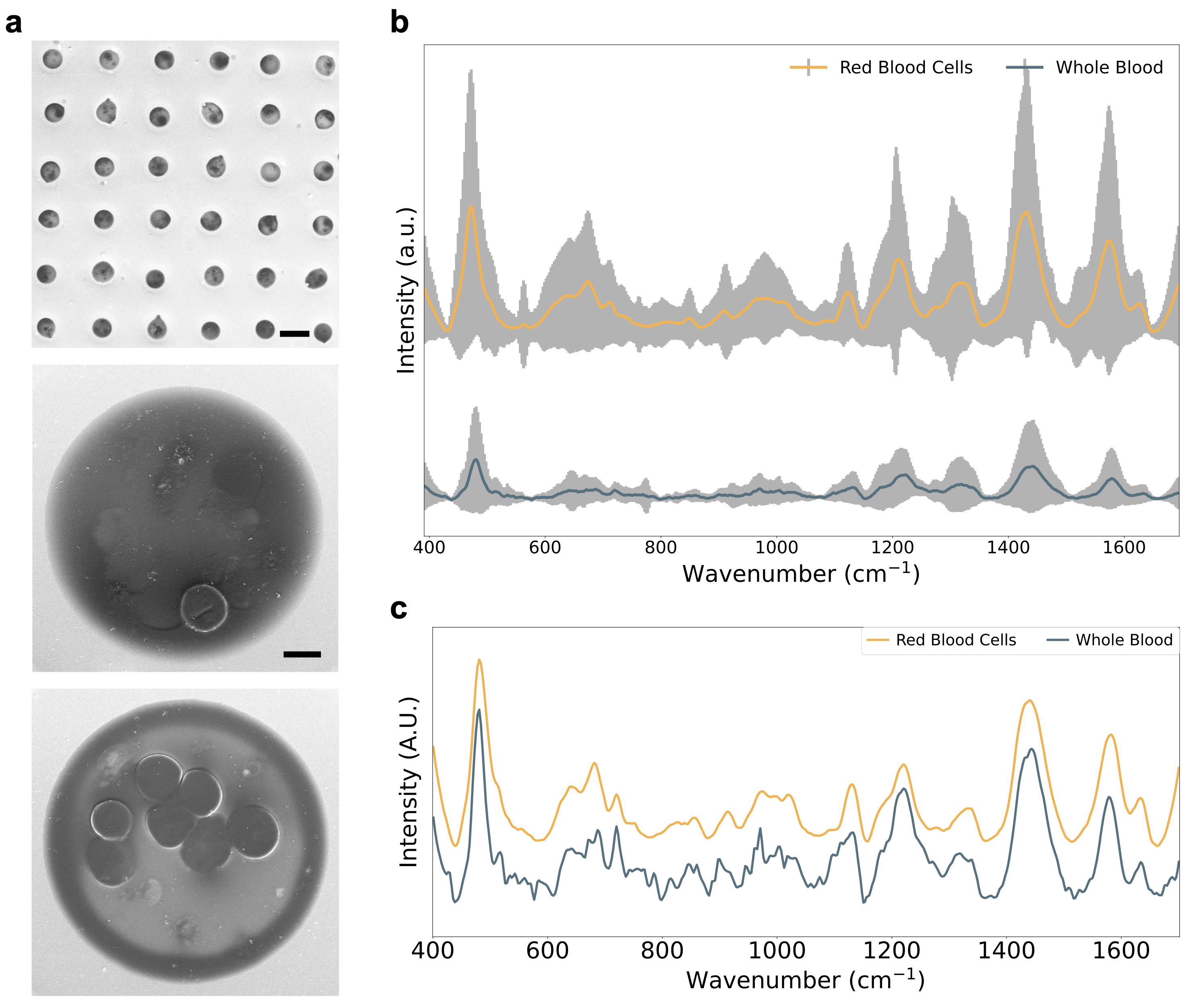}
\caption{Data from droplets print from mouse whole blood mixed in with GNRs. Mouse whole blood was purchased with the addition K2EDTA as an anti-coagulating agent for processing and shipment. Blood was then diluted 1:9 with our EDTA solution (1:9 ratio v/v of  Invitrogen UltraPure 0.5 M EDTA, Invitrogen 15575020 in DI water) for a final RBC count of 1.1 e9 cells/mL, closely approximating the RBC count of samples printed with RBC mixtures. GNRs at a concentration matching that all other cellular samples was added to the solution and then samples were printed onto a gold-coated slide and Raman spectra were collected. \textbf{a,} SEMs showing stable printing of solutions from diluted mouse whole blood mixed with GNRs. Top image shows a 6x6 grid of droplets as a subset of the hundreds printed. The remaining two SEMs show two representative droplets from the sample with 1 and 7 RBC, respectively. Scale bars are 50 $\mu$m and 5 $\mu$m. \textbf{b,} Mean spectra standard deviation of 100 spectra collected from samples printed from mouse RBCs and mouse whole blood, both diluted in aqueous EDTA to a final RBC count of ~1e9 cells/mL mixed with GNRs. Spectra are shown after baseline correction \textbf{c,} Normalized mean spectra from data shown in \textbf{b}. Normalized means show that while the whole blood exhibits similar peaks to that of the RBCs, there are many more minor peaks in the spectrum, highlighting expected sample complexity from the whole blood sample.}\label{fig}
\end{figure}

\begin{figure}[H]%
\centering
\includegraphics[width=1\textwidth]{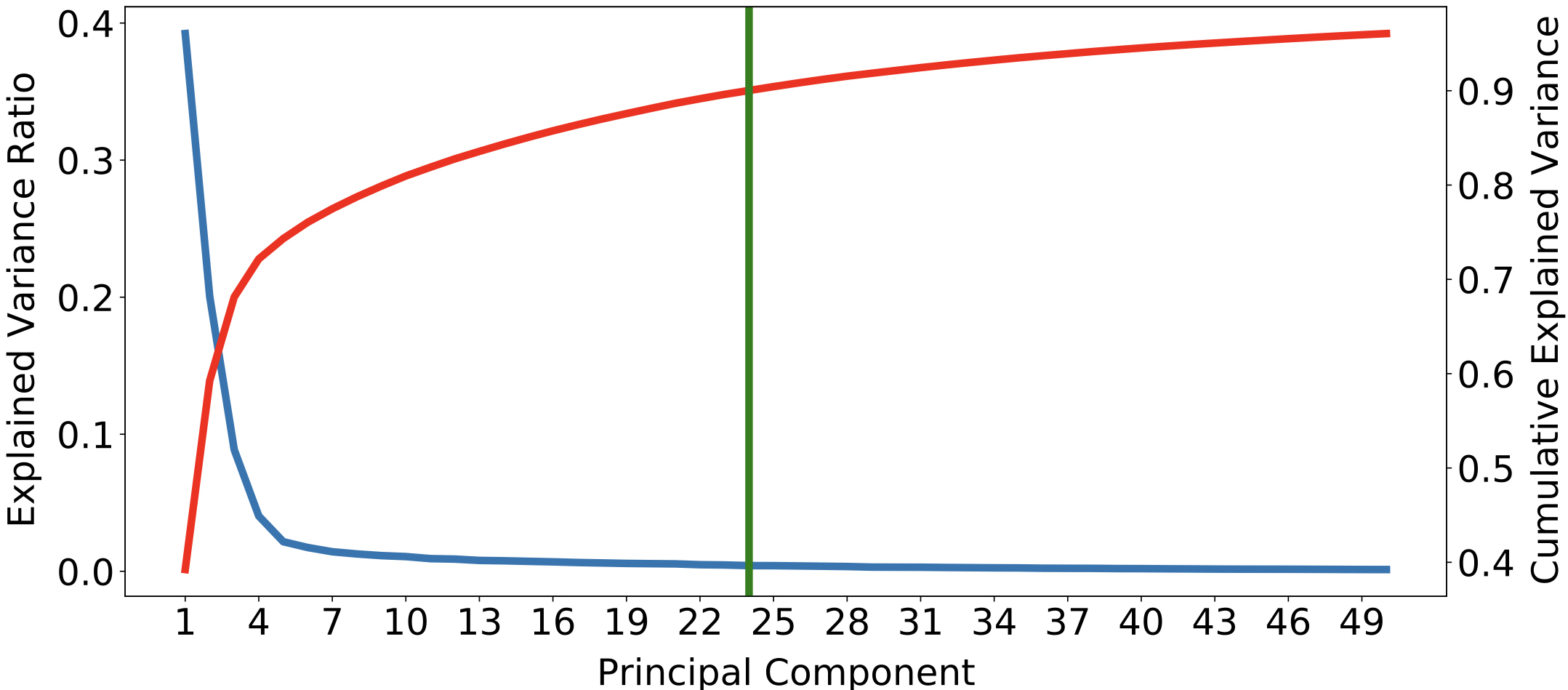}
\caption{Plot of the percentage of variance attributed to each principal component and the cumulative explained variance over 50 components. The green line indicates the number of PCA components necessary to capture 90\% of all explained variance in our samples. For all 300 spectra from our single cell-line droplets, we demonstrate that we can account for at least 90\% of all variance with 24 components generated from all 508 wavenumber features in our spectra.}\label{fig}
\end{figure}

\begin{figure}[H]%
\centering
\includegraphics[width=1\textwidth]{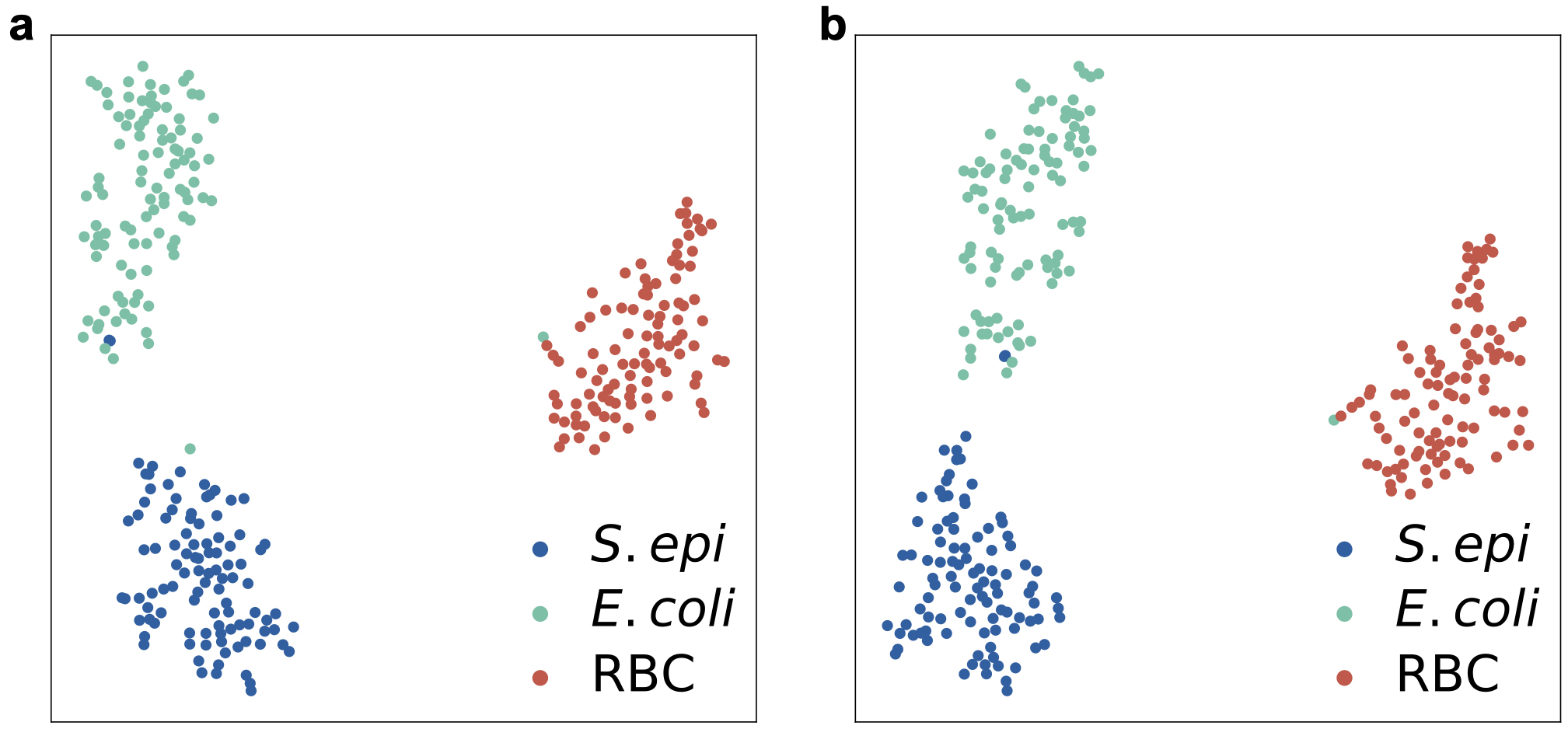}
\caption{Plots showing a 2-component, t-distributed stochastic neighbor embedding projection (t-SNE) with perplexity = 10 across our 3 single cell-line classes. Data is plotted \textbf{a,} with data inclusive of all wavenumber features and \textbf{b,} after performing a 24-component PCA for dimensionality reduction. Plots show relative clustering of our classes and minimal variation to clustering after dimensionality reduction.}\label{fig}
\end{figure}

\begin{figure}[H]%
\centering
\includegraphics[width=1\textwidth]{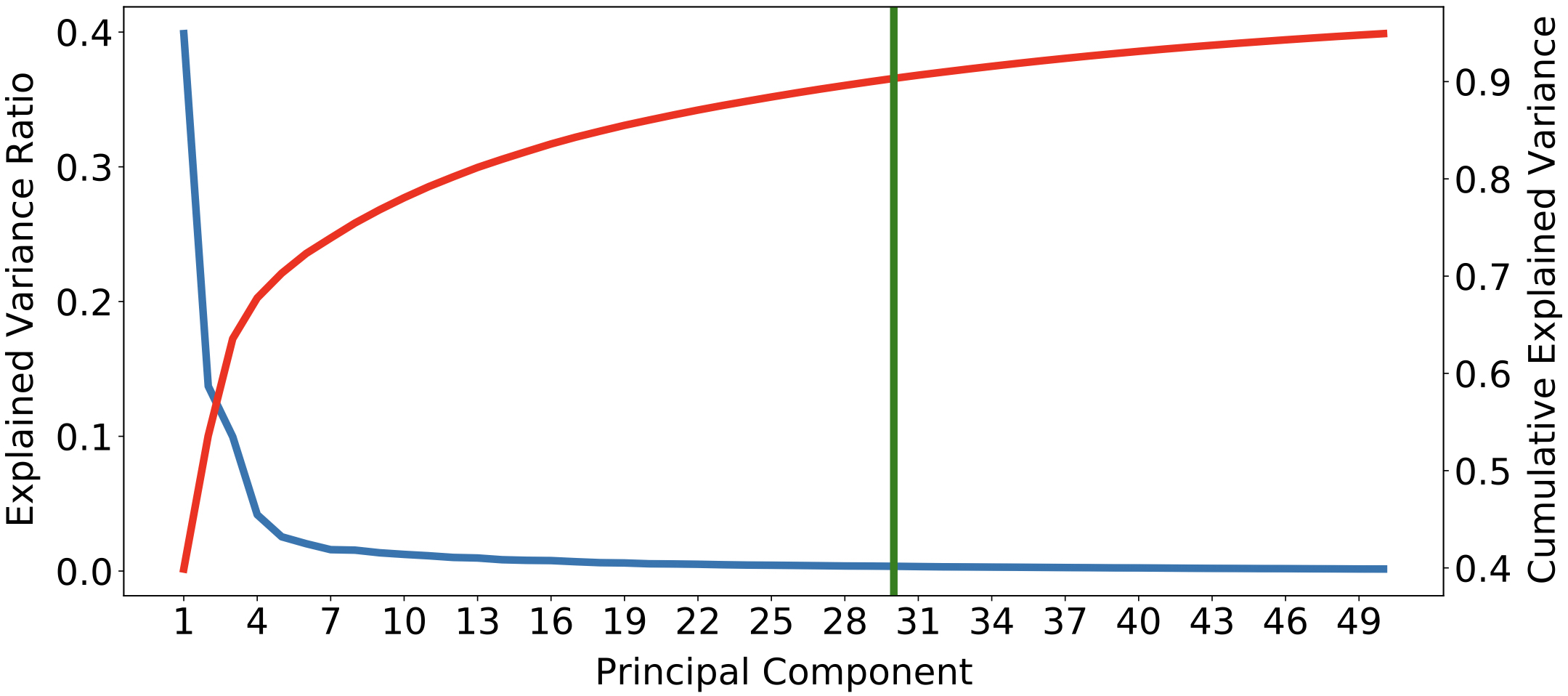}
\caption{Plot of the percentage of variance attributed to each principal component and the cumulative explained variance over 50 components. The green line indicates the number of PCA components necessary to capture 90\% of all explained variance in our samples. For all 600 spectra from our 3 single cell-line droplet classes and our 3 cell mixture classes, we demonstrate that we can account for at least 90\% of all variance with 30 components generated from all 508 wavenumber features in our spectra.}\label{fig}
\end{figure}

\begin{figure}[H]%
\centering
\includegraphics[width=1\textwidth]{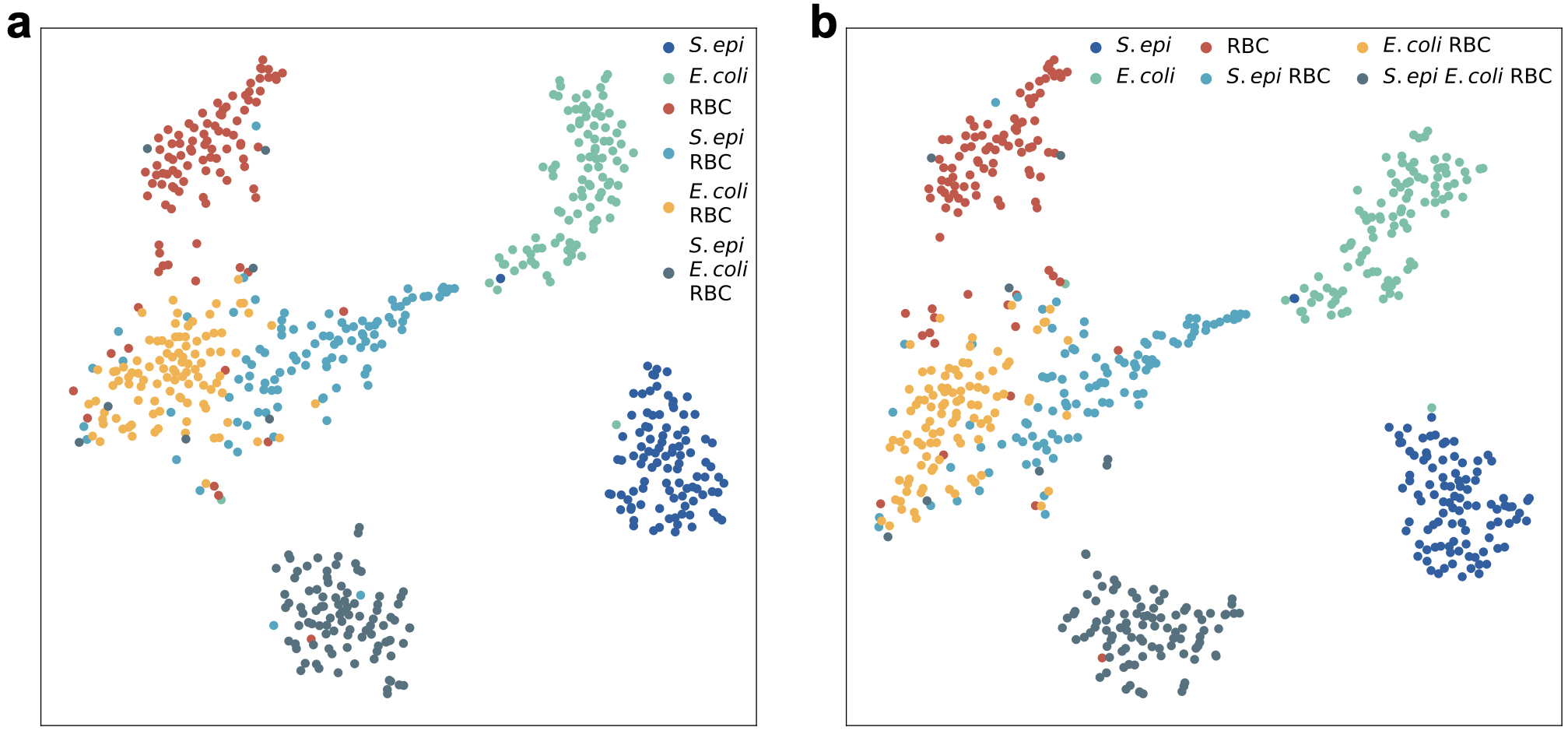}
\caption{Plots showing a 2-component, t-distributed stochastic neighbor embedding projection (t-SNE) with perplexity = 10 across all 6 of our classes. Data is plotted \textbf{a,} with data inclusive of all wavenumber features and \textbf{b,} after performing a 30-component PCA for dimensionality reduction. Plots show relative clustering of our classes and minimal variation to clustering after dimensionality reduction.}\label{fig}
\end{figure}

\begin{figure}[H]%
\centering
\includegraphics[width=0.6\textwidth]{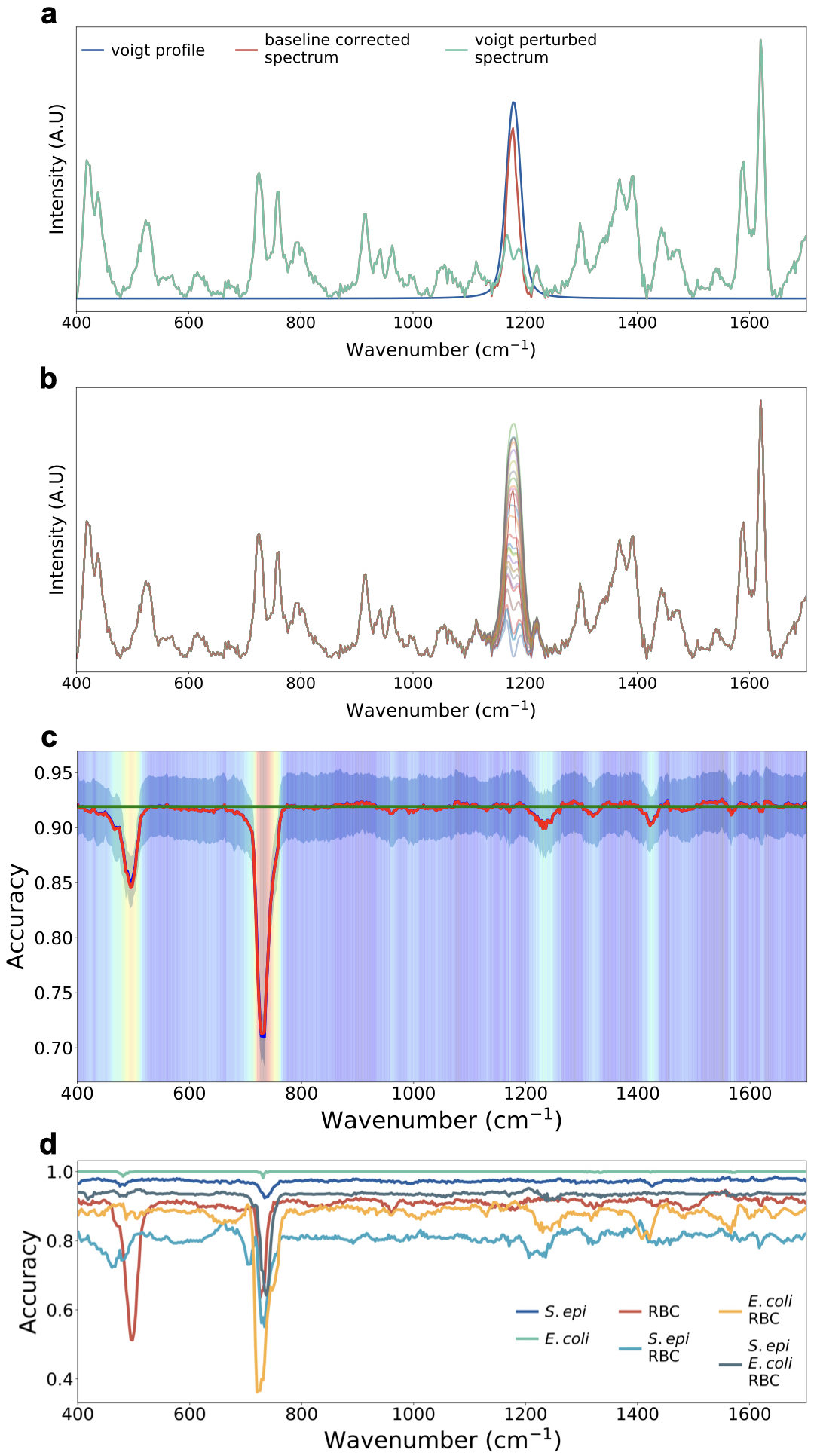}
\caption{Feature Validation. We perform feature validation on our spectra to determine which wavenumbers and spectral bands are most important for our classifier. We take our 600 spectra across all 6 cellular classes and split the samples using a stratified shuffle split into an 80:20 train/test split. For each spectrum in our training set, we iteratively perturb the spectrum at each wavenumber. After each perturbation, we calculate the classification accuracy and compare with our baseline accuracy.  Every wavenumber of each spectrum in the test set is perturbed 5 times and all results are averaged for our final feature extraction. Spectra were perturbed with a normalized Voigt profile. Line width chosen to roughly match peak widths seen in our spectra. \textbf{a,} plot showing an example Voight curve (blue), unperturbed example spectrum from our dataset (red), and perturbed spectrum (green). \textbf{b,} Voigt profile intensities were chosen through random sampling of all spectra in our training set. Plot shows an example of a spectrum from our sample set with 100 different perturbations. \textbf{c,} Heatmap highlighting feature validation performed to determine relative weight of spectral wavenumbers in our Random Forest classification. Heatmap is overlaid with a plot of mean and standard deviation of the perturbed classification accuracy (red) and f1 score (blue) calculated across all trials. Mean accuracy is plotted in green. Wavenumbers with lower accuracies are shown to be critical features, as random perturbations in these regions are highly correlated with decreases in classification accuracy. \textbf{d,} Plot of the mean classification accuracy broken down into accuracies across each of our cellular and mixture classes. }\label{fig}
\end{figure}

\clearpage
\begin{figure}[H]%
\centering
\includegraphics[width=1\textwidth]{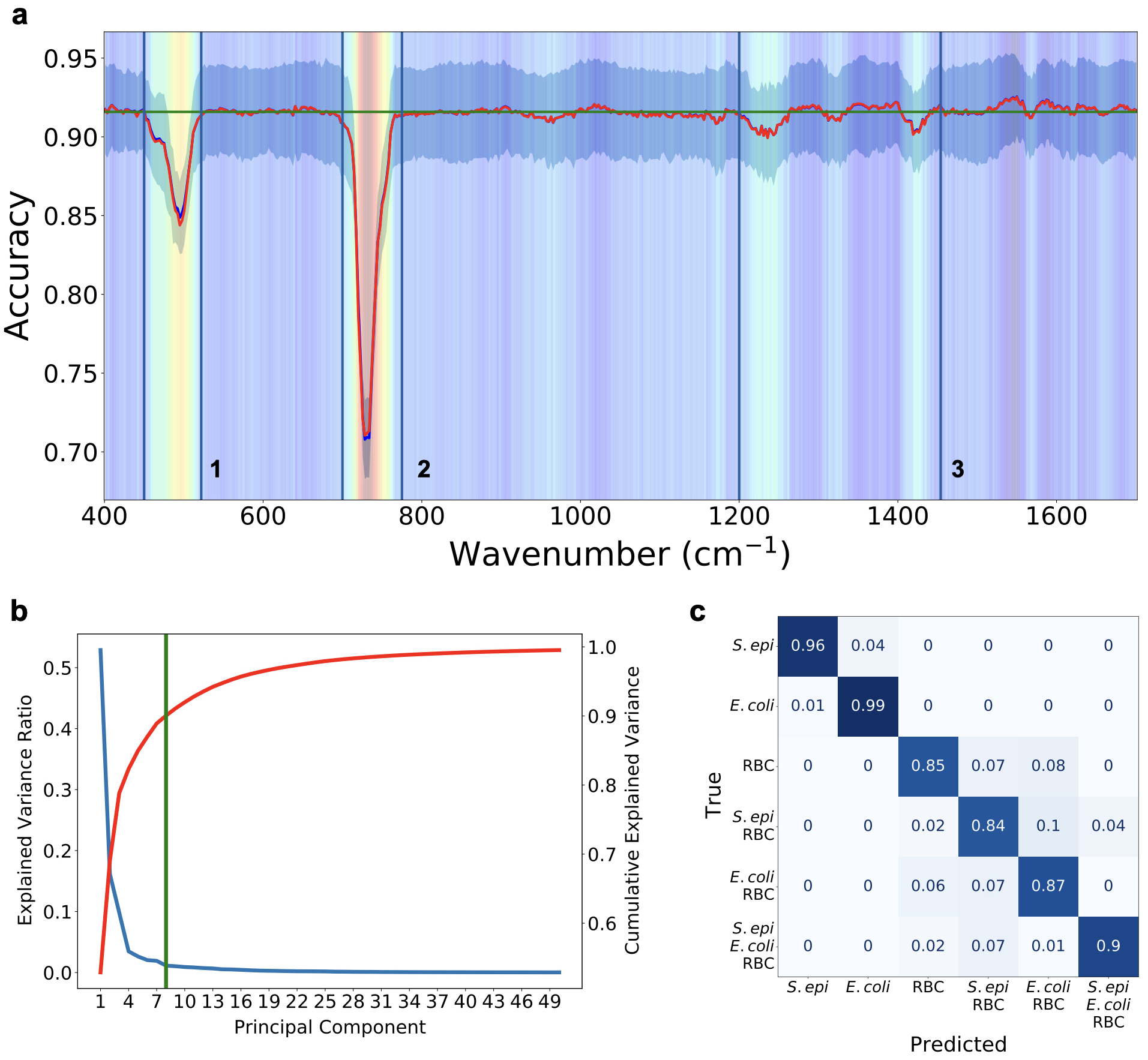}
\caption{Classification using spectral feature bands of interest evaluated across 600 spectra collected from single cell-line droplets of \textit{S. epi}, \textit{E. coli}, and mouse RBCs mixed with GNRs, and our 3 cell mixtures. \textbf{a,} Heatmap presented in Supplementary Fig. 22, highlighting feature importance calculations performed to determine relative weight of spectral wavenumbers in our random forest classification. Heatmap is overlaid with 3 bands representing key spectral bands used by our classifier. We further demonstrate that these bands are primarily responsible for our classification accuracies by preprocessing our spectra by removing spectral features outside these bands (420-522 cm\textsuperscript{-1}, 700-775 cm\textsuperscript{-1}, 1200-1454 cm\textsuperscript{-1}). We then reduced the dimensionality of our remaining features using an 8-component PCA as previously reported. \textbf{b,} Plot of the percentage of variance attributed to each principal component and the cumulative explained variance over 50 components. The green line indicates the number of PCA components necessary to capture 90\% of all explained variance in our samples. For this sample set taking only specific wavenumber bands from our spectra, we demonstrate that we can account for at least 90\% of all variance with only 8 components generated from all 508 wavenumber features in our spectra. \textbf{c,} Finally, we use our previously described random forest classifier on our samples and perform a stratified K-fold cross validation of our classifier’s performance across 10 splits. Results are plotted on a normalized confusion matrix. We show that we achieve $\ge$ 81\% classification accuracy across all samples as compared with the $\ge$ 87\% classification accuracy achieved when evaluating the entire spectra window from 400-1700 cm\textsuperscript{-1}. These results further validate our feature recognition model. Furthermore, they pave the way for future development of low cost POC systems by demonstrating that the use of low-cost spectrometers with limited spectral windows may be possible for such diagnostic work.}\label{fig}
\end{figure}

\begin{figure}[H]%
\centering
\includegraphics[width=1\textwidth]{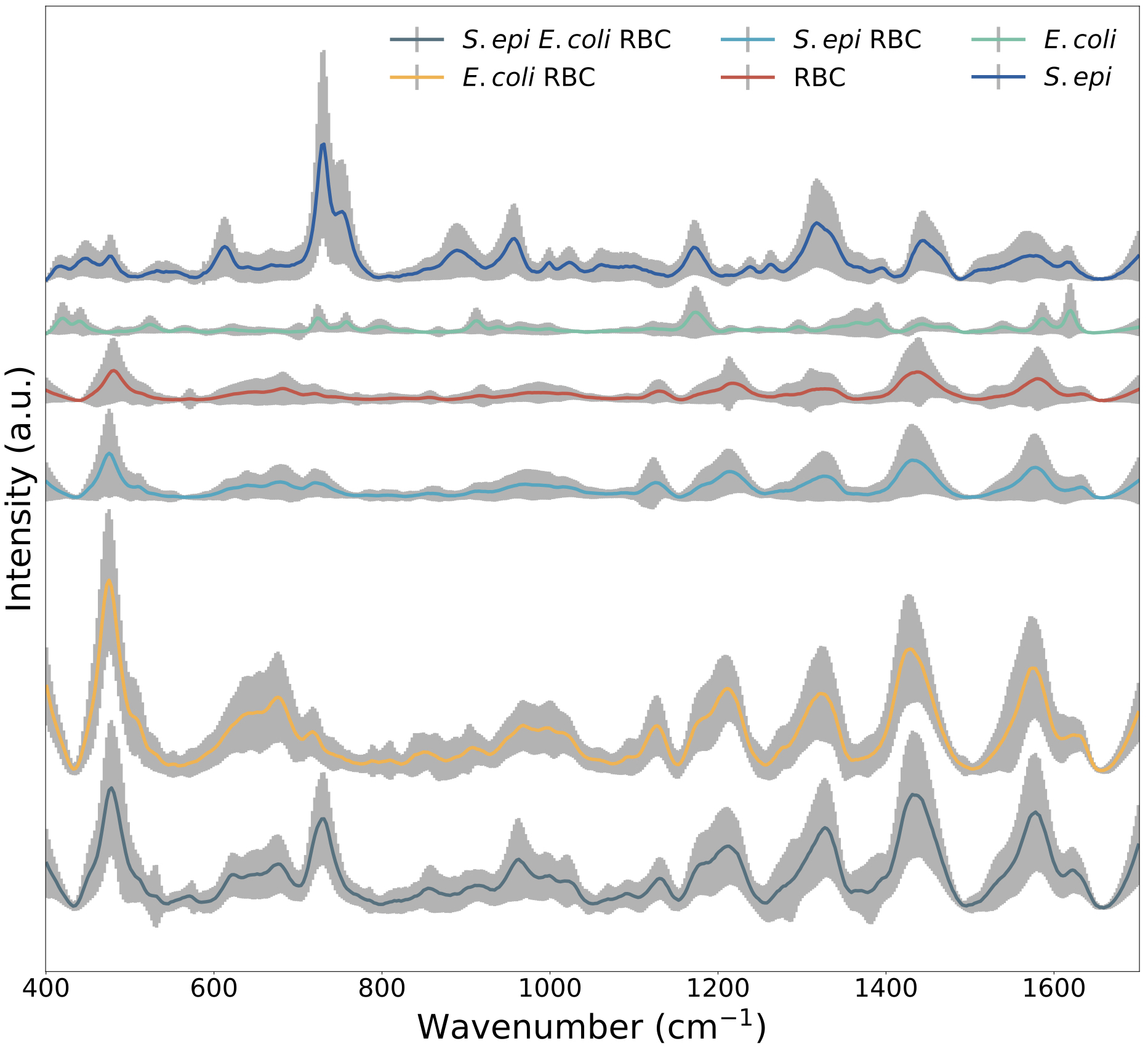}
\caption{Plot showing the mean and standard deviation of SERS spectra taken from droplets printed from our 6 droplet classes: three single-cell line classes (\textit{S. epi}, \textit{E. coli}, and RBCs) and three mixture classes (equal mixtures of \textit{S. epi} and RBCs, \textit{E. coli} and RBCs, and \textit{S. Epi}, \textit{E. coli}, and RBCs) all diluted to a final concentration of 1e9 cells/mL of each cell type in our aqueous EDTA solution and mixed with GNRs. Spectra were collected from 100 droplets for each class.}\label{fig}
\end{figure}

\begin{figure}[H]%
\centering
\includegraphics[width=0.9\textwidth]{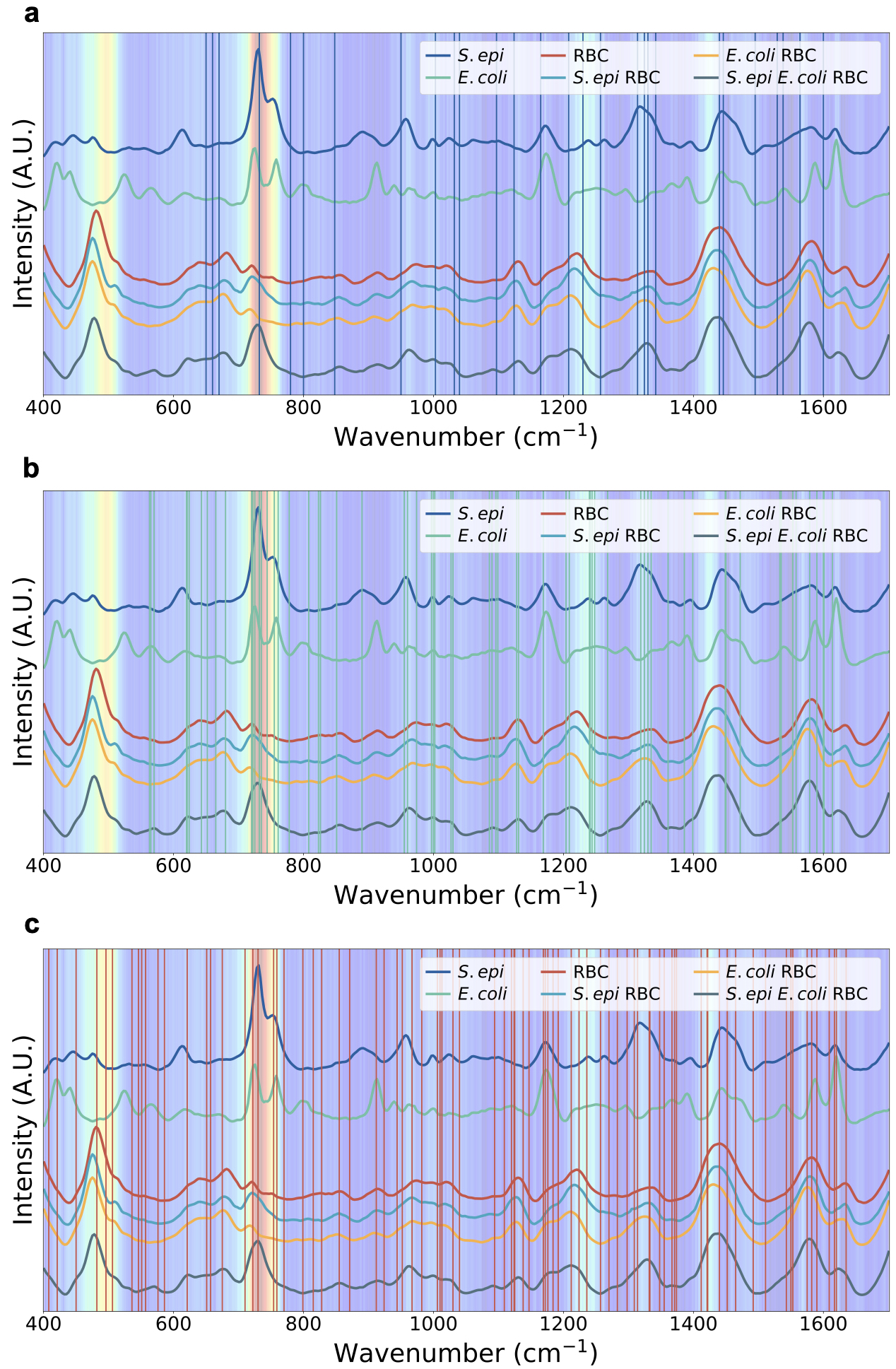}
\caption{Heatmap highlighting feature extraction performed to determine relative weight of spectral wavenumbers in our random forest classification. Heatmap is overlaid with the mean SERS spectra of 100 measurements each, taken from single droplets printed from three cell lines (\textit{S. epi}, \textit{E. coli}, and RBCs)  and three mixtures (\textit{S. epi} and RBCs, \textit{E. coli} and RBCs, and \textit{S. epi}, \textit{E. coli}, and RBCs) mixed with GNRs. Wavenumbers representative of biological peaks of dried and liquid SERS of \textbf{a,} \textit{S. epi}, \textbf{b,} \textit{E. coli}, and \textbf{c,} RBCs previously reported in the literature are plotted as vertical lines \cite{Tadesse2020-lm,Su2015-wa,Moghtader2018-lh,Witkowska2019-ox,Wang2010-hr,Sivanesan2014-er,Choi2020-fu,Zhou2014-ns,Drescher2013-ac,Premasiri2012-vb,Reokrungruang2019-qj,Paccotti2018-eo}.}\label{fig}
\end{figure}

\renewcommand{\tablename}{Supplementary Table}
\newpage
\begin{landscape}
 \begin{longtable}{p{1.5cm} p{1.5cm} p{5cm}|p{1.5cm} p{1.5cm} p{5cm}}
 \caption{Tentative band assignments of the SERS spectra of \textit{S. epi}, \textit{E. coli}, RBCs as reported in literature.\label{tab1}}\\
 \hline
 Peak Pos. (cm\textsuperscript{-1}) & Cell Type & Peak Assignment & Peak Pos. (cm\textsuperscript{-1}) & Cell Type & Peak Assignment\\
 \hline
 \endfirsthead
 
  \hline
 Peak Pos. (cm\textsuperscript{-1}) & Cell Type & Peak Assignment & Peak Pos. (cm\textsuperscript{-1}) & Cell Type & Peak Assignment\\
 \hline
 \endhead

 408 & RBC & $\delta(C_{\beta}C_aC_b)4 + \delta(C_{\beta}Me)$\cite{Premasiri2012-vb} & 1124 & \textit{S. epi} & $\nu(PO_2)$\cite{Paccotti2018-eo}\\
 421 & RBC & $\delta$(Fe–O–O)\cite{Drescher2013-ac} &  & RBC &  $\nu$13 or  $\nu$42\cite{Premasiri2012-vb}\\
 450 & RBC & Fe–O\cite{Drescher2013-ac} & 1125 & RBC & Proteins: C–N stretch and C–C stretch\cite{Drescher2013-ac}\\
 482 & RBC & $\gamma$12\cite{Drescher2013-ac} & 1128 & \textit{E. coli} & Amide III, adenine, polyadenine, and DNA\cite{Zhou2014-ns}\\
 496 & RBC & $\gamma$12\cite{Premasiri2012-vb} & 1138 & RBC &  $\nu$(C$_\beta$-Methyl),$\nu$5\cite{Drescher2013-ac}\\
 506 & RBC & Fe–O, Proteins: S–S stretch\cite{Drescher2013-ac} & 1147 & RBC & Proteins, lipids: C–N and C–C stretch\cite{Drescher2013-ac}\\
 536 & RBC & $\gamma$21, $\nu$25, Proteins: S–S stretch, skeletal deform\cite{Drescher2013-ac} & 1165 & \textit{S. epi} & Tyr, Phe, amide III\cite{Paccotti2018-eo}\\
 546 & RBC &  $\nu$(Fe–O–O)\cite{Drescher2013-ac} & 1169 & RBC &  $\nu$30\cite{Premasiri2012-vb}\\
 551 & RBC &  $\nu$49 or  $\nu$(Fe–O-O)\cite{Premasiri2012-vb} &  & \textit{E. coli} & 12-methyltetradaconic acid or 15-methylpalmitic acid or acetoacetate\cite{Su2015-wa}\\
 557 & RBC &  ${\nu}(Fe-O-O)$\cite{Drescher2013-ac} & 1172 & RBC &  ${\nu}$(pyr half-ring)$_{asym}$, ${\nu}30$\cite{Drescher2013-ac}\\
 
 565 & \textit{E. coli} & C-S-S-C\cite{Witkowska2019-ox} & 1176 & RBC & C-H bending, Tyr\cite{Reokrungruang2019-qj}\\
 563 & \textit{E. coli} & Carbohydrates\cite{Zhou2014-ns} & 1184 & RBC & Thr: CH$_3$ rocking\cite{Drescher2013-ac}\\
 570 & \textit{E. coli} & Carbohydrates\cite{Wang2010-hr} & 1192 & RBC & Thr: CH$_3$ rocking\cite{Drescher2013-ac}\\
 576 & RBC &  $\nu$(Fe–O$_2$)\cite{Drescher2013-ac} & 1204 & \textit{E. coli} & Phe\cite{Paccotti2018-eo}\\
 586 & RBC &  $\nu$48\cite{Premasiri2012-vb} &  1208 & \textit{S. epi} & Phe\cite{Paccotti2018-eo}\\
 620 & \textit{E. coli} & Phe\cite{Witkowska2019-ox} & 1209 &\textit{E. coli} & Aromatic amino acids in proteins\cite{Wang2010-hr}\\
 621 & \textit{E. coli} & C-C twisting mode of Phe\cite{Paccotti2018-eo} & 1212 & RBC &  $\delta$(C$_m$H),  $\nu$13 or  $\nu$42\cite{Drescher2013-ac} \\
  & RBC &  $\nu$12\cite{Premasiri2012-vb} & 1224 & RBC & $\delta$(C$_m$H)\cite{Drescher2013-ac} \newline $\nu$13 or  $\nu$42\cite{Drescher2013-ac,Premasiri2012-vb}\\
 624 & \textit{E. coli} & Aromatic ring skeleton\cite{Zhou2014-ns} & 1230 & \textit{S. epi} &  $\nu$(PO$_{2-}$), amide III\cite{Paccotti2018-eo}\\
 643 & \textit{E. coli} & Guanine ring breathing\cite{Paccotti2018-eo} & 1235 & \textit{E. coli} & Vibration of N-H\cite{Moghtader2018-lh}\\
 650 & \textit{S. epi} & Guanine ring breathing\cite{Paccotti2018-eo} & 1236 & RBC & Trp: ring\cite{Drescher2013-ac}\\
 651 & RBC & Cys: C–S stretch\cite{Drescher2013-ac} & 1240 & \textit{E. coli} & Amide III\cite{Witkowska2019-ox}\\
 652 & \textit{E. coli} & $\delta$(COO-)\cite{Zhou2014-ns} & 1241 & \textit{E. coli} &  $\nu$(PO$_{2-}$), amide III\\
 657 & RBC & $\delta$(pyr deform)$_{sym}$,  $\nu$7\cite{Drescher2013-ac} & 1248 & \textit{E. coli} & CH$_2$ stretching\cite{Su2015-wa}\\
 659 & \textit{E. coli} & Guanine (C-S)\cite{Su2015-wa} & 1257 & \textit{S. epi} & Amide III\cite{Paccotti2018-eo}\\
 660 & \textit{S. epi} & Guanine, thymine ring breathing\cite{Paccotti2018-eo} & & RBC & Glu: CH$_2$ wag; proteins, lipids: amide III, $\delta$(CH$_2$/CH$_3$)\cite{Drescher2013-ac}\\
 665 & \textit{E. coli} & NAG (N-acetylglucosamine)\cite{Paccotti2018-eo} & 1268 & \textit{E. coli} & $\delta$(CH$_2$) Amide III\cite{Zhou2014-ns}\\
 
 670 & \textit{S. epi} & NAG (N-acetylglucosamine)\cite{Paccotti2018-eo} & 1270 & RBCs & Proteins, lipids: amide III, $\delta$(CH$_2$/CH$_3$)\cite{Drescher2013-ac}\\
 675 & RBC &  $\nu$7\cite{Premasiri2012-vb} & 1283 & RBC & $\delta$(C$_m$H),  $\nu$21\cite{Drescher2013-ac}\\
 680 & \textit{E. coli} & Adenine\cite{Wang2010-hr} & 1298 & RBC &  $\delta$(C$_m$H),  $\nu$21\cite{Drescher2013-ac}\\
 710 & RBC &  $\nu$11\cite{Premasiri2012-vb} & 1309 & RBC &  $\nu$21\cite{Premasiri2012-vb}\\
 720 & \textit{E. coli} & Adenine in Flavin adenine dinucleotide (FAD) and Nicotinamide Adenine Dinucleotide (NAD)\cite{Witkowska2019-ox} & 1314 & \textit{S. epi} & Guanine, CH$_2$ twist (lipids)\cite{Paccotti2018-eo}\\
 722 & \textit{E. coli} & Adenine\cite{Su2015-wa} &  & RBC & Phe, Glu, Ser, Met, His: CH$_2$ wag\cite{Drescher2013-ac}\\
  & RBC & Amino acids: $\delta$(COO-)\cite{Drescher2013-ac} & 1319 & \textit{E. coli} & Guanine, CH$_2$ twist (lipids)\cite{Paccotti2018-eo}\\
 725 & \textit{E. coli} & Adenine ring breathing\cite{Paccotti2018-eo} & 1324.5 & \textit{S. epi} & Protein and carboxylic stretches\cite{Choi2020-fu}\\
 730 & RBC & Amino acids: $\delta$(COO-)\cite{Drescher2013-ac} &  & \textit{E. coli} & Protein and carboxylic stretches\cite{Choi2020-fu}\\
 731 & \textit{S. epi} & Adenine part of the flavin derivatives or glycosidic ring mode of polysaccharide\cite{Sivanesan2014-er} & 1330 & \textit{S. epi} & Adenine part of the flavin derivatives or glycosidic ring mode of polysaccharides\cite{Sivanesan2014-er}\\
  & \textit{E. coli} & Adenine part of the flavin derivatives or glycosidic ring mode of polysaccharides\cite{Sivanesan2014-er} &  & \textit{E. coli} & Adenine part of the flavin derivatives or glycosidic ring mode of polysaccharides\cite{Sivanesan2014-er} \newline
 $\nu$(NH$_2$) adenine, polyadenine, DNA\cite{Zhou2014-ns} \newline
CH$_2$/CH$_3$ wagging mode in purine bases of nucleic acids\cite{Witkowska2019-ox}\\

 732.5 & \textit{S. epi} & purine ring-breathing modes\cite{Choi2020-fu} & 1332 & RBC &  $\nu$41\cite{Premasiri2012-vb}\\
 735 & \textit{E. coli} & adenine, glycosidic ring mode\cite{Zhou2014-ns} & 1333 & RBC &  $\nu$(pyr half-ring)$_{sym}$, Proteins: $\delta$(CH); CH$_2$, CH$_3$ wag\cite{Drescher2013-ac}\\
 744 & \textit{E. coli} & B$_{1g}$ heme vibration (cytochrome c)\cite{Paccotti2018-eo} & 1335 & \textit{E. coli} & CH$_2$ deformation or Trp\cite{Su2015-wa} \newline
Protein twisting (CH$_2$ and CH$_3$),  $\nu$(NH$_2$) adenine\cite{Moghtader2018-lh,Paccotti2018-eo}\\
 754 & RBC &  $\nu$15\cite{Premasiri2012-vb} \newline Porphyrin ring breathing\cite{Reokrungruang2019-qj} & 1342 & \textit{S. epi} & Protein twisting (CH$_2$ and CH$_3$),  $\nu$(NH$_2$) adenine\cite{Paccotti2018-eo}\\
 755 & \textit{E. coli} & Trp ring breathing\cite{Paccotti2018-eo} & 1348 & RBC & Glu, Asp, Asn, Gln: CH$_2$ sciss, Ala, Leu, Val, Ile: CH$_3$ deform\cite{Drescher2013-ac}\\
 759 & RBC &  $\nu$(pyr breathe),  $\nu$15\cite{Drescher2013-ac} & 1355 & \textit{E. coli} & Ch deformation vibrations\cite{Moghtader2018-lh}\\
 761 & \textit{E. coli} & Ring I deformation\cite{Wang2010-hr} &  & RBC & Glu, Asp, Asn, Gln: CH$_2$ sciss, Ala, Leu, Val, Ile: CH$_3$ deform\cite{Drescher2013-ac}\\
 770 & RBC & Trp: indole sym. breathe\cite{Drescher2013-ac} & 1361 & \textit{E. coli} & =CH in plane (lipid) or amide III (protein)\cite{Wang2010-hr}\\
 778 & \textit{E. coli} & DNA/RNA ring breathing (cytosine/thymine)\cite{Paccotti2018-eo} & 1367 & RBC & Half ring stretching, porphyrinReokrungruang2019-qj\\
 780 & \textit{S. epi} & DNA/RNA ring breathing (cytosine/thymine)\cite{Paccotti2018-eo} & 1368 & \textit{E. coli} &  $\nu$(COO-) and $\delta$(C-H) proteins\cite{Paccotti2018-eo}\\
 799 & RBC &  $\nu$(pyr breathe),  $\nu$6\cite{Drescher2013-ac} & 1371 & RBC & $\nu$4\cite{Premasiri2012-vb}\\
 800 & \textit{S. epi} & DNA/RNA ring breathing & 1374 & RBC &  $\nu$(pyr half-ring)$_{sym}$,  $\nu$4, Proteins, lipids: $\delta$(CH$_3)_{sym}$\cite{Drescher2013-ac}\\
 
 808 & \textit{E. coli} &  $\nu$(CN) Tyr, Val\cite{Zhou2014-ns} & 1386 & \textit{E. coli} & $\delta$(CH$_3$) symmetrical\cite{Paccotti2018-eo}\\
 815 & RBC & Ser: $\gamma$(COO-)\cite{Drescher2013-ac} & 1399 & \textit{E. coli} & C-O-O- stretching in amino acids\cite{Paccotti2018-eo}\\
 823 & \textit{E. coli} & Different C-N stretch\cite{Wang2010-hr} & 1412 & RBC &  $\nu$(pyr quarter-ring)$_{sym}$,  $\nu$20, Proteins: COO- sym stretch\cite{Drescher2013-ac}\\
 826 & \textit{E. coli} & V$_a$(O-P-O) str.\cite{Paccotti2018-eo} & 1421 & RBC &  $\nu$(C$\_alpha$C$_m)_{sym}$\\
 828 & RBC & $\gamma$(CmH)\cite{Drescher2013-ac} & 1422 & RBC &  $\nu$28\cite{Premasiri2012-vb}\\
 848 & \textit{S. epi} & Thymine\cite{Paccotti2018-eo} & 1440 & \textit{S. epi} & protein or lipid\cite{Choi2020-fu}\\
 851 & \textit{E. coli} & Thymine\cite{Paccotti2018-eo} &  & RBC & CH$_2$ deformation - lipid, protein\cite{Reokrungruang2019-qj}\\
 855 & RBC & $\gamma$-Porphyrin\cite{Drescher2013-ac} & 1446 & \textit{S. epi} & Scissoring (fatty acids, phospholipids, and mono- and oligo-saccharides); CH$_2$/CH$_3$ deformation\cite{Paccotti2018-eo}\\
 871 & RBC & $\gamma$(C$_m$H)\cite{Drescher2013-ac}, $\gamma$10 & 1449 & \textit{E. coli} & Scissoring (fatty acids, phospholipids, and mono- and oligo-saccharides); CH$_2$/CH$_3$ deformation\cite{Paccotti2018-eo}\\
 912 & RBC & Glu, Ile, Thr, Lys: C–C stretch\cite{Drescher2013-ac} & 1450 & \textit{E. coli} & CH$_2$/CH$_3$ deformation of proteins and lipids\cite{Witkowska2019-ox}\\
 924 & RBC & Amino acids: C-COO- stretch\cite{Drescher2013-ac} & 1452 & RBC & $\delta$(CH$_2$/CH$_3$), Lipids:$\delta$(CH$_2$/CH$_3$)\cite{Drescher2013-ac}\\
 944 & RBC &  $\nu46$\cite{Premasiri2012-vb} & 1465 & RBC &  $\nu(C_{\alpha}C_m)_{sym}$, ${\nu}3$, Lipids:$\delta(CH_2/CH_3)$\cite{Drescher2013-ac}\\
 950 & \textit{S. epi} &  $\nu(CH_3)$ of proteins (a-helix)\cite{Paccotti2018-eo} & 1472 & \textit{E. coli} & CH$_2$ deformation of the protein molecules\cite{Su2015-wa,Moghtader2018-lh}\\
 
 952 & RBC & $\gamma(C_aH=)$\cite{Drescher2013-ac} & 1492 & RBC & trp:indole ring bend, indole CH bend\cite{Drescher2013-ac}\\
 955 & \textit{E. coli} &  $\nu(CH_3)$ of proteins (a-helix)\cite{Paccotti2018-eo} \newline
 $\nu$(CN)\cite{Zhou2014-ns} \newline C=C deformation\cite{Witkowska2019-ox} & 1495 & \textit{S. epi} & $\delta(CH_2)$\cite{Paccotti2018-eo}\\
 960 & \textit{E. coli}\cite{Su2015-wa} & $\delta$(C=C) or tyrosine & 1511 & RBC &  $\nu$38\cite{Premasiri2012-vb}\\
 967 & RBC & Proteins: C–C stretch\cite{Drescher2013-ac} & 1529 & \textit{S. epi} & Amide II of proteins, N-acetyl related bands (amide II)\cite{Paccotti2018-eo}\\
 974 & \textit{E. coli} & C=C deformation\cite{Wang2010-hr} & 1533 & \textit{E. coli} & Amide II of proteins, N-acetyl related bands (amide II)\cite{Paccotti2018-eo}\\
 982 & RBC & $\gamma(C_aH=)$\cite{Drescher2013-ac} & 1535 & \textit{E. coli} & adenine, cytosine, and guanine\cite{Paccotti2018-eo}\\
 997 & \textit{E. coli} & Phe or glucose\cite{Su2015-wa} & 1538 & \textit{S. epi} & Amide II of proteins\cite{Paccotti2018-eo}\\
 1000 & \textit{E. coli} & Phe\cite{Paccotti2018-eo} & 1543 & RBC &  $\nu$11\cite{Premasiri2012-vb}\\
 1002 & \textit{S. epi} & Phe\cite{Paccotti2018-eo} & 1550 & RBC &  $\nu$(C$_\beta$C$_\beta$), $\nu$11, Proteins, lipids: amide II, Trp\cite{Drescher2013-ac}\\
 1006 & RBC & Phe: indole asymmetric ring breathe\cite{Drescher2013-ac} & 1553 & \textit{E. coli} & Amide II of proteins\cite{Paccotti2018-eo}\\
 1010 & RBC &  $\nu$45\cite{Premasiri2012-vb} &  & RBC & N-H, Trp\cite{Reokrungruang2019-qj}\\
 1013 & RBC & Trp: indole asymmetric ring breathe\cite{Drescher2013-ac} & 1558 & \textit{E. coli} & CH$_2$ deformation\cite{Wang2010-hr}\\
 1027 & \textit{E. coli} & A Ring stretching or (C-H) deformation\cite{Su2015-wa} & 1564 & \textit{S. epi} & Amide II of proteins, guanine/adenine\cite{Paccotti2018-eo}\\
 
 1030 & \textit{E. coli} & Phe: C-H in plane bending\cite{Paccotti2018-eo} & 1575 & RBC & skeletal mode, trp\cite{Drescher2013-ac}\\
  & RBC & Phe: in-plane ring CH deform\cite{Drescher2013-ac} & 1579 & \textit{E. coli} & Guanine, adenine, trp (proteins)\cite{Reokrungruang2019-qj}\\
 1032 & \textit{S. epi} & Phe: C-H in plane bending\cite{Paccotti2018-eo} & 1582 & RBC & C-C asymmetric stretching, hemoglobin\cite{Reokrungruang2019-qj}\\
 1040 & \textit{S. epi} & $\nu$(CC) aromatic ring\cite{Paccotti2018-eo} & 1590 & \textit{E. coli} & Phe, hydroxyproline, Tyr\cite{Witkowska2019-ox}\\
  & RBC &  $\delta(=C_bH_2)_{asym}$\cite{Drescher2013-ac} &  & RBC &  $\nu(C_{\alpha}C_m)_{asym}$, $\nu$37, Phe, Tyr\cite{Drescher2013-ac}\\
 1056 & \textit{E. coli} & Stretching vibration of C-C in alkanes\cite{Moghtader2018-lh} & 1600 & \textit{S. epi} & Protein and carboxylic stretches\cite{Choi2020-fu}\\
 1086 & \textit{E. coli} & Phe\cite{Su2015-wa} &  & \textit{E. coli} & Protein and carboxylic stretches\cite{Choi2020-fu}\\
 1090 & \textit{E. coli} & C-C skeletal and C-O-C stretching from glycosidic link\cite{Witkowska2019-ox} & 1601 & \textit{E. coli} & Tyr, C-N stretching vibration\\
 1094 & RBC & $=C_{2vinyl}H$\cite{Drescher2013-ac} & 1609 & RBC &  $\nu(C_a=C_b)$, $\nu(C=C)_{vinyl}$\cite{Drescher2013-ac}\\
 1096 & \textit{S. epi} & $\nu_s(PO_2)$\cite{Paccotti2018-eo} & 1614 & \textit{E. coli} & adenine, guanine (ring stretching)\cite{Wang2010-hr}\\
 1097 & \textit{E. coli} & $\nu_s(PO_2)$\cite{Paccotti2018-eo} & 1617 & RBC & C=C trp, tyrosine\cite{Reokrungruang2019-qj}\\
 1099 & \textit{E. coli} & carbohydrates, C-C, C-O, -C-OH\cite{Wang2010-hr} & 1620 & RBC &  $\nu(C=C)_{vinyl}$\cite{Premasiri2012-vb}\\
 1109 & RBC & Proteins, lipids: C–N and C–C stretch\cite{Drescher2013-ac} & 1635 & RBC &  $\nu(C_{\alpha}C_m)_{asym}$, $\nu10$\cite{Drescher2013-ac}\\
 1120 & RBC &  $\nu5$\cite{Premasiri2012-vb} &  &  & \\

 \end{longtable}
 
\label{Note: }{\textbf{Note: }The Raman shift can vary slightly, depending on cell culture media, bacterial strain, and Raman substrate as well as acquisition parameters including excitation wavelength and temperature\cite{Drescher2013-ac}, and as such, the peak assignments are often treated as approximations\cite{Moghtader2018-lh}.}
\newline
\newline
\label{Note: } {\textbf{Abbreviations: }pyr, pyrrol; deform, deformation; sym, symmetric; asym, asymmetric; Cys, cysteine; Lys, lysine; Glu, glutamic acid; Ile, isoleucine; Phe, phenylalanine; Met, methionine; His, histidine; Asp, aspartic acid; Asn, asparagine; Gln, glutamine; Ala, alanine; Leu, leucine; Val, valine; Trp, tryptophan; Ser, serine; Thr, threonine; breathe, breathing; sciss, scissoring; stretch, stretching; wag, wagging. Vibrations: $\nu$, valence; $\delta$, deformation, $\gamma$, deformation (out of plane)\cite{Drescher2013-ac}.}

\end{landscape}

\begin{figure}[H]%
\centering
\includegraphics[width=0.9\textwidth]{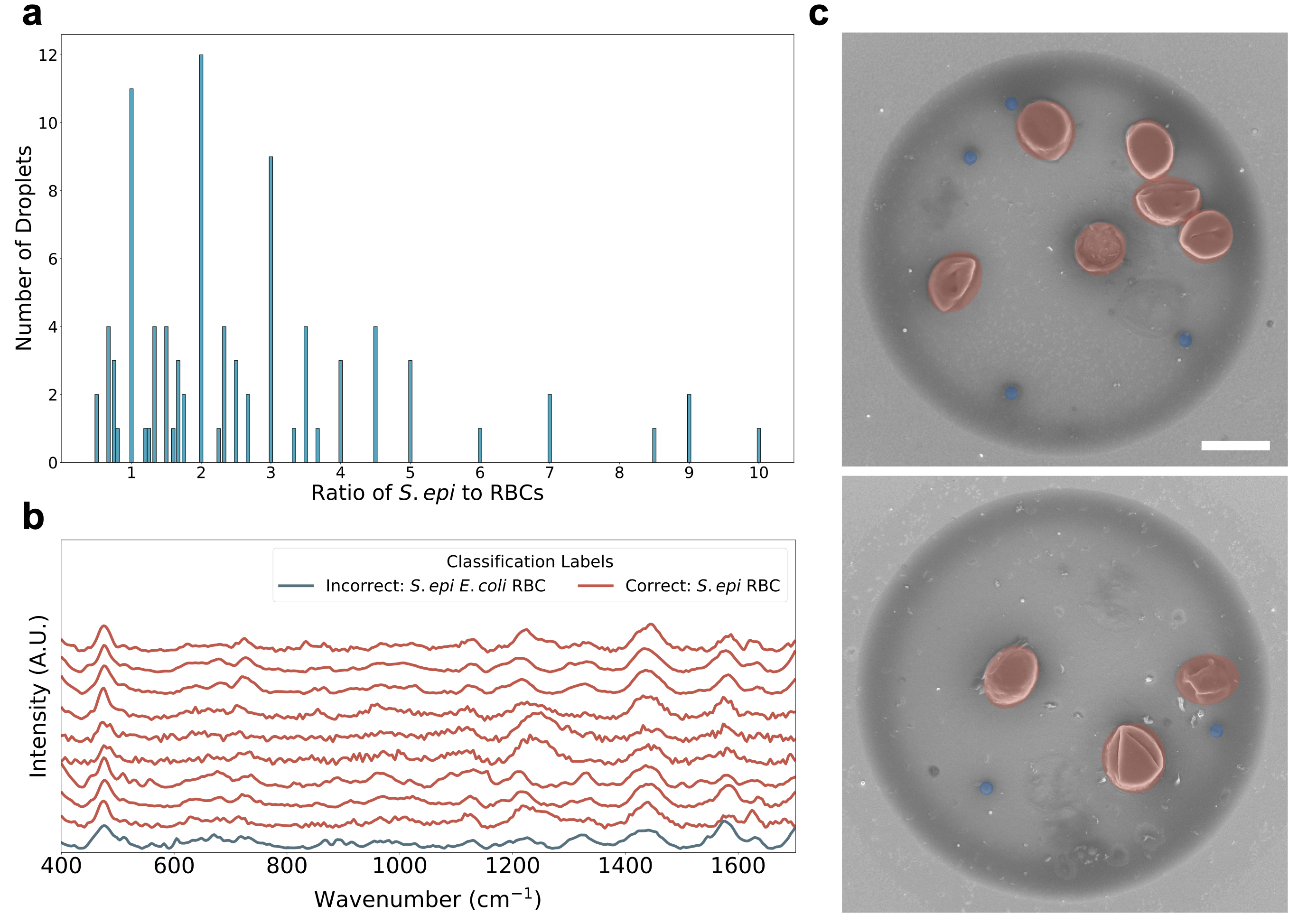}
\caption{We further analyze samples printed from mixtures of \textit{S. epi} bacteria, RBCs, and GNRs to understand how our ML algorithms would handle droplets printed with a greater number of RBCs as compared to \textit{S. epi}. For the 100 droplets analyzed in Figure 4 from the main text, we count the number of \textit{S. epi} bacteria and RBCs and \textbf{a,} plot ratio of \textit{S. epi}:RBCs in each of the droplets. Of note, 8 droplets had 0 RBCs and therefore are not included on the plot. \textbf{b,} There were 10 droplets with a greater number of RBCs than \textit{S. epi} in each droplet. For these droplets, we train our ML algorithm on the remaining 590 spectra in our 6 cellular classes. We then test it on these 10 spectra. The pre-processed spectra are plotted above, with the color indicating the classifier prediction. We got a 90\% classification accuracy, with 1 spectra falsely classified as containing both \textit{S. epi} and \textit{E. coli} bacteria along with RBCs. \textbf{c,} We show two representative SEMs from these 10 droplets, with the RBCs and \textit{S. epi} false colored in red and blue, respectively. The scale bar is 5 $\mu$m. These results are indicative of what we might see in a more clinically relevant sample with a greater concentration of RBCs than bacteria. We see that even in our limited dataset, we get comparable classification accuracies to samples with greater bacterial numbers, showing promise for our platform’s ability to detect lower bacterial concentrations.}\label{fig}
\end{figure}

\begin{figure}[H]%
\centering
\includegraphics[width=0.9\textwidth]{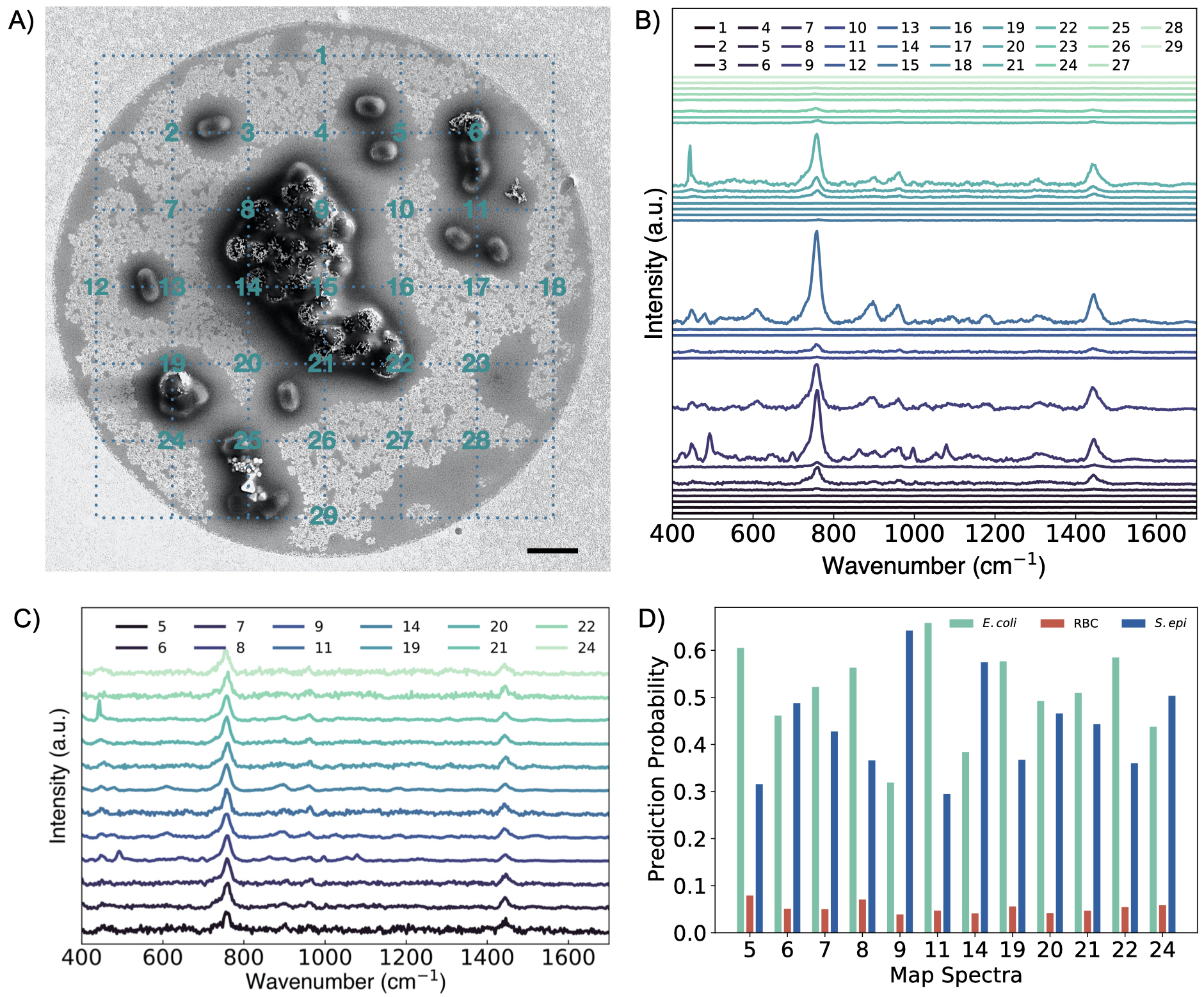}
\caption{Raman mapping of droplets printed from \textit{S. epi}, \textit{E. coli}, and GNR mixtures with a focal spot size of 0.7 $\mu$m chose to allow for single cell interrogation. \textbf{a,} SEM of a single droplet with an overlay showing the 29 XY coordinates of spectra collection points. Scale bar is 2 $\mu$m.  \textbf{b,} Waterfall plot showing the Raman spectra collected at each XY location. Spectra were preprocessed with a baseline fit. \textbf{c,} Waterfall plot showing the 12 remaining spectra after further preprocessing. First, spectra below an intensity count of 150 were removed to eliminate spectra taken from the substrate in locations without cells. Subsequently, remaining spectra were normalized to a zero mean and unit variance using the Scikit-learn python library Standard Scaler function \cite{Pedregosa2011-rk}.  \textbf{d,} Processed spectra were then evaluated using a SVM, Scikit-learn,\cite{Pedregosa2011-rk} optimized and trained on our known sample set of 300 droplets (Figure 3c). The plot shows the prediction probabilities across each of our 3 cell types, predicted for each processed spectrum.}\label{fig}
\end{figure}

\begin{figure}[H]%
\centering
\includegraphics[width=0.9\textwidth]{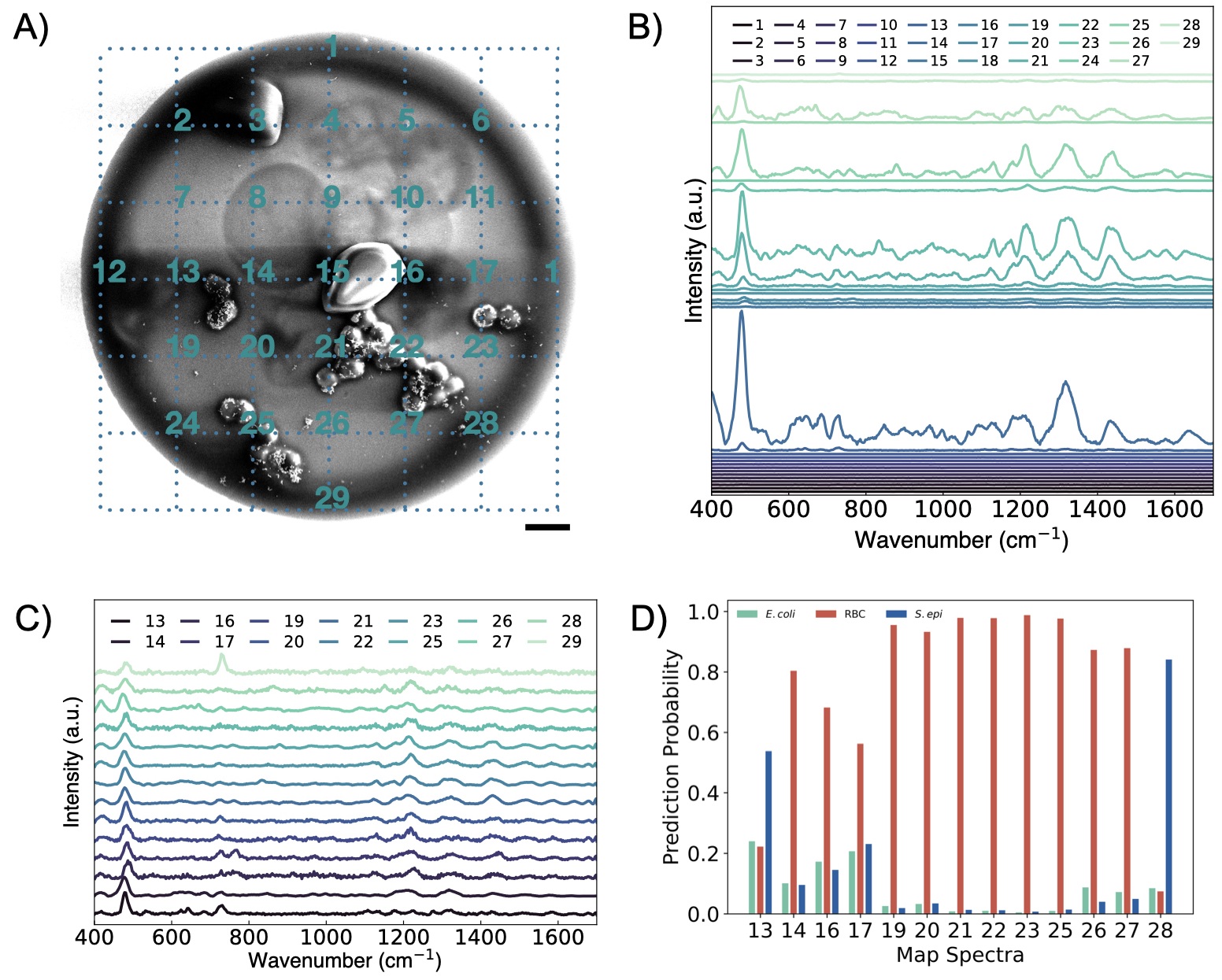}
\caption{Raman mapping of droplets printed from \textit{S. epi}, RBCs, and GNR mixtures with a focal spot size of 0.7 $\mu$m chose to allow for single cell interrogation. \textbf{a,} SEM of a single droplet with an overlay showing the 29 XY coordinates of spectra collection points. Scale bar is 2 $\mu$m.  \textbf{b,} Waterfall plot showing the Raman spectra collected at each XY location. Spectra were preprocessed with a baseline fit. \textbf{c,} Waterfall plot showing the 14 remaining spectra after further preprocessing. First, spectra below an intensity count of 150 were removed to eliminate spectra taken from the substrate in locations without cells. Subsequently, remaining spectra were normalized to a zero mean and unit variance using the Scikit-learn python library Standard Scaler function \cite{Pedregosa2011-rk}.\textbf{d,} Processed spectra were then evaluated using a SVM, Scikit-learn,\cite{Pedregosa2011-rk} optimized and trained on our known sample set of 300 droplets (Figure 3c). The plot shows the prediction probabilities across each of our 3 cell types, predicted for each processed spectrum.}\label{fig}
\end{figure}

\newpage
\begin{figure}[H]%
\centering
\includegraphics[width=.85\textwidth]{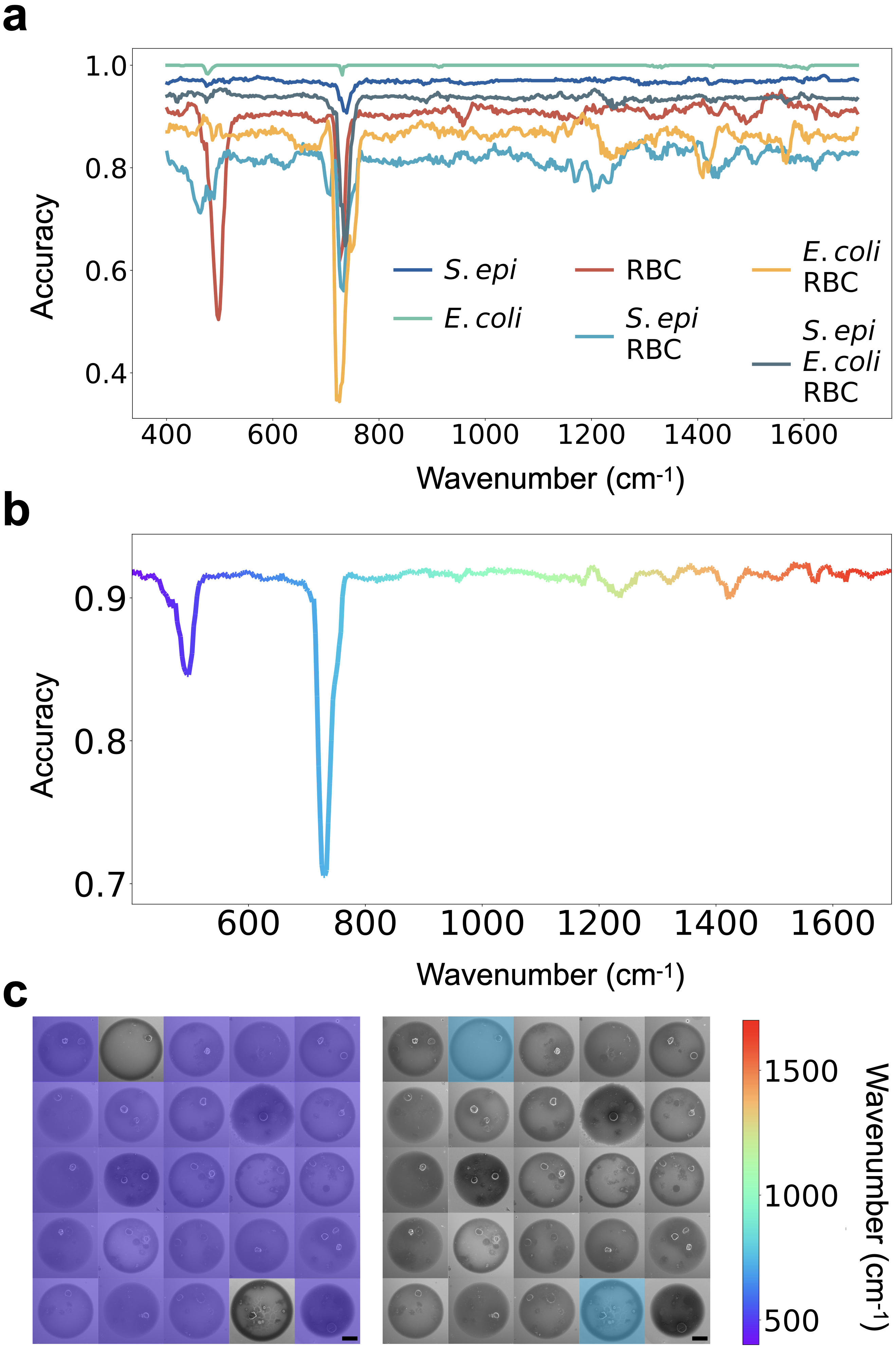}
\caption{Future vision for Raman hyperspectral imaging. \textbf{a,} Calculated plot of the classification accuracy for each independent sample class, highlighting wavenumber importance for each individual class. Wavenumbers with lower accuracies are shown to be critical features as random perturbations are highly correlated with decreases in classification accuracy. \textbf{b,}Calculated plot of the mean classification accuracy across all 6 sample classes, calculated across all perturbation trials. Plot is shown as a rainbow spectrum to correspond with simulated hyperspectral images shown in \textbf{c}. Plot is identical to that shown in the main text Fig. 4d. \textbf{c,} Simulated image showing a grid array of 25 droplet SEM. 23 droplets contain only RBCs and GNRs while 2 droplets contain a mix of \textit{S. epi} bacteria, RBCs, and GNRS. The image on the left shows the droplets containing only RBCs lighting up when interrogated with 816 nm light,  while the image on the right shows droplets containing a mixture of bacteria and RBCs lighting up when interrogated with light at 833 nm. Chosen wavelengths correspond to accuracy dips associated with each droplet class. Scale bar is 10 $\mu$m.}\label{fig}
\end{figure}

\newpage
\begin{figure}[H]%
\centering
\includegraphics[width=1\textwidth]{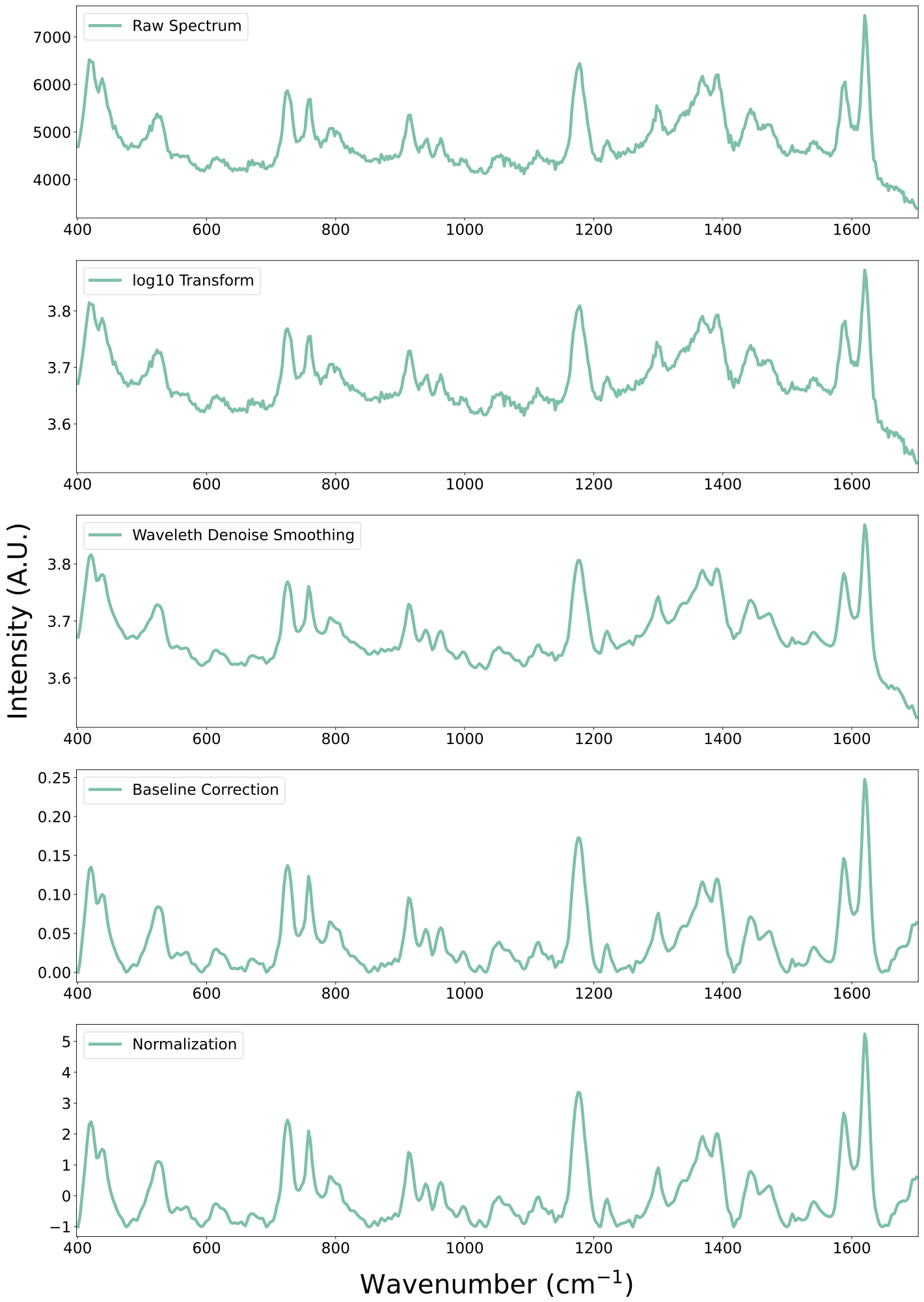}
\caption{Spectral Preprocessing. Plot showing a sample spectra taken from our dataset of spectra collected from droplets printed with \textit{E. coli} bacteria and GNRs. Plot shows (from top to bottom) the raw spectrum, the spectrum after a log\textsubscript{10} transformation, spectrum after smoothing using a wavelet denoising, spectrum with baseline correction, and finally the normalized spectrum.}\label{fig}
\end{figure}

%%===========================================================================================%
%% If you are submitting to one of the Nature Portfolio journals, using the eJP submission   %
%% system, please include the references within the manuscript file itself. You may do this  %
%% by copying the reference list from your .bbl file, paste it into the main manuscript .tex %
%% file, and delete the associated \verb+\bibliography+ commands.                            %
%%===========================================================================================%%
\bibliography{SI_bibliography}% common bib file
